\documentclass[aps,prb,twocolumn,nofootinbib,superscriptaddress]{revtex4-2}

\usepackage[a-1b]{pdfx}

\usepackage{dcolumn}
\usepackage{bm}
\usepackage{multirow}

\usepackage[normalem]{ulem}

\usepackage{amsmath}
\usepackage{amsthm}
\usepackage{amstext}
\usepackage{amssymb}
\usepackage{mathrsfs}
\usepackage{amsfonts}
\usepackage{amsbsy} 

\usepackage[all,matrix,cmtip]{xy}

\usepackage{color}

\usepackage{hyperref}
\hypersetup{
    colorlinks = true,
    allcolors = blue,
}

\definecolor{red}{rgb}{1,0,0}
\definecolor{blue}{rgb}{0,0,1}
\definecolor{dblue}{rgb}{0,0,0.4}
\definecolor{green}{rgb}{0,1,0}
\definecolor{black}{rgb}{0,0,0}
\definecolor{white}{rgb}{1,1,1}

\definecolor{brn}{rgb}{.8,.4,.0}
\definecolor{redo}{rgb}{1,.5,.0}
\definecolor{ddgrn}{rgb}{0,0.4,0}
\definecolor{dgrn}{rgb}{0,0.55,0}
\definecolor{dbl}{rgb}{0,0,0.5}

\usepackage[bbgreekl]{mathbbol}

\newcommand{\onebb}{\mathbb{1}}

\newcommand{\Z}{\mathbb{Z}}

\newcommand{\R}{\mathbb{R}}

\newcommand{\ii}{\hspace{1pt}\mathrm{i}\hspace{1pt}}
\newcommand{\ee}{\hspace{1pt}\mathrm{e}}

\newcommand{\<}{\langle} 
\renewcommand{\>}{\rangle} 

\newcommand{\Rf}[1]{Ref.~\onlinecite{#1}}

\newcommand{\ie}{{\it i.e.~}} 
\newcommand{\eg}{{\it e.g.~}}

\newcommand{\bpm}{\begin{pmatrix}}
\newcommand{\epm}{\end{pmatrix}}
\newcommand{\bmm}{\begin{matrix}}
\newcommand{\emm}{\end{matrix}}

\newcommand{\cA}{ {\cal A} }

\newcommand{\cD}{ {\cal D} }

\newcommand{\cH}{ {\cal H} }

\newcommand{\cR}{ {\cal R} } 
\newcommand{\cS}{ {\cal S} }

\usepackage{euscript}

\newcommand\eD           {\EuScript{D}}

\newcommand{\al}{\alpha}

\newcommand{\ga}{\gamma}

\newcommand{\om}{\omega}

\newcommand{\si}{\sigma}

\newcommand{\frmbox}[1]{\begin{center}\fbox{\parbox{3.3in}{\parindent=0pt #1}}\end{center}}

\usepackage{draft}

\usepackage{booktabs}
\usepackage{enumitem}
\usepackage{graphicx}
\usepackage{caption}
\usepackage{subfigure}
\usepackage{relsize}
\usepackage{booktabs}

\usepackage{adjustbox}

\usepackage{letltxmacro}
\LetLtxMacro{\Oldcite}{\cite}
\renewcommand{\cite}[2][]{\mbox{\Oldcite[#1]{#2}}} 

\newcommand{\ax}{e_x}
\newcommand{\ay}{e_y}
\newcommand{\az}{e_z}
\newcommand{\ak}{e_t}
\newcommand{\au}{e_{xt}}
\newcommand{\av}{e_{yt}}
\newcommand{\aw}{e_{zt}}
\newcommand{\aX}{m_x}
\newcommand{\aY}{m_y}
\newcommand{\aZ}{m_z}
\newcommand{\aU}{m_{xt}}
\newcommand{\aV}{m_{yt}}
\newcommand{\aW}{m_{zt}}
\newcommand{\aI}{f_{xt}}
\newcommand{\aJ}{f_{yt}}
\newcommand{\aK}{f_{zt}}
\newcommand{\aL}{f_x}
\newcommand{\aM}{f_y}
\newcommand{\aN}{f_z}
\newcommand{\aT}{s_t}
\newcommand{\aD}{s_t'}

\newcommand{\sx}{\begin{pmatrix}0 & 1\\1 & 0\end{pmatrix}}

\newcommand{\sz}{\begin{pmatrix}1 & 0\\0 & -1\end{pmatrix}}
\newcommand{\Sz}{\begin{pmatrix}\tfrac12 & 0\\0 & -\tfrac12\end{pmatrix}}

\newcommand{\isz}{\begin{pmatrix}\ii & 0\\0 & -\ii\end{pmatrix}}
\newcommand{\misz}{\begin{pmatrix}-\ii & 0\\0 & \ii\end{pmatrix}}
\newcommand{\isx}{\begin{pmatrix}0 & \ii\\\ii & 0\end{pmatrix}}
\newcommand{\isy}{\begin{pmatrix}0 & 1 \\ -1 & 0\end{pmatrix}}

\newcommand{\cRep}{\cR\mathrm{ep}}

\begin{document}
\title{Emanant and emergent symmetry-topological-order from
low-energy spectrum }

\author{Zixin Jessie Chen}
\affiliation{Department of Physics, Massachusetts Institute of Technology,
Cambridge, Massachusetts 02139, USA}

\author{\"Omer M. Aksoy}
\affiliation{Department of Physics, Massachusetts Institute of Technology,
Cambridge, Massachusetts 02139, USA}

\author{Cenke Xu}
\affiliation{Department of Physics, University of California, Santa Barbara, CA
93106, USA }

\author{Xiao-Gang Wen}
\affiliation{Department of Physics, Massachusetts Institute of Technology,
Cambridge, Massachusetts 02139, USA}

\begin{abstract}
    Low-energy  emanant and emergent symmetries can be anomalous, higher-group, or
non-invertible. A way to systematically capture the properties of such symmetries is through
the topological orders in one-higher dimension, known as symmetry topological orders (symTOs).
Consequently, identifying the emergent or emanant symmetry of a system is not
simply a matter of determining its group structure, but rather of computing the
corresponding symTO.  In this work, we develop a method to compute the symTO of
1+1D systems by analyzing their low-energy spectra under closed boundary
conditions with all possible symmetry twists. Following this approach, we show
that the gapless antiferromagnetic (AF) spin-$1/2$
Heisenberg model possesses an exact emanant symTO corresponding to the $D_8$
quantum double $\eD(D_8)$, when the global symmetry is 
restricted to the $\mathbb{Z}_2^x \times \mathbb{Z}_2^z$
subgroup of the $SO(3)$ spin-rotation symmetry and lattice translations. 
Moreover, this model exhibits an emergent $SO(4)$ symmetry, 
whose exact components are described jointly by automorphisms of  
$\eD(D_8)$ and the $SO(3)$ spin-rotations.  Using the condensable algebras of the
emanant symTO, we
further identify several other phases that may be accessible by modifying
interactions among low-energy excitations:  (1) a gapped dimer phase, connected
to the AF phase via an $SO(4)$ rotation, (2) a commensurate collinear
ferromagnetic phase that breaks translation by one site with a $\omega \sim k^2$ mode, (3) an incommensurate, translation-symmetric
ferromagnetic phase featuring both $\omega \sim k^2$ and $\omega \sim k$ modes,
(4) and an incommensurate ferromagnetic phase that breaks translation by one site
with both $\omega \sim k^2$ and $\omega \sim k$ modes.

\end{abstract}

\maketitle

\setcounter{tocdepth}{1}
{\small \tableofcontents }

\section{Introduction} \label{sec:intro}

Symmetry is one of the most important properties of a physical system: it
constrains the low-energy behavior, the possible nearby phases, 
and the transitions between them. Given a lattice model,
the corresponding low-energy effective theory may exhibit more symmetries, 
which can be \emph{emanant} and/or \emph{emergent}~\cite{CS221112543}. 
An emanant symmetry originates from an exact lattice symmetry, 
and can take a very different form in the low-energy effective theory; 
an emergent
symmetry is an additional symmetry that is present only at low energies.
In recent years,
it has become clear that emanant and emergent symmetries can go beyond ordinary symmetries
\cite{MT170707686,W181202517,BZ250113866}: they can combine ordinary (0-form) symmetries (described by
groups), higher-form and higher-group symmetries
\cite{NOc0605316,NOc0702377,KT13094721,GW14125148}, anomalous ordinary
symmetries \cite{H8035,CGL1314,W1313,KT14030617,WS170302426,MT170707686},
anomalous higher symmetries
\cite{KT13094721,GW14125148,TK151102929,T171209542,P180201139,DT180210104,BH180309336,ZW180809394,WW181211968,WW181211967,GW181211959,WW181211955},
non-invertible 0-form symmetries (in 1+1D)
\cite{PZh0011021,CSh0107001,FSh0204148,FSh0607247,FS09095013,DR11070495,BT170402330,CY180204445,TW191202817,KZ191213168,I210315588,Q200509072},
non-invertible higher (``algebraic higher'') symmetries
\cite{KZ200308898,KZ200514178,HV201200009,KZ211101141,CS211101139,BT220406564,FT220907471},
and/or non-invertible gravitational anomalies
\cite{KW1458,FV14095723,M14107442,KZ150201690,KZ170200673,JW190513279},
including anomaly-free/anomalous non-invertible higher-symmetries described by
fusion higher-categories
\cite{JW191213492,KZ200308898,KZ200514178,FT220907471}. These generalized
symmetries go beyond groups, and even beyond higher groups and standard anomaly
notions, yet they can all be described by topological orders in one-higher
dimension. In fact, topological orders with a gapped boundary classify all
finite generalized symmetries (up to holo-equivalence \cite{JW191213492,KZ200514178},
also known as Morita equivalence) in one lower dimension. We therefore refer to
such bulk data as \emph{symmetry-topological order} (symTO) or
\emph{symmetry-topological field theory} (symTFT).

A connection between boundary symmetry and bulk topological order was first
observed in \Rf{LW0605}, where the topological entanglement entropy was shown
to originate from a boundary conservation law rooted in the bulk topological
order; this was later confirmed numerically \cite{YS13094596}. A systematic
symmetry/topological-order correspondence was developed via the holographic
picture of emergent non-invertible gravitational anomalies
\cite{KZ150201690,KZ170501087,JW190513279} and of dualities
\cite{FT180600008,PV190906151}, leading to a holographic understanding of
generalized symmetries
\cite{JW191213492,LB200304328,KZ190504924,KZ200514178,
	GK200805960,AS211202092,FT220907471}.
This holographic point of view has been applied to understand 
gapped and gapless phases of matter with generalized 
symmetries~\cite{TW191202817,KZ200514178,CW220506244, MT220710712, CW221214432,wen2023classification,BBP231217322,BW240300905,CAW240505331,BIT240509754,BT240505302,LZ240612151,W240805801,BPS240805266,ACS240805585,BS240902166,PLA240918113,BST250220440,BW250312699,AW250321764,W250313685,PAL250702036,STZ250705350}.

With this holographic viewpoint, determining a system’s low-energy
emergent or emanant symmetry is no longer just a computation of a group
structure. Instead, one needs to identify the corresponding symTO in one-higher
dimension. This requires a new type of calculation and is the main focus of
this paper. As an illustrative example, we study the spin-$1/2$ antiferromagnetic
Heisenberg chain with $SO(3)$ spin-rotation and translation symmetries, and compute its
low-energy symTO (equivalently, the associated braided tensor category).

Because the symTO for the continuous spin-rotation symmetry $SO(3)$ is not yet
fully developed, we first tackle a simpler problem by focusing 
on the discrete $\Z_2^x$, $\Z_2^y$, and $\Z_2^z$ symmetries generated by
$\pi$-rotations about $S_x,S_y,S_z$. Thus, we consider the spin-$1/2$ XYZ
model with translation symmetry near the isotropic limit:
\begin{align} H \;=\; \sum_{j=1}^L \bigl(J_x S^x_j S^x_{j+1} + J_y S^y_j S^y_{j+1} +
J_z S^z_j S^z_{j+1}\bigr), \end{align} with $J_x,J_y,J_z \sim 1$ (\ie
$|J_x-J_y|,\ |J_y-J_z|,\ |J_x-J_z| \ll J_x,J_y,J_z$). The lattice symmetry of
the XYZ model is $\Z_2^x \times \Z_2^z \times \Z_L$, where $\Z_L$ denotes
translations.

We compute the low-energy emanant symmetry at energies $\ll J_x,J_y,J_z\sim 1$.
The emanant symmetry includes the discrete subgroup $\Z_2^x\times \Z_2^z\subset SO(3)$
and an additional low-energy emanant $\Z_2^t$ symmetry. 
The latter $\Z^t_2$ symmetry emanates from the $\Z_L$ lattice translations.
This is because when $J_x,J_y,J_z\sim 1$ 
the low-energy excitations are centered around 
quasi-momenta $k=0$ and $k=\pi$, and as a result translations by 
two-lattice spacing act trivially. However, the emanant
symmetry is not simply described by a symTO corresponding to 
ordinary $\Z_2^x\times \Z_2^z \times \Z_2^t$ symmetry. Instead, we find
that the emanant symmetries are encoded by the $D_8$ quantum double,
$\eD(D_8)$. This conclusion follows from the low-energy spectra of the model
under various symmetry-twist boundary conditions. Our method for extracting
symTO from low-energy spectra is general and applies to other 1+1D system.

What does it mean for the low-energy effective theory of the XYZ model to have
an exact symmetry described by the symTO $\eD(D_8)$? A useful viewpoint is that
a symTO corresponds to a collection of symmetries which form a holo-equivalence
class. The low-energy dynamics is constrained by every symmetry in this class.
The holo-equivalence class of $\eD(D_8)$ contains a $\Z_2^x\times \Z_2^z\times
\Z_2^{zt}$ symmetry with a type-III mixed anomaly (here $\Z_2^{zt}$ is
generated by the combined action of $\Z_2^z$ and $\Z_2^t$).  Therefore, the
emanant symmetry of the XYZ model is $\Z_2^x\times \Z_2^z\times \Z_2^{zt}$ with
a type-III mixed anomaly.

Although the emanant symmetry of the XYZ chain can be described by a group with
an anomaly, the holographic $\eD(D_8)$ symTO point of view is still very
useful.  It allows us to use the Lagrangian condensable algebras of $\eD(D_8)$
to systematically obtain all possible gapped phases with the same symmetry. 
Using the symTO $\eD(D_8)$  we identify 11 gapped phases.

The spin-$1/2$ $SO(3)$-symmetric Heisenberg chain likewise exhibits the emanant
symTO $\eD(D_8)$ at low energies. Within $\eD(D_8)$, we identify an $S_3$ subgroup of
its $S_4$ automorphism group that permutes the $\Z_2^x$, $\Z_2^y$, and $\Z_2^z$
symmetries. This $S_3$ is realized as a subgroup of the $SO(3)$ spin
rotations in the microscopic lattice model and is an exact symmetry.

The full $S_4$ automorphism is not an exact emanant symmetry of the Heisenberg
chain. Nevertheless, it emerges as an approximate symmetry that becomes
increasingly accurate at low energies. Combining the exact $SO(3)$
spin-rotation symmetry with this emergent $S_4$, one recovers the well-known
emergent anomalous $SO(4)$ symmetry of the $SO(3)$ Heisenberg chain
\cite{affleck-1987,sandvik-2018}. From the symTO viewpoint, this emergent
anomalous $SO(4)$ symmetry must contain the anomaly-free emanant $SO(3)$
spin-rotation symmetry together with the emanant symTO $\eD(D_8)$. Since
$\eD(D_8)$ can be viewed as a realization of $\Z_2^x\times \Z_2^z\times
\Z_2^{zt}$ with a type-III mixed anomaly, this suggests that
\frmbox{the full emanant symmetry of the $SO(3)$ Heisenberg chain is
$SO(3)\times \Z_2^t$ with a mixed anomaly, which constitutes the exact part of
the emergent anomalous $SO(4)$ symmetry.}

We can refine this statement by relating the anomalies of the emergent $SO(4)$
and the emanant $SO(3)\times \Z_2^t$ symmetries. 
The group $SO(4)$ can equivalently be written  as $(SU_R(2)\times SU_L(2))/\Z_2$, 
whose anomaly is characterized by a cocycle in $H^3(SO(4);\R/\Z)=H^3(SU_L(2)\times
SU_R(2)/\Z_2;\R/\Z)=\Z^L\times \Z^R$. Thus $SO(4)$ anomalies are labeled by
$(k_L,k_R)\in \Z^2$. The $SO(3)\times \Z_2^t$ anomaly lies in $H^3(SO(3)\times
\Z_2^t;\R/\Z)=\Z\times \Z_2$, where the $\Z_2$ factor is due to the mixed
anomaly; we label accordingly these anomalies by $(k,\sigma)\in \Z\times \Z_2$.
Under this restriction, an $SO(4)$ anomaly $(k_R,k_L)$ reduces to an $SO(3)\times
\Z_2^t$ anomaly as
\begin{align}
\label{eq:anomaly map}
(k,\sigma)=(k_R+k_L,\ k_R \!\!\!\pmod{2}). 
\end{align}

In our case, the emergent $SO(4)$ symmetry of the Heisenberg chain carries
anomaly $(k_R,k_L)=(1,-1)$, which gives rise to the $\mathfrak{su}(2)_1\times
\overline{\mathfrak{su}(2)}_1$ Kac-Moody algebra describing decoupled gapless right- and
left-movers. Under the map \eqref{eq:anomaly map}, this reduces
to an $SO(3)\times \Z_2^t$ anomaly $(k=0,\sigma=1)$. The antiferromagnetic Heisenberg
chain can also be described by a $(1+1)$D nonlinear sigma model with target space
$S^2$ and a topological $\Theta$-term; the $\Z_2$ nature of $\sigma$ precisely
matches the two symmetry-preserving choices $\Theta = 0,\pi$.

An important application of the emanant and emergent symmetries described above
is to constrain the phases neighboring the gapless phase of the Heisenberg
chain. Knowing the symTOs of the emanant and emergent symmetries allows us to
use condensable algebras in these symTOs to systematically enumerate the
neighboring gapped and gapless phases. We carry out this analysis for the XYZ
chain with symTO $\eD(D_8)$ in Sec.~\ref{sec:d8}, and apply it to the $SO(3)$
Heisenberg chain and its neighboring phases in Sec.~\ref{neighbor}.

\section{Heisenberg model and its phase diagram} \label{sec:model}

\subsection{Definitions and symmetries}

We consider the quantum spin-$1/2$ antiferromagnetic Heisenberg chain
\begin{subequations}
\label{eq:Ham Heisenberg}
\begin{equation}
H = \sum_{j=1}^L \mathbf S_j \cdot \mathbf S_{j+1},
\end{equation}
where the components $\mathbf S_j=(S_j^x, S_j^y, S_j^z)$ satisfy the algebra
\begin{align}
\left[S^a_j, S^b_k\right]=\mathrm{i}\, \delta_{jk} \epsilon^{abc}S^c_j,
\end{align}
\end{subequations}
which satisfy the periodic boundary conditions on the spin operators, \ie,
$S^a_{j+L} \equiv S^a_j$ for $a=x,y,z$.  Later on, we will also consider
boundary conditions with discrete internal symmetry twists.

While the Hamiltonian \eqref{eq:Ham Heisenberg} has continuous $SO(3)$ spin-rotation symmetry,
we will focus on the Abelian subgroup $\Z^x_2\times\Z^z_2\subset SO(3)$ generated by the
$\pi$-rotations around $x$ and $z$ axes.
The elements of this subgroup are then represented by the unitary operators
\begin{align}
R_a = \prod_{j=1}^{L} e^{\mathrm{i} \pi S^a_j} =  \prod_{j=1}^{L} \mathrm{i}\sigma^a_j,
\end{align}
where $\sigma^a_j$ are the Pauli matrices with $a=x,y,z$.

In addition to these internal symmetries, the Hamiltonian \eqref{eq:Ham Heisenberg} has the translation
symmetry implemented by the unitary operator $T: S^a_j \mapsto S^a_{j+1}$,
which generates the group $\Z_L$.
As we shall see, these subsymmetries capture the most important universal features
of the low-energy excitations.

Because each translation unit cell hosts a half-odd-integer spin, the Lieb-Schultz-Mattis (LSM) theorem~\cite{lieb-1961, oshikawa-2000}
and its extensions~\cite{OT180808740,OTT200406458,YO201009244} preclude a non-degenerate gapped
ground state that is simultaneously invariant under  $\Z^x_2\times\Z^z_2\times \Z_L$ symmetry.
In the language of field theory, this ingapability condition can be thought of resulting from a mixed anomaly between $\Z_L$ translation symmetry and the internal
$\Z^x_2\times\Z^z_2$ symmetry~\cite{CCR17050389,CS221112543,AMT230800743,seifnashri-2024,
PAL250702036}, which we refer to as the LSM anomaly.

\subsection{Adjacent gapped phases}
\label{subsec:adjacent gapped phases}

Consistently with the LSM anomaly, Hamiltonian \eqref{eq:Ham Heisenberg}
has a gapless spectrum. Its low-energy properties are described by the
$SU(2)_1$ Wess-Zumino-Witten (WZW) conformal field theory (CFT), \ie,
the compact boson CFT with central charge $c=1$
at the self-dual radius $R=\sqrt{2}$ \cite{eggert-1996,Okamoto-1992}.
The Hamiltonian density for this CFT is
\begin{equation} \label{eq:h-luttinger}
\cH= \frac{v}{2} \, \left[\frac{1}{K}\,(\partial_x\theta)^2 + K\,(\partial_x \phi)^2\right]
\end{equation}
where  $\theta$ and $\phi$ are dual compact boson fields with $\theta\sim\theta + \sqrt{2\pi}$,
$\phi\sim\phi + \sqrt{2\pi}$, and the coupling $K=1$.

\begin{figure}[t]
\subfigure[]{\includegraphics[width=0.47\columnwidth]{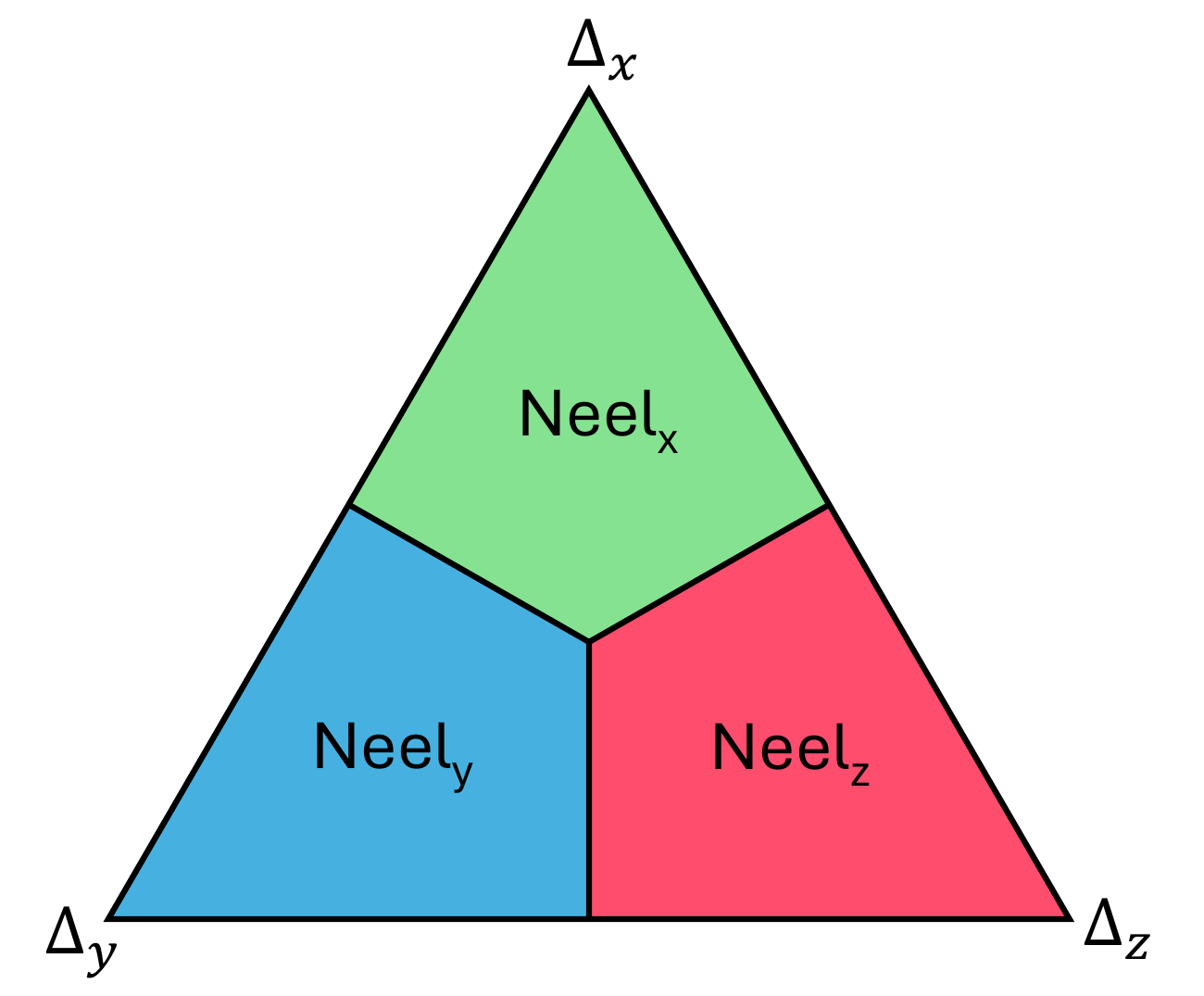}}
\hfill
\subfigure[]{\includegraphics[width=0.47\columnwidth]{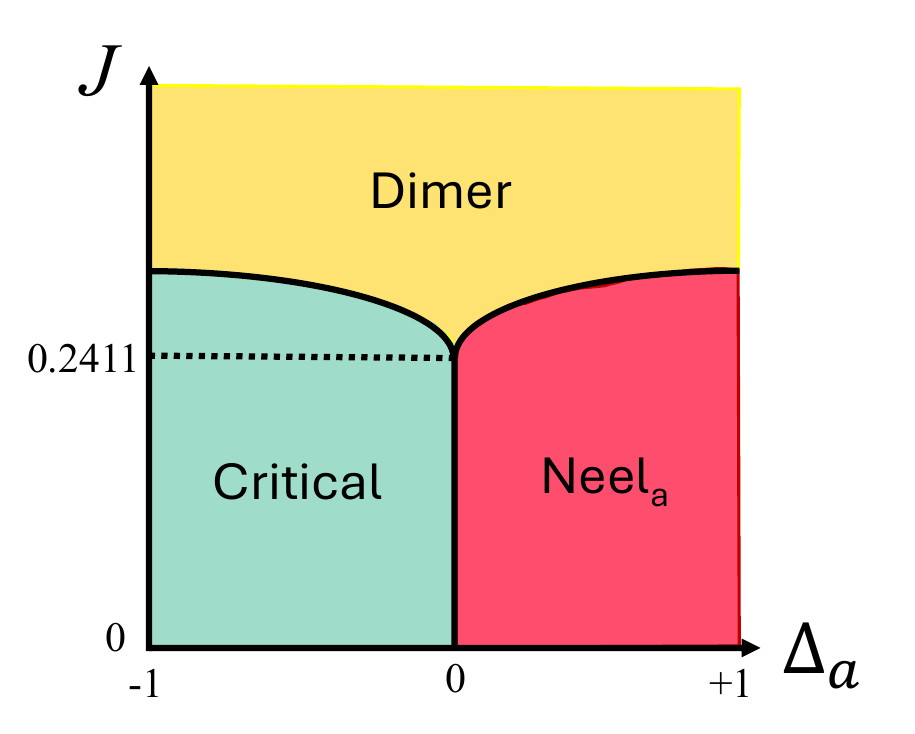}}
\caption{Two-dimensional slices of the phase diagram of antiferromagnetic Heisenberg chain \eqref{eq:Ham Heisenberg} perturbed by the
deformations \eqref{eq:NNN deformation} and \eqref{eq:pert-neel-1}. (a) Three N\'eel phases at the slice $J=0$ and $\sum_a \Delta_a \leq 1$, with $0\leq \Delta_a$ and the center of the triangle
being the point $\Delta_a=0$. (b) Dimer and $\text{N\'eel}_a$ phases at the two-dimensional slice with $J, \Delta_a \neq 0$.}
\label{fig:phase}
\end{figure}

Because of the underlying LSM anomaly, any gapped phase stabilized by symmetric deformations
breaks the $\Z^x_2\times\Z^z_2$ internal or $\Z_L$ translation symmetries spontaneously, see Table~\ref{tab:phases}.
There are four such adjascent gapped symmetry-breaking phases~\cite{affleck-1987, affleck-1989, nersesyan-1993, mudry-2019},
as shown in Fig.~\ref{fig:phase}.

\begin{table}[t]
\begin{tabular}{l|c|c|c|c|c}
\toprule
Phase & $\mathbb Z_2^{x}$ & $\mathbb Z_2^{y}$ & $\mathbb Z_2^{z}$ & $\Z_L$ &  GS degeneracy \\
\hline
N\'eel$_x$ & $\checkmark$ & $\times$ & $\times$ & $\times$  & 2 \\
N\'eel$_y$ & $\times$ & $\checkmark$ & $\times$ & $\times$  & 2 \\
N\'eel$_z$ & $\times$ & $\times$ & $\checkmark$ & $\times$ & 2 \\
Dimer      & $\checkmark$ & $\checkmark$ & $\checkmark$ & $\times$ & 2 \\
\bottomrule
\end{tabular}
\caption{Spontaneous symmetry breaking patterns in the four gapped phases, where each checkmark indicates that the symmetry is unbroken.}
\label{tab:phases}
\end{table}

\paragraph{Dimer phase.}
Tuning up the antiferromagnetic next-nearest-nerighbor exchange term
\begin{align}
\label{eq:NNN deformation}
\Delta H = J \sum_{j=1}^L \mathbf S_j \cdot \mathbf S_{j+2}, \quad J>0
\end{align}
generates at low-energy limit the perturbation
\begin{align}
\label{eq:NNN coupling at CFT}
\delta \mathcal{H} \sim \lambda_\phi \cos\bigl(\sqrt{8\pi} \, \phi(x)\bigr), \quad  \lambda_\phi < 0,
\end{align}
with scaling dimension $\Delta = 2/K$, which is marginal when $K=1$.
The field theory for the spin-1/2 Heisenberg chain can also be written as
\begin{align}
\label{eq:WZWfield}
S = S_{\rm WZW} + \lambda \int dx d\tau {\bm J}_L \cdot {\bm J}_R
\end{align}
$S_{\rm WZW}$ is the WZW model action for the $SU(2)_1$ CFT. There is a critical coupling $J_c\sim 0.2411$, and both $\lambda_\phi$ and $\lambda$ are proportional to $J - J_c$. When $J > J_c$ ($J < J_c$), $\lambda$ is marginally relevant (irrelevant). Since $\lambda$ is the only term that breaks the $SO(4)$ symmetry of the WZW theory, at the critical coupling $J = J_c$, the system is tuned to the point where the $SO(4)$ symmetry emerges even at short distance, as the leading $SO(4)$ breaking term is tuned to zero, and the other terms breaking the $SO(4)$ symmetry are irrelevant.

The $SO(4)$ symmetry of Eq.~\eqref{eq:WZWfield} with $\lambda = 0$ will become manifest in the ED data to be discussed in the next few sections. In fact the model has an even larger $O(4)$ symmetry, whose improper $\mathbb Z_2$ rotation corresponds to the reflection of the chain. In field theory this improper $\mathbb Z_2$ exchanges ${\rm SU(2)}_L$ and ${\rm SU(2)}_R$. Since it is not clear how to insert a flux of reflection, the improper $\mathbb Z_2$ will not be discussed on the equal footing as the rest of the symmetry.


When $J>J_c$, the perturbation \eqref{eq:NNN coupling at CFT} opens a gap in the spectrum
and pins the value of the field $\phi$ to the values $\phi=0,\,\sqrt{2\pi}/2$.
There are two degenerate ground states that are distinguished by the non-vanishing expectation value of the
dimer order parameter
\begin{equation} \label{eq:op-dimer}
m_d = \frac{1}{L} \sum_{j=1}^L (-1)^j \ \mathbf{S}_j \cdot \mathbf{S}_{j+1}.
\end{equation}
This order parameter preserves the internal $\Z^x_2\times\Z^z_2$ symmetry while breaking the $\Z_L$
translation symmetry by one lattice site.

\paragraph{N\'eel phases.}
The remaining three gapped phases are stabilized by the easy-axis anisotropy couplings
\begin{equation}
	\label{eq:pert-neel-1}
\Delta H_a = \Delta_a \sum_{j=1}^L (S^a_j S^a_{j+1} + J S^a_j S^a_{j+2}),
\end{equation}
with $a=x,y,z$, which preserve the $\Z^x_2\times\Z^z_2\times\Z_L$ symmetry while breaking the $\mathrm{SO(3)}$ spin-rotation symmetry
down to a $\mathrm{U}(1)_a$ symmetry along $a$-axis. For $\Delta_a>0$, these perturbations open a gap in the spectrum with
twofold degenerate ground states that are distinguished by the non-vanishing expectation
value of the order parameters
\begin{equation} \label{eq:op-neel}
m_a = \frac{1}{L} \sum_{j=1}^L (-1)^j \ S^a_j,
\end{equation}
which preserves the $\Z^a_2$ while breaking both $\Z^b_2$ ($b\neq a)$
and $\Z_L$ symmetries spontaneously. Importantly, the composition $R_b\, T$ of the spontaneously
broken spin-flip operators $R_b$ and translation $T$ symmetries is also preserved.

At low-energy limit, the anisotropy terms \eqref{eq:pert-neel-1} have different effects for
$a=x,y,z$~\footnote{Here we follow the convention of \Rf{mudry-2019}
when identifying perturbations \eqref{eq:pert-neel-1} at the lattice level with their
continuum counterparts.}. When $\Delta_z>0$, the term \eqref{eq:pert-neel-1}
renormalizes the parameter $K$ to $K'\approx K + c\,\Delta_z$ with a positive constant $c$
and leads to the perturbation
\begin{equation} \label{eq:pert-neel-2}
\delta \mathcal{H} \sim g_a \cos\bigl(\sqrt{8\pi} \, \phi(x)\bigr), \quad g_a > 0,
\end{equation}
which is now relevant with scaling dimension $\Delta = 2/K'<2$.
This term opens a gap by pinning the field $\phi$ to the values $\phi=\sqrt{2\pi}/4,\, 3\sqrt{2\pi}/4$,
which describes the  $\text{N\'eel}_z$ phase.

When $\Delta_b>0$ ($b=x,y$), the term \eqref{eq:pert-neel-1}
renormalizes the parameter $K$ to $K'\approx K - c\,\Delta_b$
with a positive constant $c$
and leads to the relevant perturbation
\begin{equation} \label{eq:pert-neel-2}
\delta \mathcal{H} \sim g_b \cos\bigl(\sqrt{8\pi} \, \theta(x)\bigr),
\end{equation}
with scaling dimension $\Delta = 2K' <2$. The coupling $g_b<0$ ($g_b>0$) when $b=x$ ($b=y$), and, hence, the
dual field $\theta$ is pinned to the values  $\theta=0,\, \sqrt{2\pi}/2$ ($\theta=\sqrt{2\pi}/4,\, 3\sqrt{2\pi}/4$)
which describes the  $\text{N\'eel}_x$ ($\text{N\'eel}_y$) phase.

\section{Low-energy excitations from exact diagonalization} \label{sec:ed}

We aim to match the low-energy excitations of spin-$\frac12$ antiferromagnetic
Heisenberg chain with the anyon content of an appropriate symTO.  To do so, we
are going to perform exact diagonalization (ED) with $\Z^x_2\times\Z^z_2$
symmetry twisted boundary conditions for different system sizes.  This allows one to organize the low-energy excitations in terms of
irreducible representations (irreps) of appropriate global (emergent or
emanant) symmetries in distinct flux sectors.  For exact lattice symmetries,
such irreps consist of exactly degenerate states, while for emergent symmetries
they would be approximately degenerate.

\begin{figure*}[t!]
\subfigure[$L = 20$]{
\includegraphics[width=0.9\columnwidth]{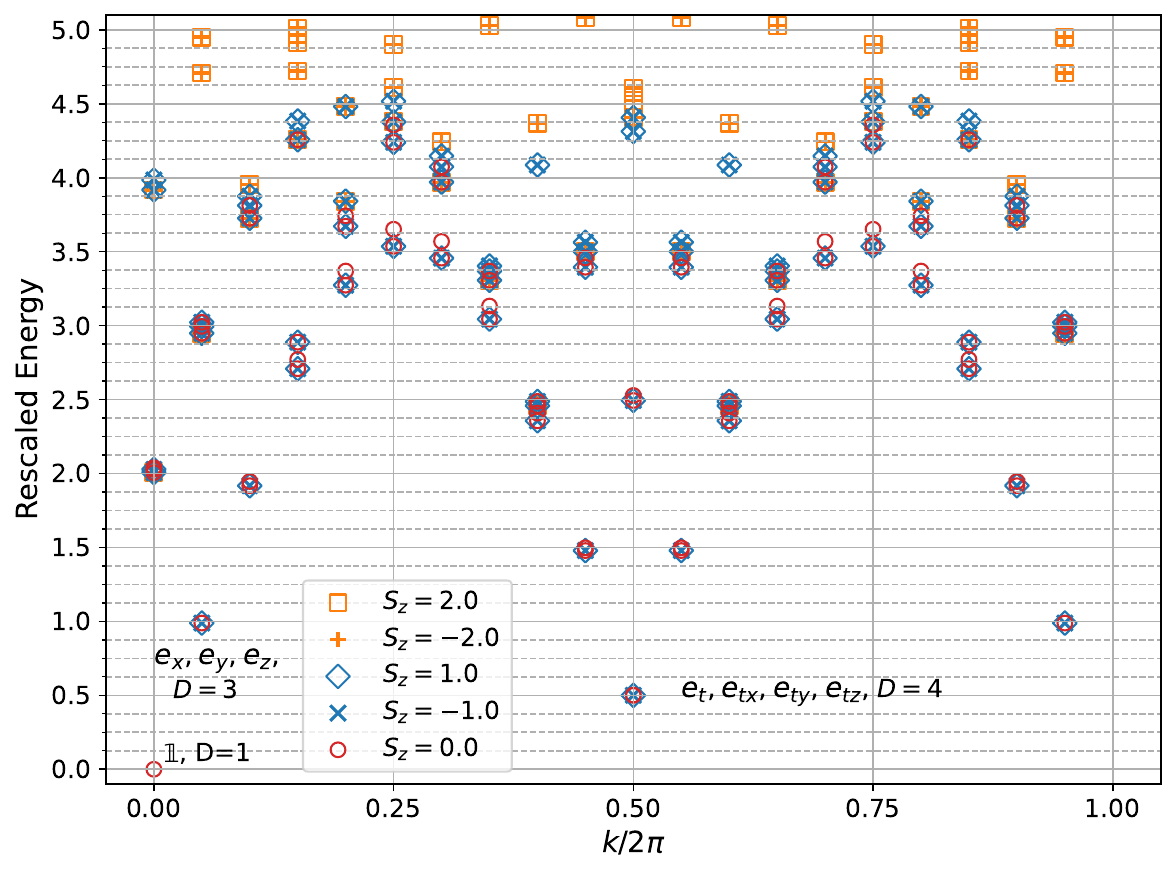}
\label{fig:pbc-a}
}
\subfigure[$L = 21$]{
\includegraphics[width=0.9\columnwidth]{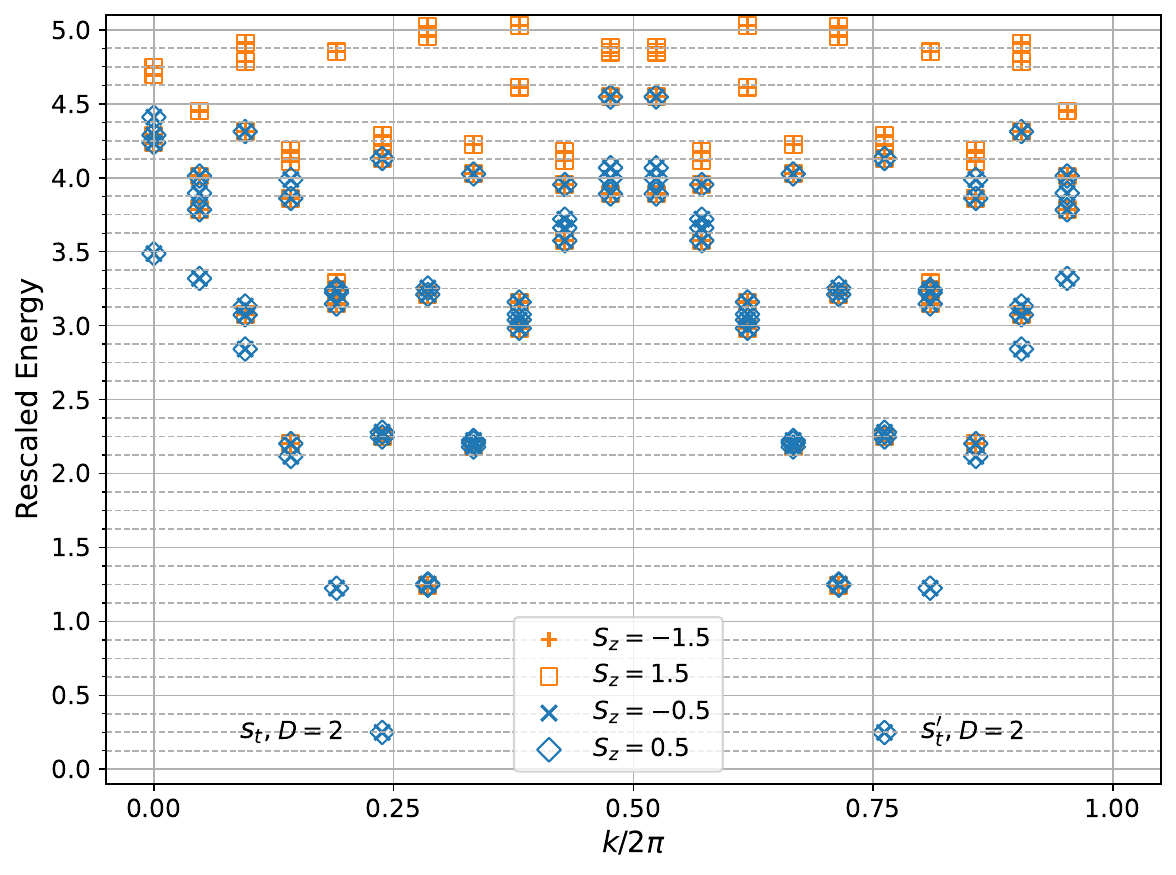}
\label{fig:pbc-b}
}

\subfigure[$L = 22$]{
\includegraphics[width=0.9\columnwidth]{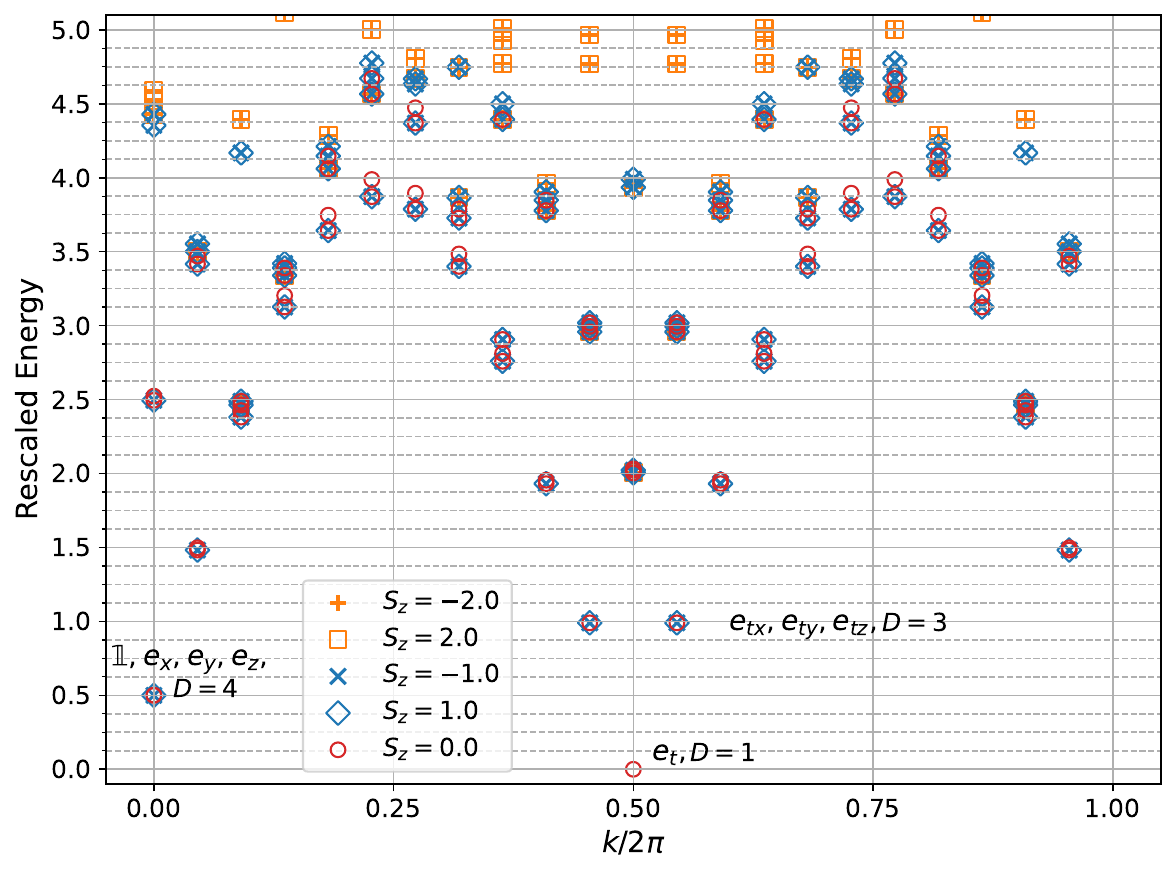}
\label{fig:pbc-c}
}
\subfigure[$L = 23$]{
\includegraphics[width=0.9\columnwidth]{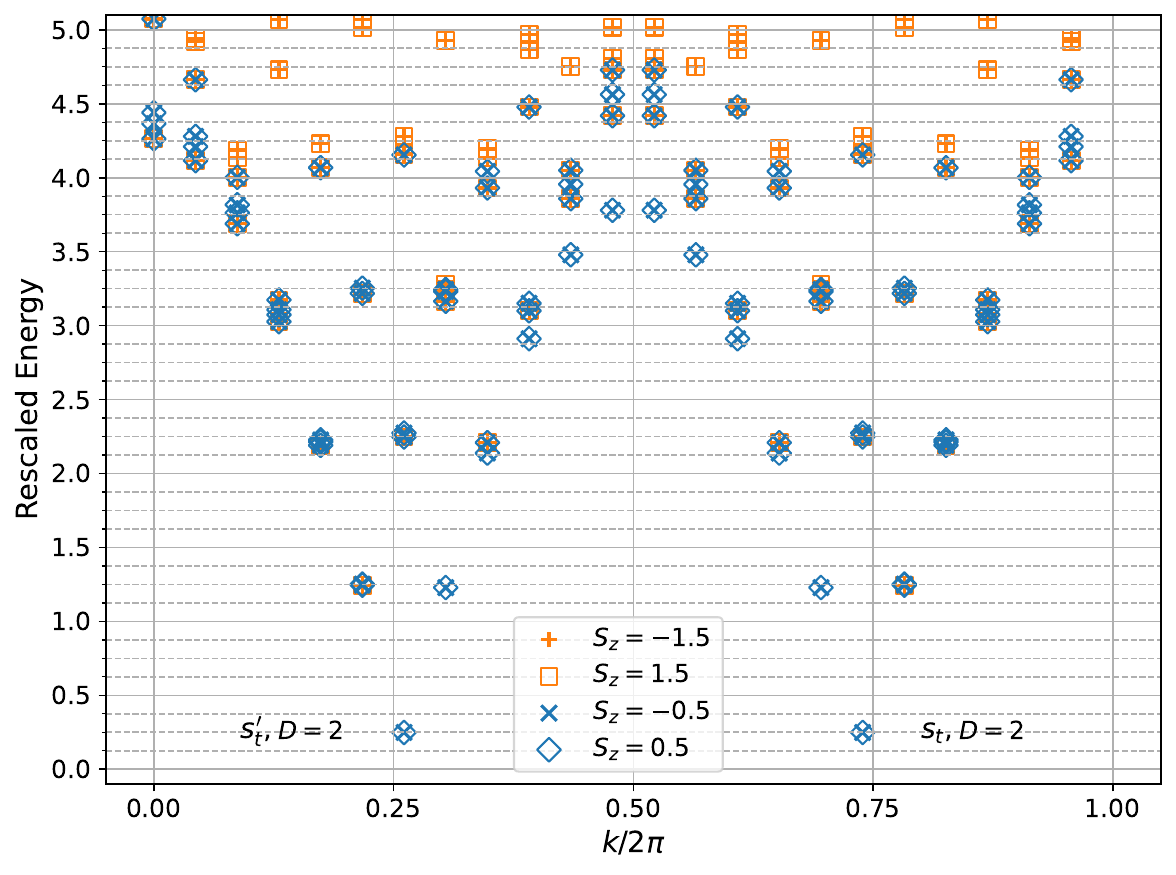}
\label{fig:pbc-d}
}

\caption{ED spectra for the Heisenberg model with PBC for $L=20,21,22,23$.
The momentum is normalized by $2 \pi$. We label the first few low-energy
eigenstates with their corresponding anyons and degeneracies (denoted as $D$).
The full lattice symmetry is $SO(3) \times \mathbb{Z}_L$. But as we noted
the lattice model at $J = J_c$ has good emergent SO(4) symmetry even at relatively
short distance, and this SO(4) symmetry leads to the nearly fourfold degeneracy
for some of the low-energy states.}
\label{fig:pbc}
\end{figure*}

Because of the LSM anomaly in the $\Z^x_2\times\Z^z_2\times\Z_L$ symmetry, the
possible multiplet structure of the low-energy irreps is already constrained.
More precisely, first we note that the internal symmetry group
$\Z^x_2\times\Z^z_2$ obeys the (projective) algebra
\begin{align}
R_x\,R_z = (-1)^L\, R_z\, R_x.
\end{align}
The dependence on the number of sites $L$ means that the internal symmetry group
is represented by non-trivial projective representation when $L$ is odd. In
this case, every eigenstate is at least doubly degenerate, \ie, the entire
spectrum can be organized into two-dimensional projective irreps of the
$\Z^x_2\times\Z^z_2$.  Similarly,  imposing $\Z^a_2$-twisted boundary conditions
by inserting a symmetry defect between sites $L$ and $1$ modifies the
translation operator as
\begin{subequations}
\begin{align}
T_a = e^{\mathrm{i} \pi S^{a}_1} T,
\end{align}
which now satisfies the algebra
\begin{align}
T^L_a = R_a,\quad
T_a\, R_b = - R_b\, T_a,\quad b\neq a.
\end{align}
\end{subequations}
This means that  (i) the global symmetry group is now changed to
$\Z^b_2\times Z^a_{2L}$ and (ii) this group is represented
projectively.

\begin{figure*}[t!]
\subfigure[$L = 20$]{
\includegraphics[width=0.9\columnwidth]{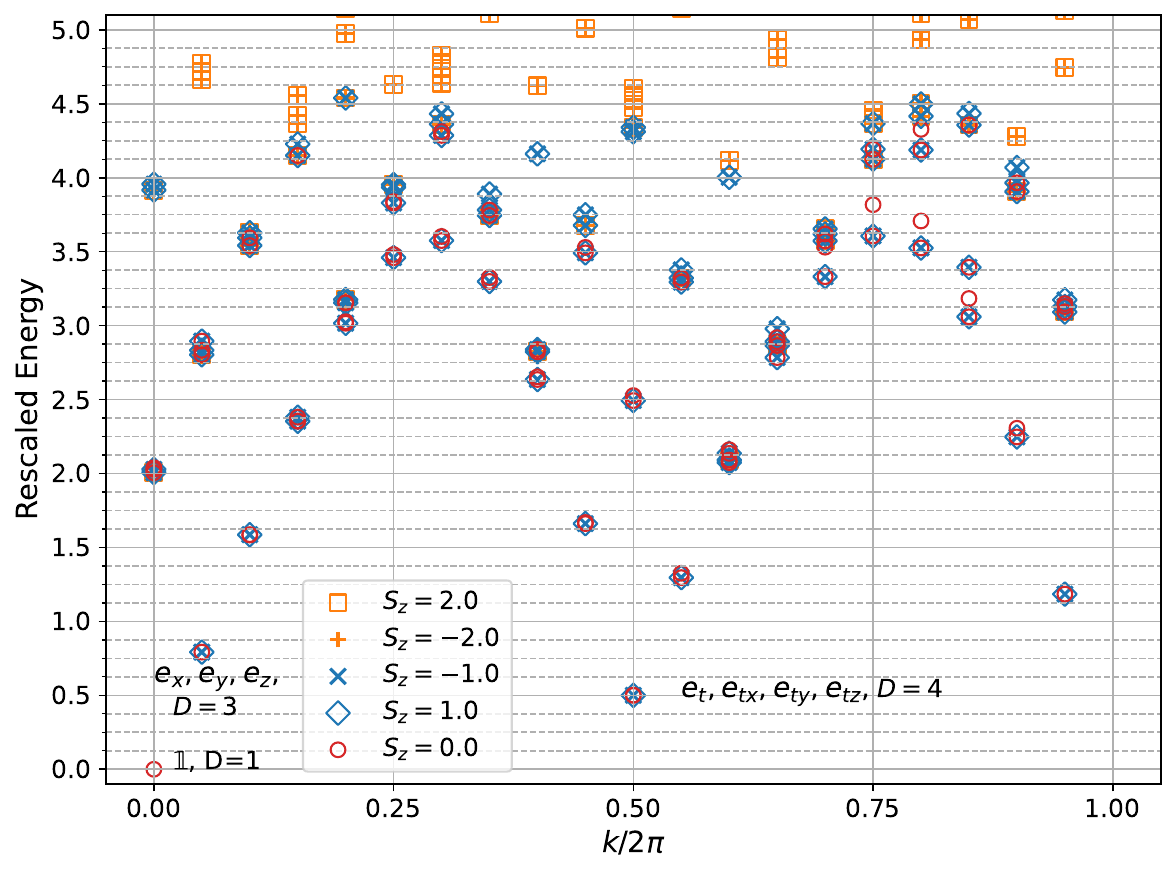}
\label{fig:pbc-chiral-a}
}
\subfigure[$L = 21$]{
\includegraphics[width=0.9\columnwidth]{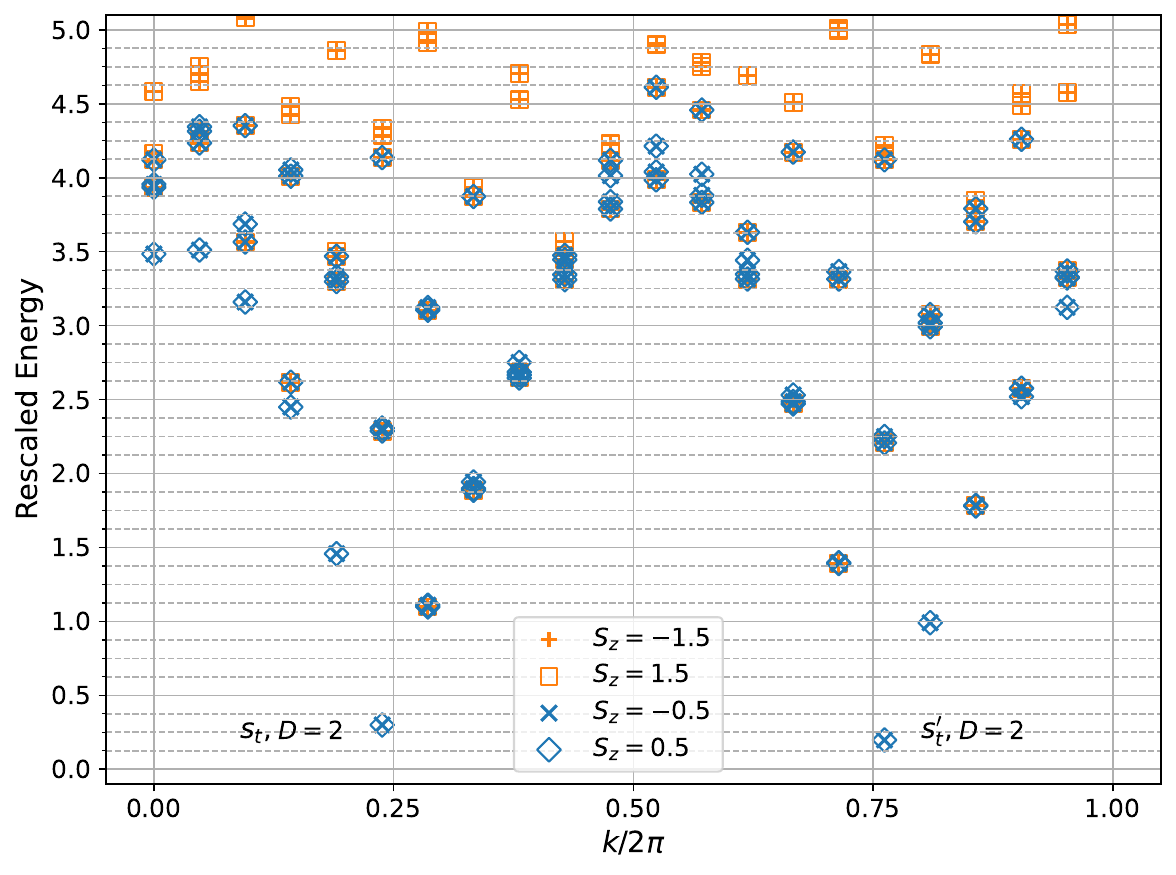}
\label{fig:pbc-chiral-b}
}

\subfigure[$L = 22$]{
\includegraphics[width=0.9\columnwidth]{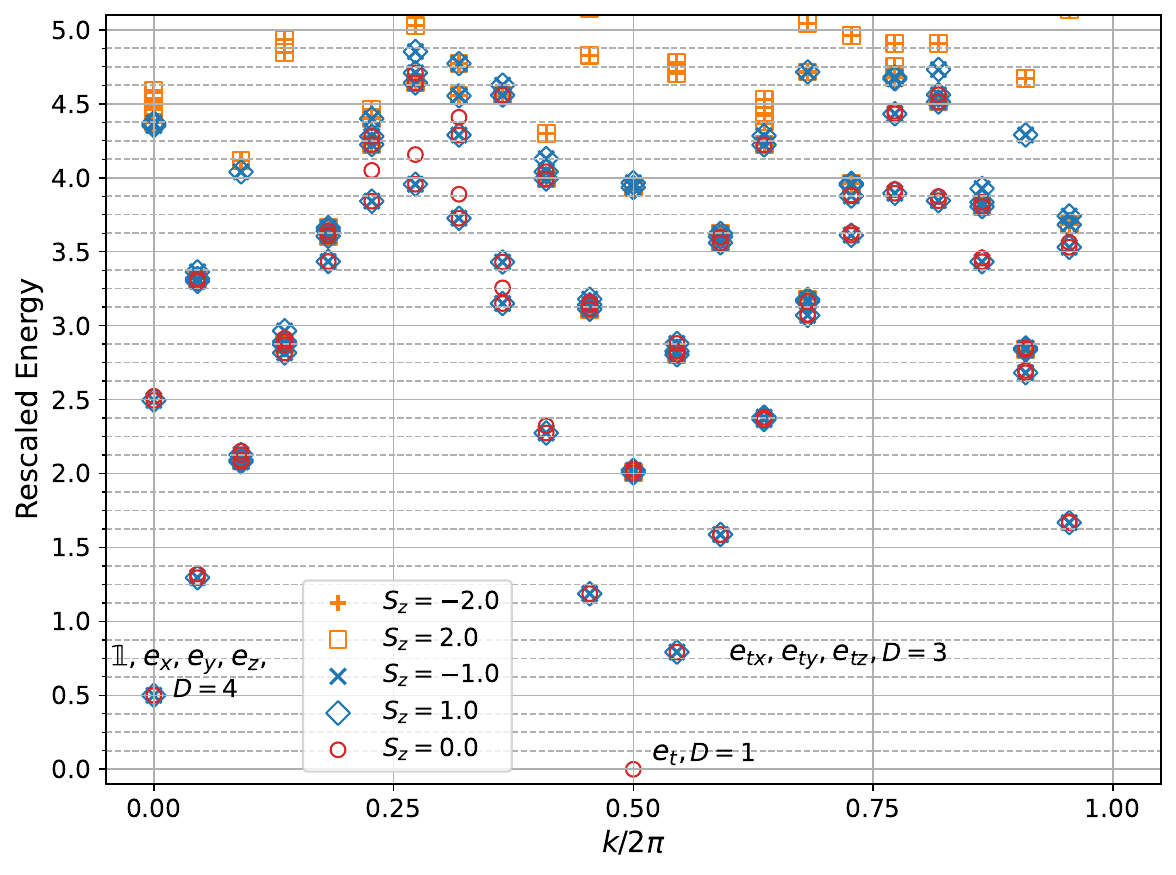}
\label{fig:pbc-chiral-c}
}
\subfigure[$L = 23$]{
\includegraphics[width=0.9\columnwidth]{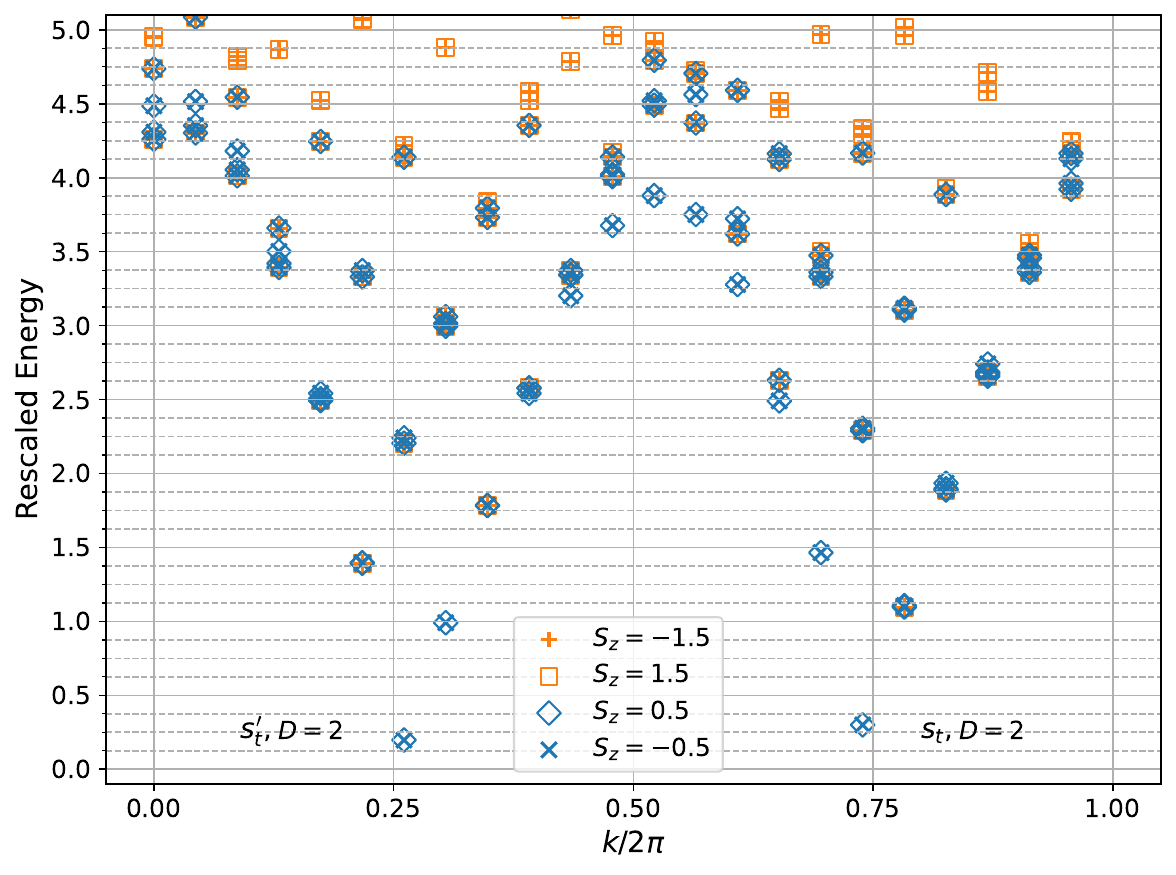}
\label{fig:pbc-chiral-d}
}

\caption{ED spectra for the Heisenberg model under PBC with a chiral perturbation for $L=20,21,22,23$.
The chiral perturbation term is $0.1\sum_j \mathbf S_j \cdot (\mathbf S_{j+1} \times \mathbf S_{j+2})$.
The degeneracy of states labeled by $(k, S_z^\text{tot})$ is preserved. However, as the improper $\Z_2$ reflection symmetry is broken,
6-fold degeneracy of states with momentum $k$ and $-k$ is lifted.}
\label{fig:pbc-chiral}
\end{figure*}

As we discussed previously, an O(4) symmetry already emerges at short distance
in the Hamiltonian we simulate. The O(4) symmetry also manifests in the ED spectrum:
some of the low energy states constitute nearly degenerate quadruple states, as they
form a vector representation of the SO(4). The extra improper $\Z_2$ rotation
(reflection) of O(4) ensures the degeneracy between states with momentum $k$ and $-k$.
We have also verified that a perturbation in the Hamiltonian that breaks
reflection will lift the degeneracy between $k$ and $-k$, but keeps the
degeneracy of the quadruple states. See Fig.~\ref{fig:pbc-chiral} for the ED spectra.

We note that under any boundary condition, translation by two sites, $j\mapsto
j+2$ commutes with all internal symmetries. As we shall see, on the low-energy
degrees of freedom the translation by two sites will flow to pure continuum
translation symmetry while there will be an exact $\Z^t_2$ symmetry that
emanate from translation by one site~\cite{JW210602069,CS221112543}. Coseting
by translation by two sites, we are going to identify the low-energy emanant
symmetry group as $\Z^x_2\times\Z^z_2\times \Z^t_2$.

\subsection{The spectrum calculation}

With the above understanding, we perform ED and organize the spectrum into the
irreps of $\Z_2^x\times \Z_2^z\times \Z_2^t $ symmetry.  More specifically, we
diagonalize the Hamiltonian
\begin{align}
\label{eq:h-pbc}
H = \sum_{j=1}^L \mathbf S_j \cdot \mathbf S_{j+1} + J\sum_{j=1}^L \mathbf S_j \cdot \mathbf S_{j+2},
\end{align}
with the coupling $J$ tuned to the value $J=0.2411$. While the low-energy properties are
described by the same CFT for $0\leq J\leq J_c$,  this particular choice tunes the lattice model to
the multicritical point at the intersection of four gapped phases described in Sec.\ \ref{subsec:adjacent gapped phases}
and makes the ED spectrum closer to the tower of excitations in the CFT.

To cover all the flux sectors of $\Z^x_2\times\Z^z_2$
symmetry we are going to impose either periodic boundary conditions (PBC)
\begin{subequations}
\begin{align}
S^a_{j+L} \equiv S^a_j,
\end{align}
or $\Z^z_2$-twisted boundary conditions (TBC)
\begin{align}
S^a_{j+L} \equiv R_z\, S^a_j\, R^\dag_z.
\end{align}
\end{subequations}
Without loss of generality, we only consider the boundary conditions
twisted by $\Z^z_2$ symmetry since the remaining non-trivial twists by elements
in $\Z^x_2\times \Z^z_2$ amounts to a permutation of labels due to the spin rotation symmetry
of Hamiltonian \eqref{eq:h-pbc}.
Under PBC, the momentum $k$ is defined by
\begin{equation} \label{eq:T}
T \ket{\psi_k} = \ee^{\ii k} \ket{\psi_k},
\end{equation}
with $k=2\pi n/L$ and $n=0,1,\cdots,L-1$.
For TBC, the appropriate quantum number $k$ is the eigenphase of the twisted
translation $T_z$,
\begin{equation} \label{eq:T-twist}
T_z \ket{\psi_k} \equiv \ee^{\ii  \pi S^z_1} T \ket{\psi_k} = \ee^{\ii k} \ket{\psi_k},
\end{equation}
with with $k=2\pi n/2L$ and $n=0,1,\cdots,2L-1$.
which commutes with the $\mathbb{Z}^z_2$-twisted Hamiltonian.

The low-energy spectra
of Hamiltonian \eqref{eq:h-pbc} is periodic in $L$ modulo $4$~\footnote{In general, one performs ED for different system size $L$ until a repetition in the spectrum patterns appears.}, see Fig.\ \ref{fig:pbc} or~\cite{JW210602069}.
For both PBC and TBC, we compute the spectra for chains with $L = 20, 21, 22,
23$, thereby sampling every congruence class $L \equiv 0, 1, 2, 3 \pmod{4}$.
We are going to further interpret the case of odd $L$ as the insertion of emanant $\Z^t_2$
symmetry flux relative to the case of even $L$~\cite{CS221112543,SS230702534,SSS240112281}.
Hence, by performing ED with these four system sizes $L$ together with both PBC and TBC
we are going to organize the spectra of Hamiltoninan \eqref{eq:h-pbc} in terms of the irreps in every possible flux sector of emanant $\Z^x_2\times\Z^z_2\times \Z^t_2$ symmetry.


\begin{figure*}[t!]
\subfigure[$L = 20$]{
\includegraphics[width=\columnwidth]{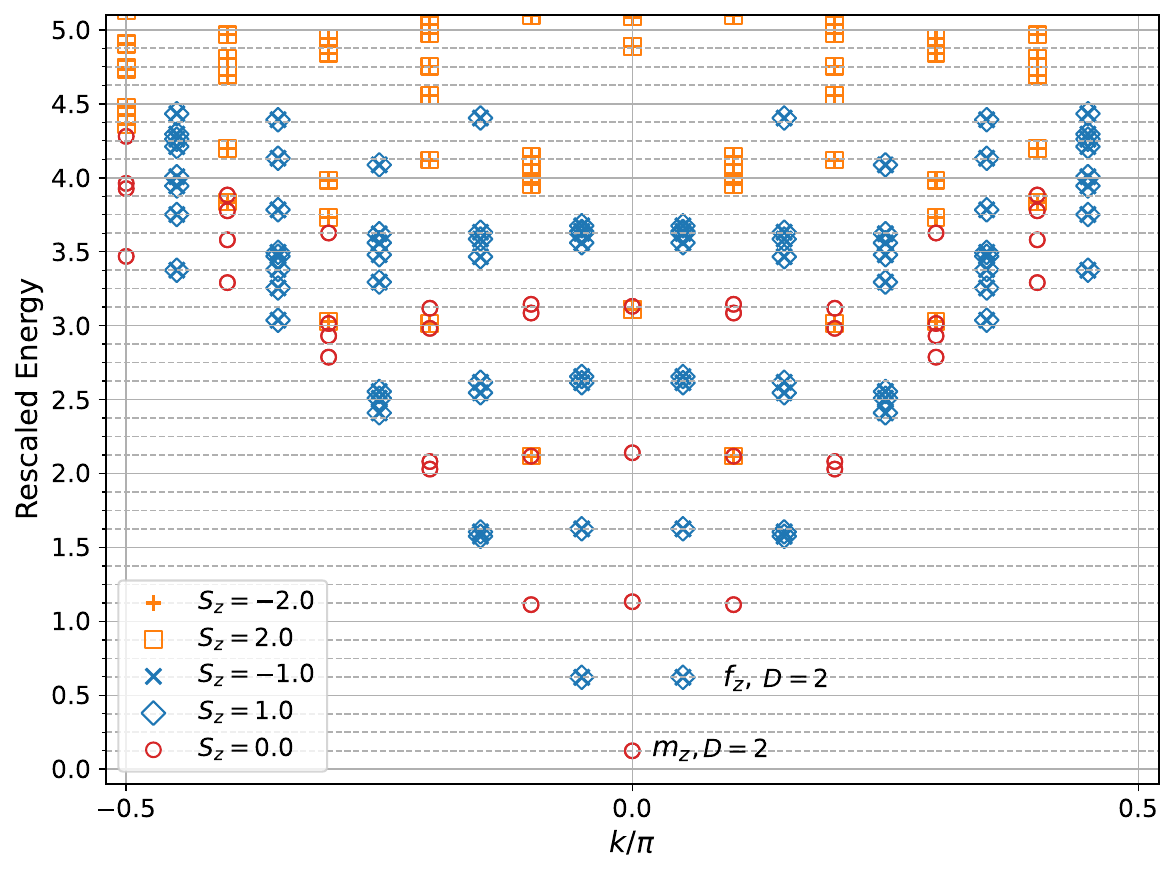}
\label{fig:tbc-a}
}\hfill
\subfigure[$L = 21$]{
\includegraphics[width=\columnwidth]{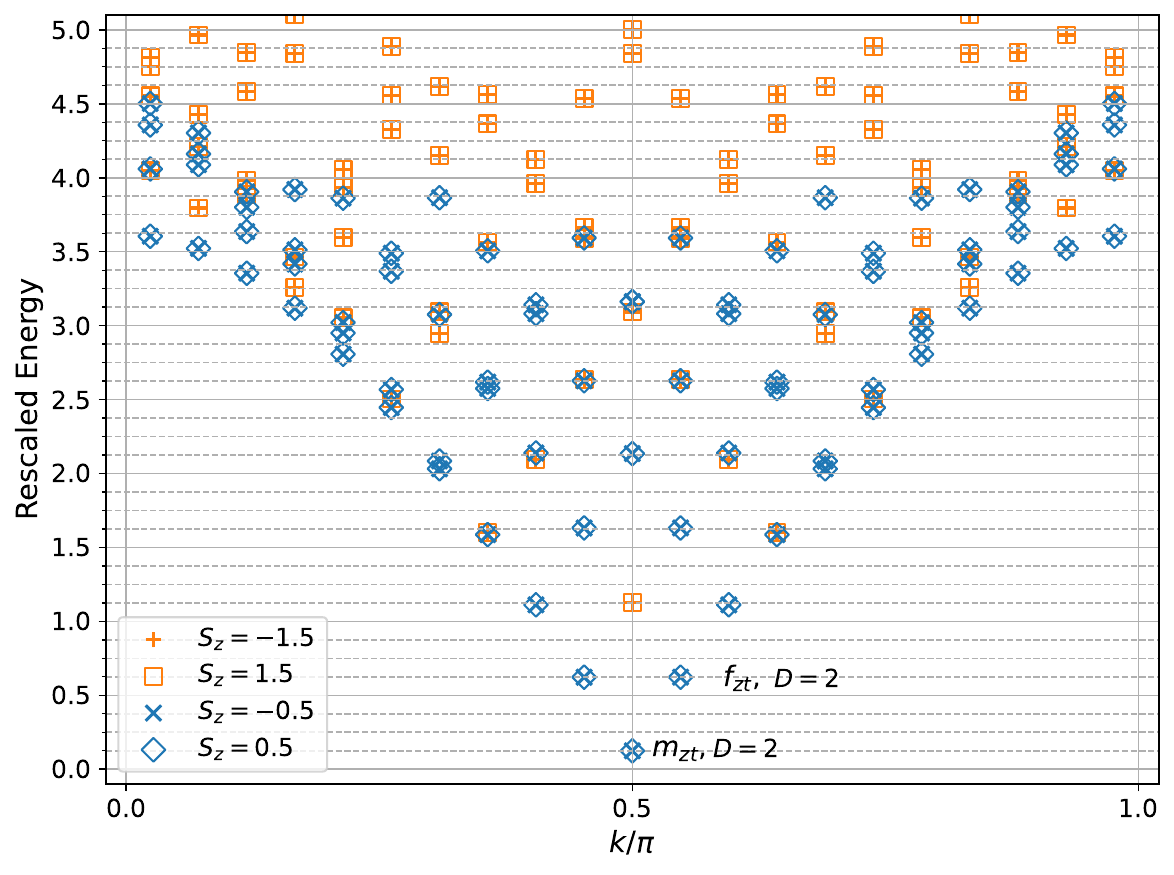}
\label{fig:tbc-b}
}

\vspace{1ex}

\subfigure[$L = 22$]{
\includegraphics[width=\columnwidth]{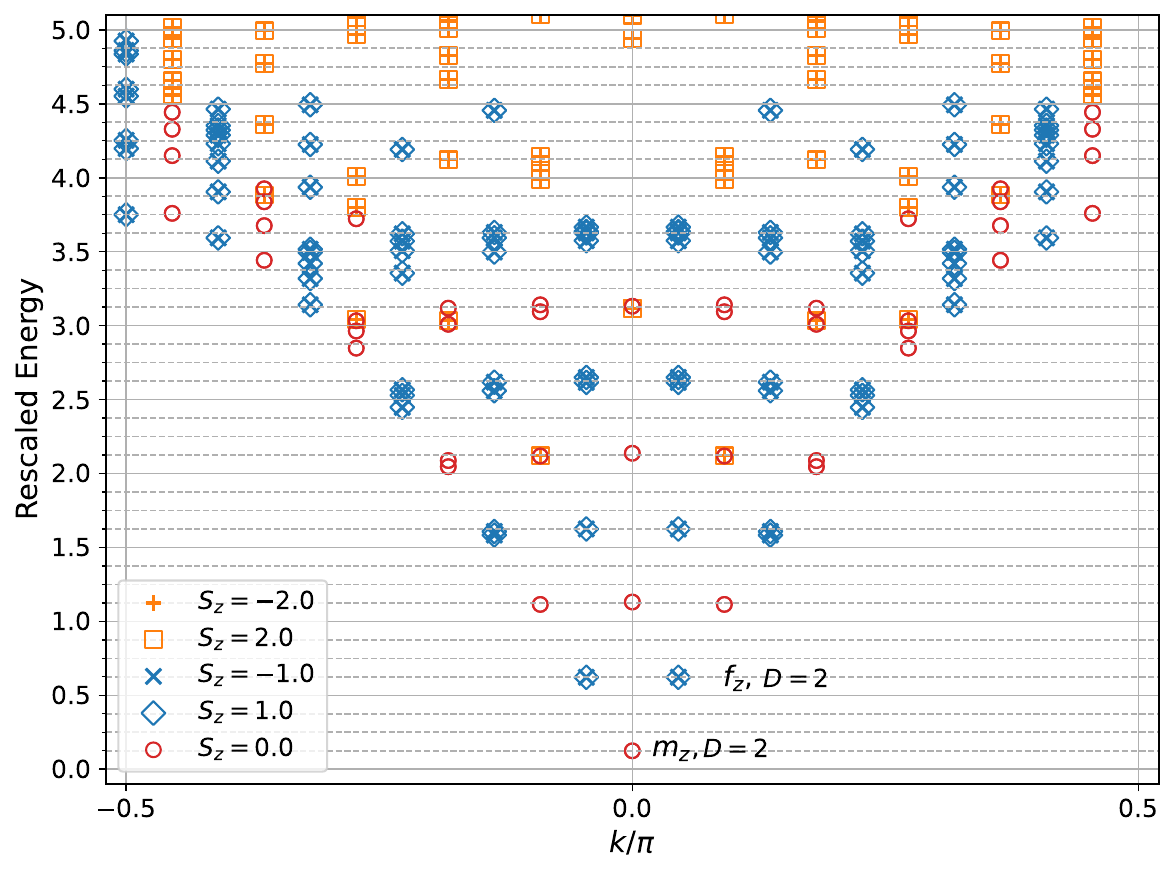}
\label{fig:tbc-c}
}\hfill
\subfigure[$L = 23$]{
\includegraphics[width=\columnwidth]{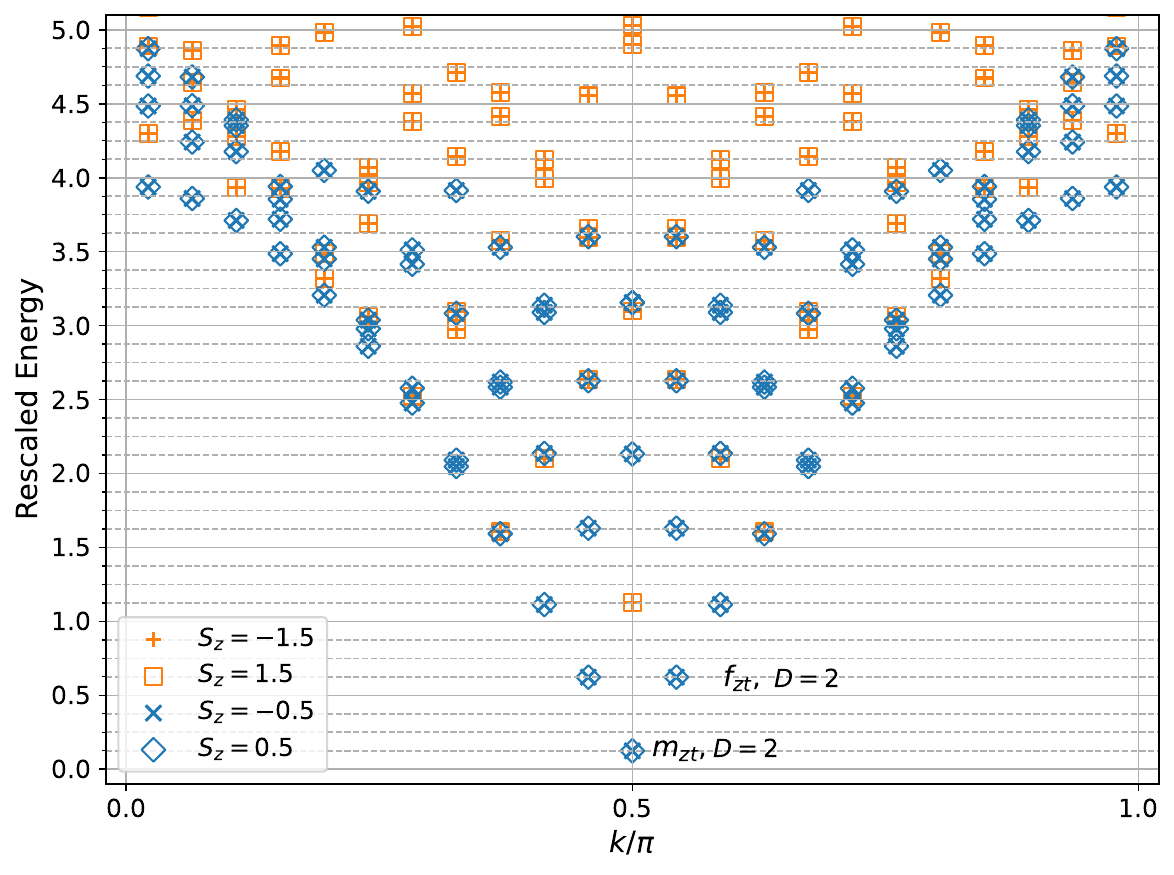}
\label{fig:tbc-d}
}
\caption{ED spectra for the Heisenberg model with $\mathbb{Z}^z_2$-twisted
boundary conditions for $L=20,21,22,23$. The first few low-energy
eigenstates are labeled with their corresponding anyons and degeneracies.
The momentum is normalized by $\pi$. The full lattice symmetry is $\frac{\mathrm{U}(1)_z \times \Z^z_{2L}}{\Z^z_2} \rtimes \mathrm{Z}^x_2$.}
\label{fig:tbc}
\end{figure*}

We also normalize the energy $E$ of each eigenstate to that of their conformal weight
usıng the CFT spectrum
\begin{equation} \label{eq:E(L)}
E(L) = \epsilon L + \frac{2 \pi v}{L} \left[\bigr(h + \bar{h} + N + \bar{N}\bigr) - \frac{c}{12} \right] + O({1}/{L^2}),
\end{equation}
where $\epsilon$ gives the energy density, $v$ denotes spinon velocity, and $c
= 1$ for the $\mathrm{SU}(2)_1$ WZW theory. The terms $(h + \bar{h})$ and $(N + \bar{N})$
give the conformal weights of the corresponding primaries and descendants,
respectively. For even $L$ under PBC, the ground state has total conformal
weight $h + \bar{h} = N + \bar{N}= 0$. Then using the groundstate energies at
$L = 20, 22, 24$, we obtain through fitting
\begin{subequations} \label{eq:fit-data}
    \begin{align}
        \epsilon &= -0.401954 \pm 0.000000, \\
        \quad v &= 1.176693 \pm 0.000131.
    \end{align}
\end{subequations}
These are bulk quantities and therefore apply to all boundary conditions.
Subtracting the extensive term and rescaling by $2 \pi v / L$ yields energy in
units of conformal weights deliver
\begin{equation} \label{eq:conformal-weight}
    \tilde{E}(L) \equiv \frac{L}{2 \pi v} [E(L) - \epsilon L] + \frac{c}{12} = h + \bar{h} + N + \bar{N},
\end{equation}
which we use in all plots of ED spectra.

\subsection{The spectra for PBC} \label{subsec:pbc}

The energy spectrum on a lattice is organized by the lattice symmetry.  For a
lattice with $L$ and PBC, the lattice symmetry group is $\Z_2^x\times
\Z_2^z\times \Z_L^t$, generated by the pairwise commuting operators $R_x$, $R_z$, and $T$.
In constrast, for a lattice with odd $L$ and PBC, the lattice symmetry group is $(\Z_2^x\times_\om
\Z_2^z)\times \Z_L^t$, generated by the operators $R_x$, $R_z$, and $T$.
Here, by subscript $\om$ we denote the 2-cocycle in the group cohomology class
$[\om]\in H^2(\Z^x_2\times\Z^z_2, \R/\Z)\cong \Z_2$
that specifies the nontrivial global projective representation of
$Z^x_2\times\Z^z_2$ due to the anticommutation
 $R_x \, R_z= -R_z\,R_x$~\footnote{From now on, we use $\times_\om$ to denote
 nontrivial projective algebra between generators of two subgroups.}.

Since the $\Z_L^t$ translation symmetry commutes with the
Hamiltonian, the momentum $k \pmod {2 \pi}$ is a good quantum number.
From the ED calculations, we observe that for even $L$, the low-energy excitations belong to two
branches centered around momentum sectors at $k = 0,\pi$ (see Fig.~\ref{fig:pbc}). The ground
state is a non-degenerate singlet with $S_z^\text{tot} = 0$. Its momentum is $k
= 0$ for $L = 0 \pmod 4$ and $k = \pi$ for $L = 2 \pmod 4$.
On even-$L$ chains, the translation $T$ and spin-flip operators $R_x,\, R_z$
all commute with one another, so every eigenstate can be
simultaneously labeled by the triplet $(R_x,  R_z, e^{\mathrm{i}k})$.

For odd $L$,  the low-energy excitations belong to two momentum sectors at $k =
\pi/2$ and $3\pi/2$ (see Fig.~\ref{fig:pbc}).
Finite-size quantization of picking $k \in \{2\pi n/L |\, n\in \mathbb{Z}_L\}$ shifts
groundstate momenta slightly off $\pi/2$ and $3\pi/2$ in the
spectra, which converge back to these values in the $L \rightarrow \infty$ limit.
In this case, the two spin-flip operators anticommute
($R_x R_z = - R_z R_x$) while both still commuting with  translations $T$.
This forces every energy level in a given momentum sector
to be at least twofold degenerate, \ie, each momentum sector consists of two-dimensional (2D)
projecive irreps of $\Z_2^x\times_\om\Z_2^z$ .
In particular, the groundstate
subspace is fourfold degenerate and realizes two copies of such 2D projective irreps
that carry spin-1/2, \ie,  $S_z^\text{tot} =\pm 1/2$.
In the groundstate subspace, we can express the symmetry operators $R_x,\, R_z,$ and $T$
as $4\times4$ matrices. We choose a basis  in which $T$ remains diagonal
and express the remaining symmetry operators $R_x$ and  $R_z$
as $2\times 2$ matrices in each momentum eigenstector, see
Table~\ref{tab:pbc}.

We note that low-energy sectors carry crystal momenta that always differ by
multiple of $\pi$. Thus the translation symmetry in the continuum limit is
generated by two-site translation $T^2$.  The one-site translation $T$, or more
precisely $T/\sqrt{T^2}$, becomes a $\Z_2^t$ internal symmetry in the continuum
limit. The different lattice size for PBC can then be viewed as different symmetry twists
of $\Z_2^t$ internal symmetry emanating from the
translation symmetry.  The fact that a different $\Z_2^t$ twist also
shifts the low-energy momenta by $k\approx \pm \pi/2$ is a signature of the fact that
the $\Z_2^t$ symmetry in the continuum is anomalous. In other words, the defect of the emanant $\Z^t_2$ symmetry acquires fractional charge of $\pm \mathrm{i}$ under itself, which is the case for anomalous $\Z_2$ 
symmetry in $1+1$D~\cite{LS190404833}.

Since the rescaled energies are in units of conformal weights, the spectra
show the existence of $\mathrm{SU}(2)_1$ WZW CFT's two Virasoro primaries in
the untwisted sector \cite{francesco-1997}. The identity $\mathbf{1}$ with $(h,
\bar{h}) = (0, 0)$ appear as the unique even-$L$ ground state, and the spinon
doublet $g^{\alpha \beta}$ with $(h, \bar{h}) = (\tfrac 1 4, \tfrac 1 4)$
appear as the first excited states in even-$L$ spectra. Furthermore, we are going to interpret
the deviation of the ground-state momenta from $ \pi/2$ and $3\pi/2$
as the topological spin of these excitations. These spins can be fractional -- see the
discussion in Section~\ref{sec:symto}.

\begin{table}[t]
\centering
\resizebox{\columnwidth}{!}{
\begin{tabular}{c|c|c|c|c|c|c}
\toprule
\(L\bmod4\) & Degeneracy & $\ee^{\ii k}$ & $R_x$ & $R_y$ & $R_z$ & $S_z^\text{tot}$ \\
\hline
$0$ & $1$ & $1$  & $+1$ & $+1$ & $+1$ & $0$ \\
\hline
\multirow{2}{*}{1}
& $2$ & $\ii$   & $\isx$ & $\isy$ & $\isz$ & $\Sz$ \\
\cline{2-7}
& $2$ & $-\ii$  & $\isx$ & $\isy$ & $\isz$ & $\Sz$ \\
\hline
$2$ & $1$ & $-1$ & $+1$  & $+1$  & $+1$  & $0$ \\
\hline
\multirow{2}{*}{3}
& $2$ & $\ii$   & $\isx$ & $\isy$ & $\isz$ & $\Sz$ \\
\cline{2-7}
& $2$ & $-\ii$  & $\isx$ & $\isy$ & $\isz$ & $\Sz$ \\
\bottomrule
\end{tabular}
}
\caption{Representations of symmetry operators in the groundstate subspace of Hamiltonian \eqref{eq:h-pbc} for fixed momentum eigensectors with PBC.}
\label{tab:pbc}
\end{table}

\subsection{The spectra for TBC}

Under the $\mathbb{Z}^z_2$-twisted boundary conditions for both $L$ even and odd,
the twisted translation
operator $T_z = \ee^{\ii  \pi S^z_1} T$ anticommutes with the internal
spin-flip symmetry generator $R_x$, \ie, $ R_x T_z = - T_z
R_x $ and satisfy $T_z^L =  R_z ,\ T_z^{2L}
=1$. Thus, the lattice symmetry group is $(\Z_2^x\times_\om \Z_{2L}^t)$,
generated by $ R_x$ and $T_z$.  Because of this projective algebra, one cannot simultaneously
diagonalize twisted translation and internal spin-flip symmetries. Instead,
we consider eigenvalues of the two-site twisted
translation $T_z^2$, which commutes with the internal spin-flip generator
$R_a$.  We therefore work in a reduced Brillouin
zone, identifying $k \equiv k + \pi$ (see Fig.  \ref{tab:tbc}), and treat the
one-site twisted translation $T_z$ as the emanant internal symmetry $\Z_2^t$.

For even $L$, only the twisted-momentum sector $k = 0$ mod $\pi$ appears.
$R_z$ commutes with both $R_x$ and $T_z$, so eigenstates
carry well-defined eigenvalues $(R_z, \ee^{\ii k})$. However,
the projective algebra
$T_z\, R_x= - R_x \, T_z$ forces every eigenenergy sector to be at
least twofold degenerate. The groundstates are twofold degenerate,
each with $S_z^\text{tot} = 0$. The first excited level has four degenerate eigenstates which split
into two distinct momenta, each of which are doubly degenerate
with $S_z^\text{tot} = \pm 1$.

For odd $L$,  only the momentum sector $k = \pi /2$  mod $\pi$ appears.
$R_z$ commutes with $T_z$ and both $R_z$ and $T_z$
anticommute with $R_x$. Thus, each level is twofold degenerate in both
$R_x$ and $R_z$, with $T_z$ and $R_z$ flipping the
$R_x$-eigenvalues and $R_x$ flipping the
$R_z$-eigenvalues. Note that because we identify $k \equiv k + \pi$,
$\ee^{\ii k} \equiv -\ee^{\ii k}$.
The twisted translation $T_z$ defines a
consistent momentum despite its anticommutation relations with spin-flip
symmetry generators $R_x$ and $R_z$.
The groundstates are twofold degenerate doublets, with $S_z^\text{tot} = \pm
1/2$. The first excited level again has four states, splitting into two
distinct momenta, each of which are doubly degenerate with $S_z^\text{tot} =
\pm 1/2$.

For both even and odd $L$, the groundstate subspace furnishes a 2D irrep of
the algebra generated by $\{R_x,\, R_z\}$. More concretely, for odd
$L$, the irrep is just realized by Pauli matrices on the doublet. For even $L$,
only $R_x$ and $T_z$ form a noncommuting pair, while $R_z$
acts trivially (proportional to identity) on the 2D groundstate subspace. The
groundstate properties of the TBC spectra are summarized in
Table~\ref{tab:tbc}.

Finally, inserting a $\mathbb{Z}^z_2$ defect selects the twisted sector
of the $c=1$ theory \cite{ginsparg-1988}. Its lowest Virasoro primary has
conformal weights $(h, \bar{h}) = (\tfrac{1}{16}, \tfrac{1}{16})$, exactly
matching the rescaled energy of the ED groundstates with TBC.

\subsection{Emanant symmetries and anomalies}
\label{typeIII}

We observe that the lattice symmetry groups depend on the system size and
the boundary conditions. As we discussed, for all these symmetry
groups, (twisted) translation by two-sites commutes with all other elements.
It is, thus, convenient to coset the lattice symmetry by the subgroup generated by two-site translations,
which reveals the emanant internal symmetry group~\cite{CS221112543}
that the low-energy degrees of freedom
transform under with possible anomalies. Identifying the emanant symmetry and its
anomaly is important as they control and constrain the dynamics of
the low-energy degrees of freedom.

\begin{table}[t]
\centering
\resizebox{\columnwidth}{!}{
\begin{tabular}{c|c|c|c|c|c|c}
	\toprule
	\(L\bmod4\) & Degeneracy & $\ee^{\ii k}$ & $R_x$ & $R_y$ & $R_z$ & $S_z^\text{tot}$ \\
	\hline
	$0$ & $2$ & $\sz$  & $\sx$ & $\sx$ & $+1$ & $0$ \\
	\hline
	$1$ & $2$ & $\misz$  & $\isx$ & $\isy$ & $\isz$ & $\Sz$ \\
	\hline
	$2$ & $2$ & $\sz$  & $\sx$ & $\sx$ & $+1$ & $0$ \\
	\hline
	$3$ & $2$ & $\misz$  & $\isx$ & $\isy$ & $\isz$ & $\Sz$ \\
	\bottomrule
\end{tabular}
}
\caption{Representations of symmetry operators in the groundstate subspace of Hamiltonian \eqref{eq:h-pbc}
with $\mathbb{Z}^z_2$-twisted boundary conditions.}
\label{tab:tbc}
\end{table}

For even $L$ with PBC, the emanant symmetry group is $\Z_2^x\times
\Z_2^z\times \Z_2^{zt}$, generated by commuting $ R_x,\, R_z,
R_z\,T$.  Since the emanant symmetry group is Abelian, the spectrum
is formed by 1D irreps.

For odd $L$ with PBC, the emanant symmetry group is $(\Z_2^x\times_\om
\Z_2^z)\times \Z_2^{zt}$, generated by $ R_x,\, R_z, R_z\,T$ with
anti-commuting $ R_x\, R_y= -R_y\, R_x $.

For even $L$ with TBC by $Z^z_2$, the emanant symmetry group is $\Z_2^z\times
(\Z_2^x\times_\om \Z_2^{zt})$, generated by $ R_x,\, R_z, R_z\,T_z$ with
anti-commuting $ R_x\, T_z = -T_z \, R_x $.

For odd $L$  with TBC by $Z^z_2$, the emanant symmetry group is $(\Z_2^x\times_\om \Z_2^z) \times \Z_2^{zt}$, generated by $ R_x,\, R_z,
R_z\,T_z, $ with anti-commuting $ R_x \, T_z = -T_z \, R_x $
and $ R_x \, R_z = - R_z \, R_x $.

All these latter three emanant symmetry groups are realized projectively, \ie,
only their 2D projective irreps appear in the spectrum~\footnote{All the symmetry groups mentioned above only have projective irrpes of dimension $2$.}. Each 2D projective
irrep can be
thought of as the 2D \emph{linear} irrep of the corresponding non-Abelian group $D_8$ that
arises from the group extension that trivialize the nontrivial second cohomology
class $[\om]$.

We identify the anomaly of the emanant symmetries as follows.
The emanant symmetry groups we have considered so far (for $L$ odd
or even with potential $\Z^z_2$ twist) consist of various $\Z_2$ subgroups.
One way to detect an anomalous $\Z_2$ subgroup is to insert
its symmetry defect and compute the ground state eigenvalue under it.
The symmetry defect of an anomalous $\Z_2$ symmetry carries fractional charge
under itself, leading to the symmetry operator squaring to $-1$ in the defect
sector~\cite{LS190404833}.
For example, consider the subgroup $\Z^z_2$. From Table \ref{tab:tbc},
one notes that for $L$ even, \ie, the case with only $\Z^z_2$ defect,
the $R_z$ operator acts as identity on the entire groundstate manifold. Thus,
this subgroup is non-anomalous. Because of the permutation symmetry, we
conclude that all $\Z^a_2$ subgroups corresponding to internal spin-flip
symmetries are non-anomalous.

Next, let's consider $\Z^{zt}_2$, which is always a subgroup of the emanant
symmetries that we considered. Inserting a $\Z^{zt}_2$ defect corresponds to the case of $L$ odd with TBC. From Table \ref{tab:tbc},
one sees that while the ground states are twofold degenerate, both degenerate states still satisfy
$(R_z\, T_z)^2=1$. In other words, the subgroup $\Z^{zt}_2$ itself does not carry a self-anomaly. In the similar way one finds that the subgroups $\Z^{xt}_2$ and
$\Z^{yt}_2$ are also anomaly free.

Finally, the subgroup $\Z^t_2$ of the emanant $\Z_2^x\times
\Z_2^z\times \Z_2^{zt}$ symmetry is anomalous as alluded before. This is because
in the presence of $\Z^t_2$ twist, the ground state acquires fractional eigenvalue
of $\mathrm{i}$ under $\Z^t_2$ subgroup, see Table \ref{tab:pbc} for odd $L$. Combining the fact that each of the subgroups $\Z^x_2$, $\Z^z_2$, and $\Z^{zt}_2$
is anomaly-free while their composition $\Z^t_2$ is anomalous, we conclude that the
emanant symmetry in the continuum limit is
described by $\Z_2^x\times \Z_2^z\times \Z_2^{zt}$ with a type-III mixed
anomaly. A consequence of this is that whenever a defect of an anomaly-free
$\Z_2$ subgroup is inserted, the remaining subgroup acts projectively on the twisted sector, which we verified in Tables \ref{tab:pbc} and \ref{tab:tbc}.

\section{Identifying SymTO $\cD(D_8)$ from sectors of low-energy spectra} \label{sec:symto}

In this section, we are going to use the low-energy spectra computed in
Sec.\ \ref{sec:ed} to compute the symTO of the gapless Heisenberg model,
which as we shall turns out to be the
quantum double $\eD(D_8)$ of the dihedral group $D_8$ (of order $8$).

We obtain the symTO matching the low-energy properties,
by organizing the spectra into the degenerate low-energy sectors
each which are labeled by an bulk anyon of the symTO.
To do so, we need to extract the following data:
the total number $N$ of anyons, the
topological spin $s$ of each anyon, and the quantum dimension $d$ of each
anyon.
\begin{itemize}[leftmargin=*]
\item
Distinct low-energy sectors are labeled by anyons
and characterized by irreducible representations of
the emanant symmetry group in
a given symmetry-twisted boundary condition.

\item
For each anyon, the quantum dimension $d$ is
the dimension of the corresponding irrep
of the emanant symmetry group.

\item
The topological spin $s$ of an anyon is encoded in the momentum
of the corresponding eigenstates.
Denoting by $k$ the crystal momentum
of a set of degenerate eigenstates labeled by an anyon,
we define the corresponding topological spin as
\begin{equation} \label{eq:topo-spin}
s = \frac{L}{2 \pi} (k - k_{\text{ref}}) \pmod{1}.
\end{equation}
Here $k_\mathrm{ref}$ is reference momentum carried by the lowest energy
excitation that the anyon state is created from, \eg, the momentum at which
a cone of excitations are centered at. For our purposes, the reference momentum
takes values in$k_{\text{ref}} \in \{0, \frac{\pi}{4}, \frac{\pi}{2}, \frac{3\pi}{4}\}$.
\end{itemize}

We can use any low-energy multiplet in the corresponding sector to
compute $d,s$ of an anyon (see Table~\ref{tab:matching-condition}).  All those
low-energy states will give rise to the same anyon data $d,s$.

The anyons in the symTO have more data such as the F symbols.
However, as we shall see the set $\{s_a, d_a \mid a=1,\cdots, N\}$
will be  sufficient for us to identify the bulk topological order.
After cataloging anyons for all distinct sectors,
we may verify that their fusion ring matches with that of the quantum double
$\eD(D_8)$, providing a consistency check on our conjecture.
In the remaining of this section,
we will list all the low-energy irreps of the emanant symmetry groups for all
TBCs.

\begin{table}[t] \centering \resizebox{\columnwidth}{!}{ \begin{tabular}{c|c}
\toprule Anyon data & Low-energy property  \\ \hline $N$ & \# of
distinct irreps in all possible TBCs  \\ $d$ & dimension of the irrep
\\ $s$ & $s = \frac{L}{2 \pi} (k - k_{\text{ref}})
\pmod{1}$   \\ \bottomrule \end{tabular} }
\caption{The correspondence between the anyon data $\{N, s, d\}$
and . All relevant information regarding anyon data $\{N, s, d\}$ can be
retrieved from the low-energy spectrum.} \label{tab:matching-condition}
\end{table}


\subsection{Charge sectors}

\begin{figure*}[htbp] \centering
\subfigure[$\mathbb{1}$: $\bigr\langle R_x, R_z,  \ee^{\ii k} \bigr\rangle = (1, \!1, \!1)$]{
\includegraphics[width=0.236\textwidth]{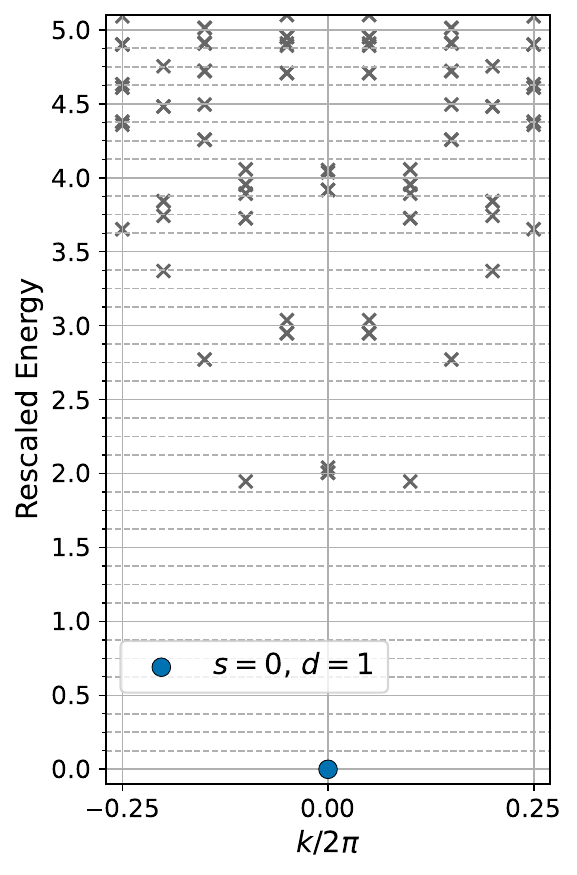}
}\hfill \subfigure[$\mathlarger{\mathlarger{e_x}}$:
$\bigr\langle R_x, R_z,  \ee^{\ii k} \bigr\rangle
= (1, \!-1, \!1)$]{
\includegraphics[width=0.236\textwidth]{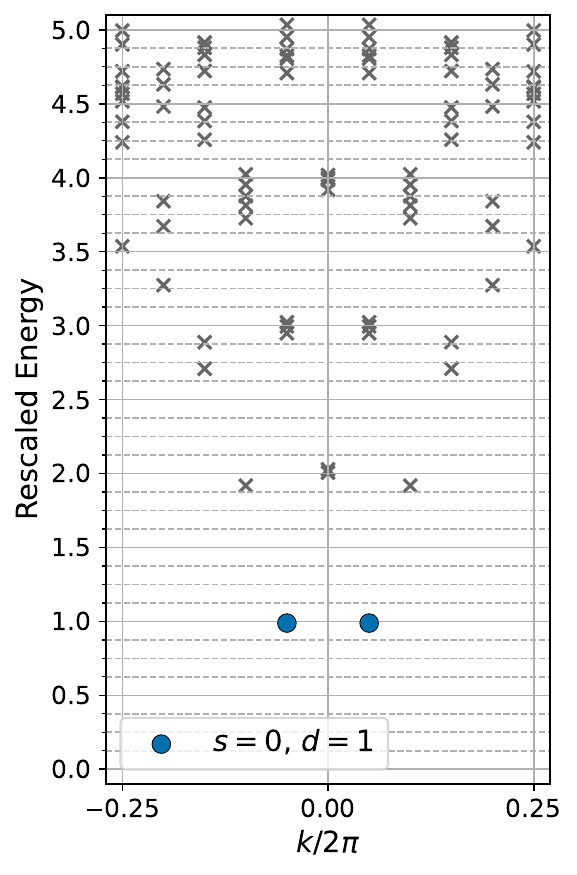}
}\hfill \subfigure[$\mathlarger{\mathlarger{e_y}}$:
$\bigr\langle R_x, R_z,  \ee^{\ii k} \bigr\rangle
= (-1, \!-1, \!1)$]{
\includegraphics[width=0.236\textwidth]{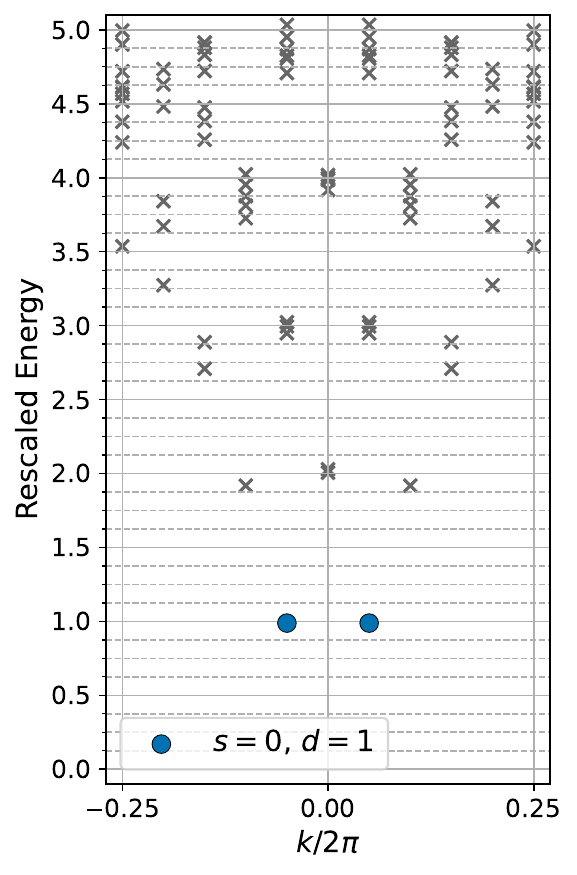}
}\hfill \subfigure[$\mathlarger{\mathlarger{e_z}}$:
$\bigr\langle R_x, R_z,  \ee^{\ii k} \bigr\rangle
= (-1, \!1, \!1)$]{
\includegraphics[width=0.236\textwidth]{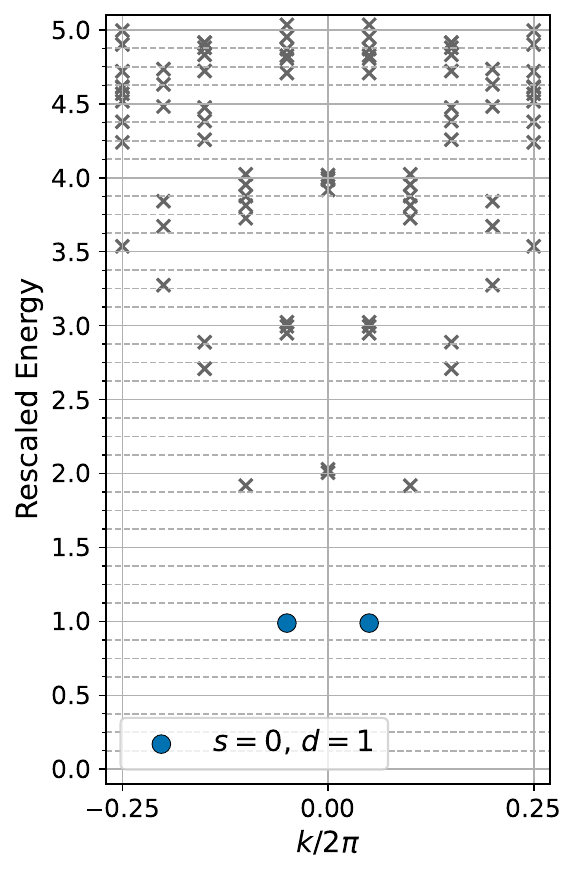}
}

\subfigure[$\mathlarger{\mathlarger{e_t}}$: $\bigr\langle R_x, R_z,  \ee^{\ii k} \bigr\rangle = (1, \!1, \!-1)$]{
\includegraphics[width=0.236\textwidth]{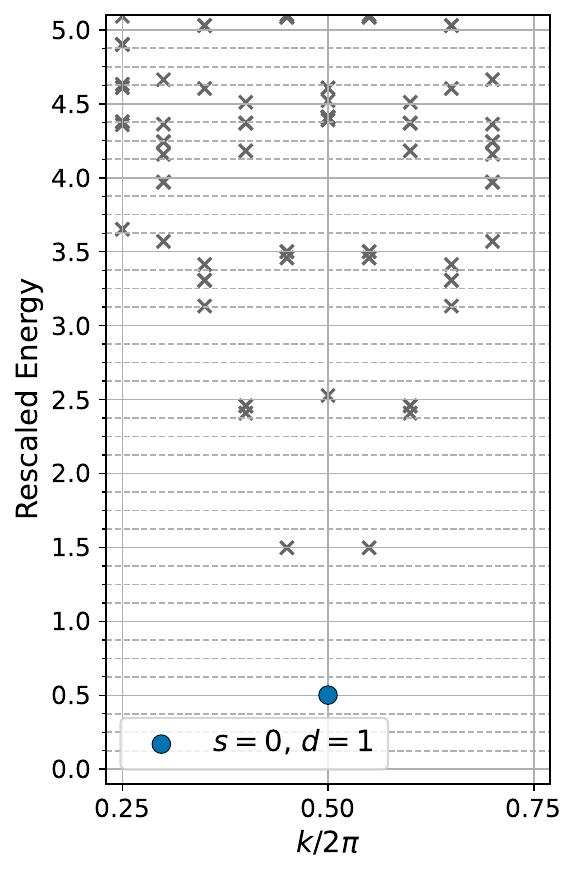}
}\hfill
\subfigure[$\mathlarger{\mathlarger{e_{xt}}}$: $\bigr\langle R_x, R_z,  \ee^{\ii k} \bigr\rangle = (1, \!-1, \!-1)$]{
\includegraphics[width=0.236\textwidth]{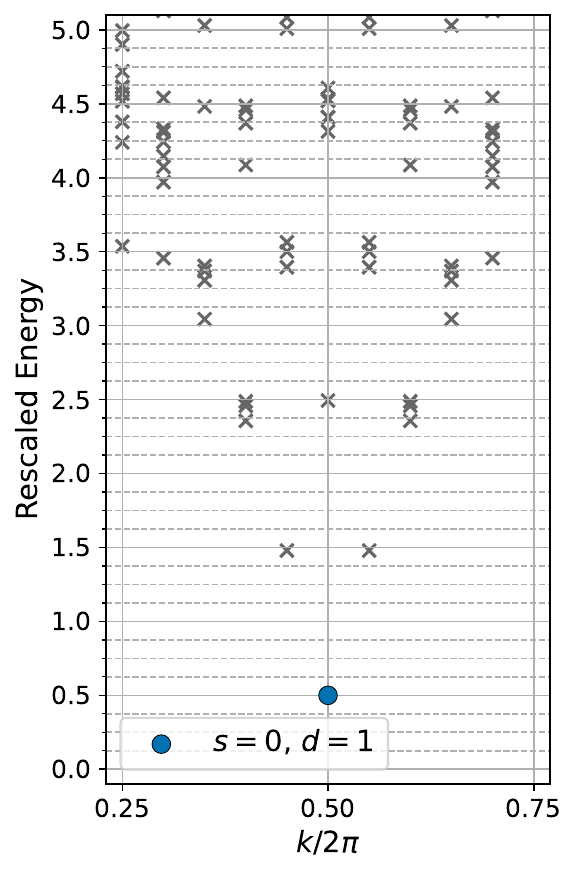}
}\hfill
\subfigure[$\mathlarger{\mathlarger{e_{yt}}}$: $\bigr\langle R_x, R_z,  \ee^{\ii k} \bigr\rangle \!= \!(-1, \!-1, \!-1)$]{

\includegraphics[width=0.236\textwidth]{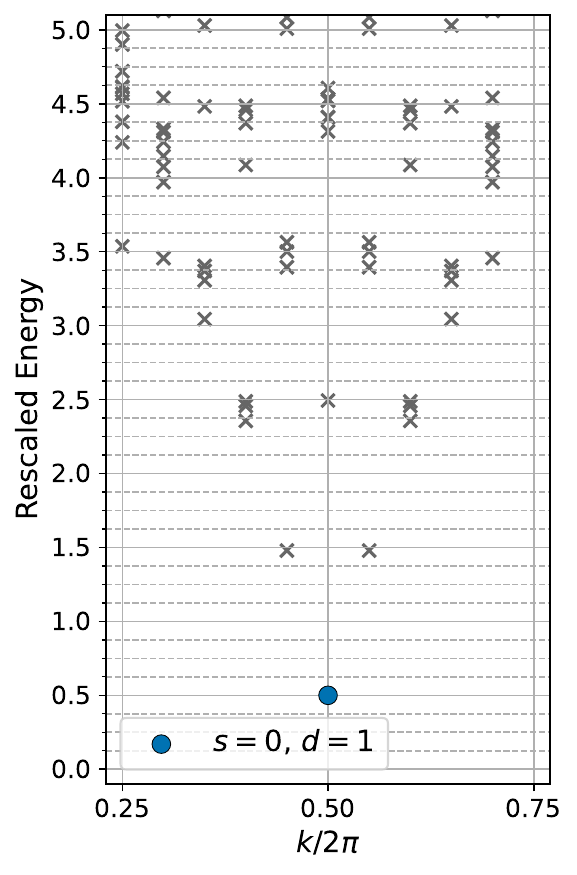}
}\hfill
\subfigure[$\mathlarger{\mathlarger{e_{zt}}}$: $\bigr\langle R_x, R_z,  \ee^{\ii k} \bigr\rangle = (-1, \!1, \!-1)$]{
\includegraphics[width=0.236\textwidth]{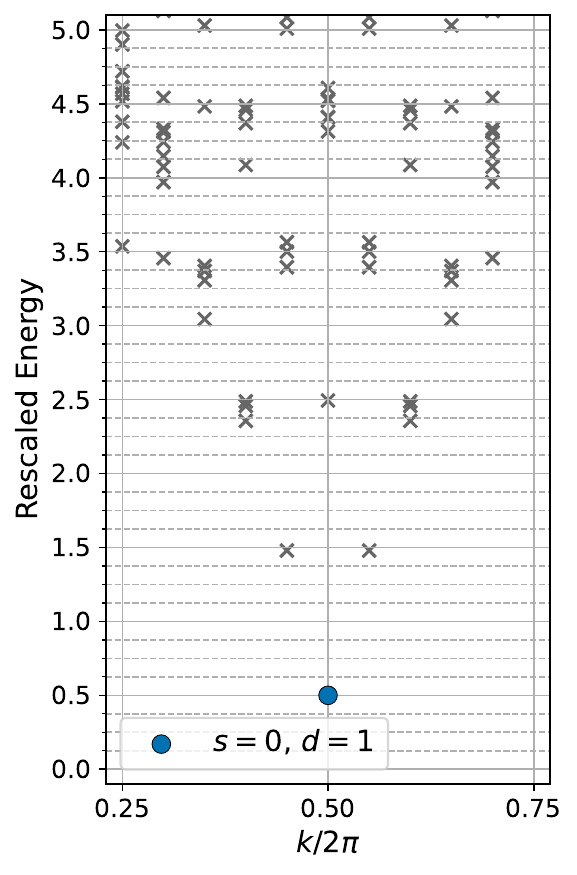}}
\caption{Charge sector anyons, evaluated at lattice size $L=20$ with PBCs.
Each spectrum is labeled with the anyon type and quantum numbers of the
projected subspace. For the corresponding irreps, see
Table~\ref{tab:charge-anyon}.} \label{fig:charge-anyons} \end{figure*}

We first analyze the pure charge sectors, which correspond to the irreps of the
emanant symmetry group $\mathbb{Z}^x_2 \times \mathbb{Z}^z_2 \times
\mathbb{Z}^{zt}_2$ without any symmetry twist.  There are in total eight 1D
irreps, labeled by $a \in \{1,x, y, z, t, xt, yt, zt\}$.  These 1D
irreps can also be labeled by
\begin{equation}
\chi(g_x, g_z, g_{zt}), \quad
g_a \in \{\pm 1\},
\end{equation}
where $g_a = + 1$ ($-1$) denotes a trivial
(nontrivial) charge under the corresponding $\mathbb{Z}^a_2$.
For each irrep one can associate a local operator that transform
accordingly under the emanant symmetry groups. For instance,
the 1D irreps $x,y,z$ correspond to the spin operators $S^{x,y,z}_j$.
Similarly, the 1D irreps $t,xt,yt,zt$
correspond to the staggered operators $(-1)^j,(-1)^jS^{x,y,z}_j$.

By bulk-boundary correspondence,
each 1D irrep labels a distinct charge anyon in the bulk
topological order. We denote a charge anyon by $e_a$, where $a \in \{x, y, z,
t, xt, yt, zt\}$ labels the  1D irreps.  For example, \begin{equation} \chi(-1,
1, 1) \leftrightarrow e_z, \quad \chi(1, -1, -1) \leftrightarrow e_x, \quad
\chi(1, 1, -1) \leftrightarrow e_{t}, \end{equation} and so on. For the trivial
charge with $\chi(1, 1, 1)$, we simply denote it as $\mathbb{1}$.

\begin{table}[t] \centering \resizebox{\columnwidth}{!}{
\begin{tabular}{c|c|c|c|c|c|c|c|c} \toprule Anyon & $s$ & $d$ & $L \mod 4$ &
$R_x$ & $R_y$ &
$R_z$ & $\ee^{\ii k}$ & irrep \\ \hline $\mathbb{1}$ & 0 & 1 &
0  & $+1$ & $+1$ & $+1$ & $+1$  & $\chi(+1, +1, +1)$ \\ \hline
$e_x$ & 0 & 1 & 0  & $+1$ & $+1$ & $-1$ & $+1$  & $\chi(+1, -1,
-1)$ \\ \hline $e_y$ & 0 & 1 & 0  & $-1$ & $+1$ & $-1$ & $+1$  &
$\chi(-1, -1, -1)$ \\ \hline $e_z$ & 0 & 1 & 0  & $-1$ & $-1$ &
$+1$ & $+1$  & $\chi(-1, +1, +1)$ \\ \hline $e_t$ & 0 & 1 & 0  & $+1$ &
$+1$ & $+1$ & $-1$  & $\chi(+1, +1, -1)$ \\ \hline $e_{xt}$ & 0 & 1 &
0  & $+1$ & $-1$ & $-1$ & $-1$  & $\chi(+1, -1, +1)$ \\ \hline
$e_{yt}$ & 0 & 1 & 0  & $-1$ & $+1$ & $-1$ & $-1$  & $\chi(-1,
-1, +1)$ \\ \hline $e_{zt}$ & 0 & 1 & 0  & $-1$ & $-1$ & $+1$ &
$-1$  & $\chi(-1, +1, -1)$ \\ \bottomrule \end{tabular} }
\caption{Relevant data
for charge anyons. The corresponding emanant symmetry group for the irreps is
$\mathbb{Z}^x_2 \times \mathbb{Z}^z_2 \times \mathbb{Z}^{zt}_2$.}
\label{tab:charge-anyon} \end{table}

To resolve these charge sectors, we consider the low-energy spectra for PBCs at
length $L  = 0 \pmod{4}$. Having PBCs ensures that none of the internal
symmetries are twisted, and lattice size of $L  = 0 \pmod{4}$ guarantees
trivial translation flux. Under the insertion of
two translation fluxes (i.e. going from $L \mod 4 = 0$ to $L \mod 4 = 2$), the vaccum state
carries a nontrivial tarnslation charge $e_t$, and the ground state anyon changes from
$\mathbb{1}$ to $e_t$, as shown in Fig.~\ref{fig:pbc}. This is a signature of the self-anomalous nature
of the translation symmetry $\mathbb{Z}^t_2$ -- the translation flux carries a fractional charge instead
of an integer one.

\begin{figure*}[htbp] \centering
\subfigure[$\mathlarger{\mathlarger{m_z}}$:
$L=20$, $R_z  =+1$]{
\includegraphics[width=0.236\textwidth]{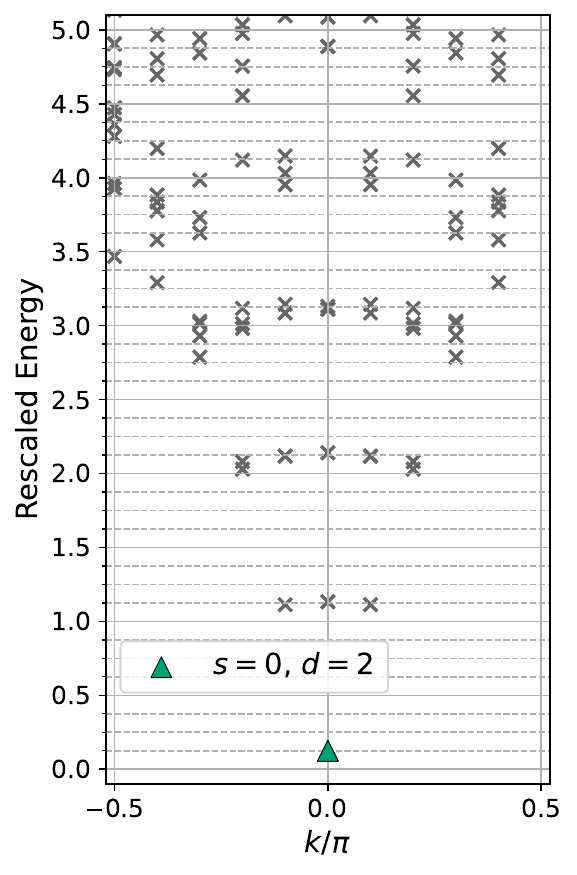}
\label{fig:flux-z} }\hfill
\subfigure[$\mathlarger{\mathlarger{f_z}}$: $L=20$,
$R_z =-1$]{
\includegraphics[width=0.236\textwidth]{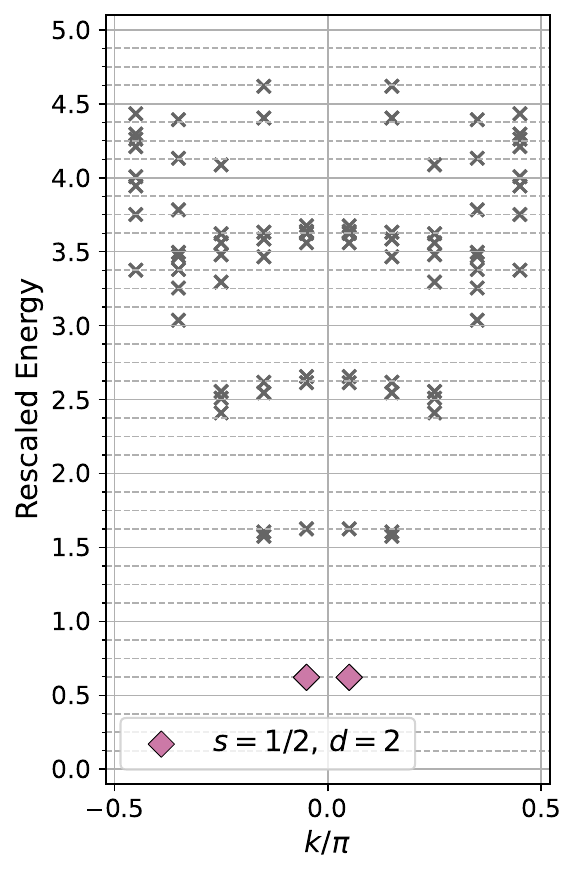}
\label{fig:dyon-z} }\hfill
\subfigure[$\mathlarger{\mathlarger{m_{zt}}}$: $L=21$,
$R_z T_z =+1$]{
\includegraphics[width=0.236\textwidth]{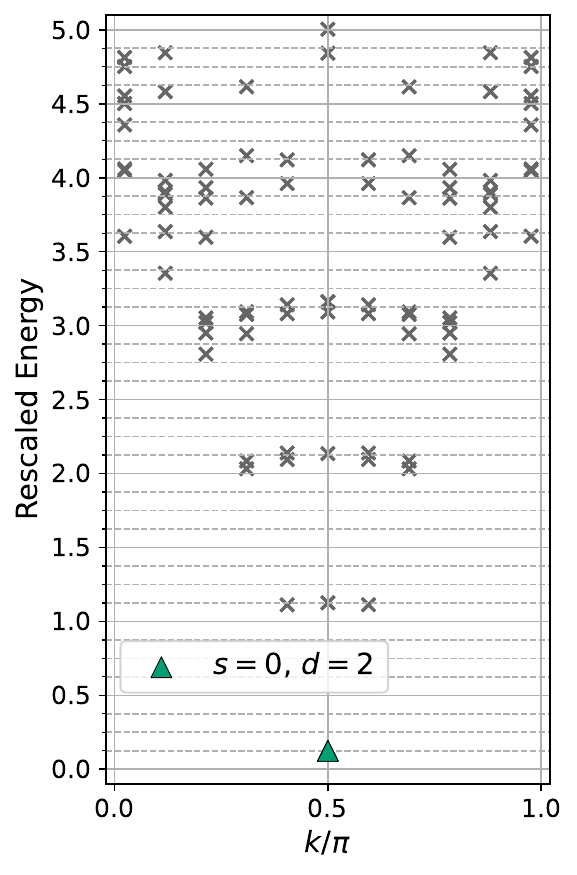}
\label{fig:flux-zt} }
\subfigure[$\mathlarger{\mathlarger{f_{zt}}}$ : $L=21$, $R_z T_z =-1$]{
\includegraphics[width=0.236\textwidth]{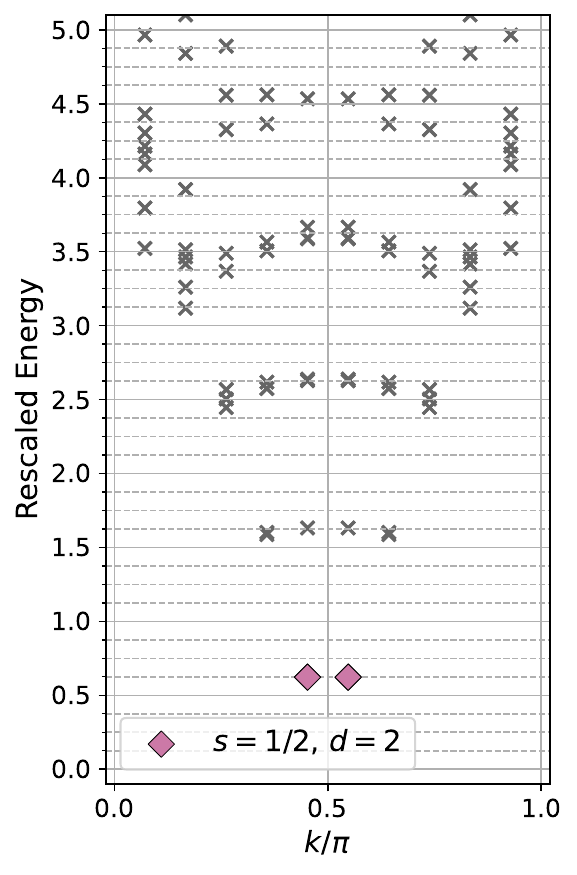}
\label{fig:dyon-zt} }

\caption{(a,b) shows the low-energy sector for $S_z$
symmetry twist and even $L$.  Each energy level is 2-fold degenerate.  (a) has
$R_z=1$, corresponding to anyon $m_z$.  (b) has $R_z=-1$,
corresponding to anyon $f_z$.  (c,d) shows the low-energy sector for $S_z$
symmetry twist and $L=$ odd. Each energy level is 2-fold degenerate. (c) has $R_z T_z = 1 $,
corresponding to the flux anyon $m_{zt}$.
(d) has $R_z T_z = -1$, corresponding to the dyon anyon $f_{zt}$.} \label{fig:fluxes-dyons}
\end{figure*}

\begin{table*}[t] \centering
\begin{tabular}{c|c|c|c|c|c|c|c|c|c} \toprule
Anyon & $s$ & $d$ & $L \mod 4$ & BCs & $R_x$ & $R_y$ & $R_z$ & $\ee^{\ii k}$ & irrep \\ \hline
$m_x$ & 0 & 2 & 0  & $\mathbb{Z}^x_2$-twisted & $+1$ & na & na & $\pm 1$ & $\pi (\mathbb{Z}^{z}_2 \times_\om \mathbb{Z}^{yt}_2)$ \\
$m_y$ & 0 & 2 & 0 & $\mathbb{Z}^y_2$-twisted & na & $+1$ & na & $\pm1$ & $\pi (\mathbb{Z}^{x}_2 \times_\om \mathbb{Z}^{zt}_2)$ \\
$m_z$ & 0 & 2 & 0 & $\mathbb{Z}^z_2$-twisted & na & na & $+1$ & $\pm1$ & $\pi (\mathbb{Z}^{y}_2 \times_\om \mathbb{Z}^{xt}_2)$ \\ \bottomrule
\end{tabular}
\hspace{0.5cm}
\begin{tabular}{c|c|c|c|c|c|c|c|c|c} \toprule
Anyon & $s$ & $d$ & $L \mod 4$ & BCs & $R_x T_x$ & $R_y T_y$ & $R_z T_z$ & $\ee^{\ii k}$ & irrep \\ \hline
$m_{xt}$ & 0 & 2 & 1 & $\mathbb{Z}^x_2$-twisted & $+1$ & na & na & $\pm\ii$ & $\pi (\mathbb{Z}^{z}_2 \times_\om \mathbb{Z}^{yt}_2)$ \\
$m_{yt}$ & 0 & 2 & 1 & $\mathbb{Z}^y_2$-twisted & na & $+1$ & na & $\pm\ii$ & $\pi (\mathbb{Z}^{x}_2 \times_\om \mathbb{Z}^{zt}_2)$ \\
$m_{zt}$ & 0 & 2 & 1 & $\mathbb{Z}^z_2$-twisted & na & na & $+1$ & $\pm\ii$ & $\pi (\mathbb{Z}^{y}_2 \times_\om \mathbb{Z}^{xt}_2)$ \\ \bottomrule
\end{tabular}
\caption{Relevant data for flux anyons.} \label{tab:flux-anyon} \end{table*}

We compute the energy spectrum for $L = 20$ with PBCs. We then
project the many-body spectrum onto eight simultaneous eigensectors of
$R_x$ with eigenvalues $\pm 1$, of $R_z$ with eigenvalues $\pm 1$,
and of $T$ with eigenvalues $\ee^{\ii k} = \pm 1$.
The projection is possible because
$R_x$, $R_z$, and $T$ commute for even $L$ with PBCs.
We label each
subspace by the triplet $( R_x, R_z, \ee^{\ii k})$,
which corresponds to irreps $g_x$, $g_z$, $g_{zt}$
of the group $\mathbb{Z}^x_2
\times \mathbb{Z}^z_2\times \mathbb{Z}^{zt}_2$, respectively.
In FIG.~\ref{fig:charge-anyons}, we show the low-energy spectra projected into
each charge sector  $L = 20$ with PBCs.

Since all irreps with PBC are 1D, all the charge anyons have quantum dimension
$d = 1$. Using Eq.~\eqref{eq:topo-spin}, we find the topological spin for all
the charge anyons to be $s = 0$. In the spectra, it is convenient to label the
lowest energy states with the anyon label $e_a$.
But all excitations within the same sector
should be identified as the same species of anyons, though the higher-energy
excitations can carry additional charges (e.g. larger momenta). The
relevant data for charge anyons, including quantum numbers, irreps, and the
emanant symmetry group, is summarized in Table~\ref{tab:charge-anyon}.

\subsection{Flux sectors}

We now turn to flux sectors of emanant symmetry $G_{\text{IR}} = \mathbb{Z}^x_2
\times \mathbb{Z}^z_2 \times \mathbb{Z}^{zt}_2$, generated by threading
symmetry fluxes around the ring (\ie imposing symmetry twisted boundary
conditions).  Because $G_{\text{IR}}$ has three independent $\mathbb{Z}_2$
components, there are in total seven nontrivial pure flux sectors.
Flux insertions probe projective representations and the underlying
cocycle of $G_{\text{IR}}$, which gives rise to the type-III anomaly, playing a
nontrivial role in determining the emanant symmetry groups and their irreps.

\subsubsection{Fluxes and dyons for $\Z^z_2$ and $\Z^{zt}_2$
symmetry twists}

\begin{table*}[t] \centering
\begin{tabular}{c|c|c|c|c|c|c|c|c|c} \toprule
Anyon & $s$ & $d$ & $L \mod 4$ & BCs & $R_x$ & $R_y$ & $R_z$ & $\ee^{\ii k}$ & irrep \\ \hline
$f_x$ & $\tfrac 1 2$ & 2 & 0 (or 2) & $\mathbb{Z}^x_2$-twisted & $-1$ & na & na & $\pm 1$ & $\pi (\mathbb{Z}^{z}_2 \times_\om \mathbb{Z}^{yt}_2)$ \\
$f_y$ & $\tfrac 1 2$ & 2 & 0 (or 2) & $\mathbb{Z}^y_2$-twisted & na & $-1$ & na & $\pm 1$& $\pi (\mathbb{Z}^{x}_2 \times_\om \mathbb{Z}^{zt}_2)$ \\
$f_z$ & $\tfrac 1 2$ & 2 & 0 (or 2) & $\mathbb{Z}^z_2$-twisted & na & na & $-1$ & $\pm 1$ & $\pi (\mathbb{Z}^{y}_2 \times_\om \mathbb{Z}^{xt}_2)$ \\ \bottomrule
\end{tabular}
\hspace{0.5cm}
\begin{tabular}{c|c|c|c|c|c|c|c|c|c} \toprule
Anyon & $s$ & $d$ & $L \mod 4$ & BCs & $R_x T_x$ & $R_y T_y$ & $R_z T_z$ & $\ee^{\ii k}$ & irrep \\ \hline
$f_{xt}$ & $\tfrac 1 2$ & 2 & 1 (or 3) & $\mathbb{Z}^x_2$-twisted & $-1$ & na & na & $\pm \ii$ & $\pi (\mathbb{Z}^{z}_2 \times_\om \mathbb{Z}^{yt}_2)$ \\
$f_{yt}$ & $\tfrac 1 2$ & 2 & 1 (or 3) & $\mathbb{Z}^y_2$-twisted & na & $-1$ & na & $\pm \ii$ & $\pi (\mathbb{Z}^{x}_2 \times_\om \mathbb{Z}^{zt}_2)$ \\
$f_{zt}$ & $\tfrac 1 2$ & 2 & 1 (or 3) & $\mathbb{Z}^z_2$-twisted & na & na & $-1$ & $\pm \ii$ & $\pi (\mathbb{Z}^{y}_2 \times_\om \mathbb{Z}^{xt}_2)$ \\ \bottomrule
\end{tabular}
\caption{Relevant data for dyons.} \label{tab:dyons} \end{table*}

\begin{figure*}[htbp] \centering \subfigure[$\mathlarger{\mathlarger{s_t}'}$: $L=21$]{
\includegraphics[width=0.48\textwidth]{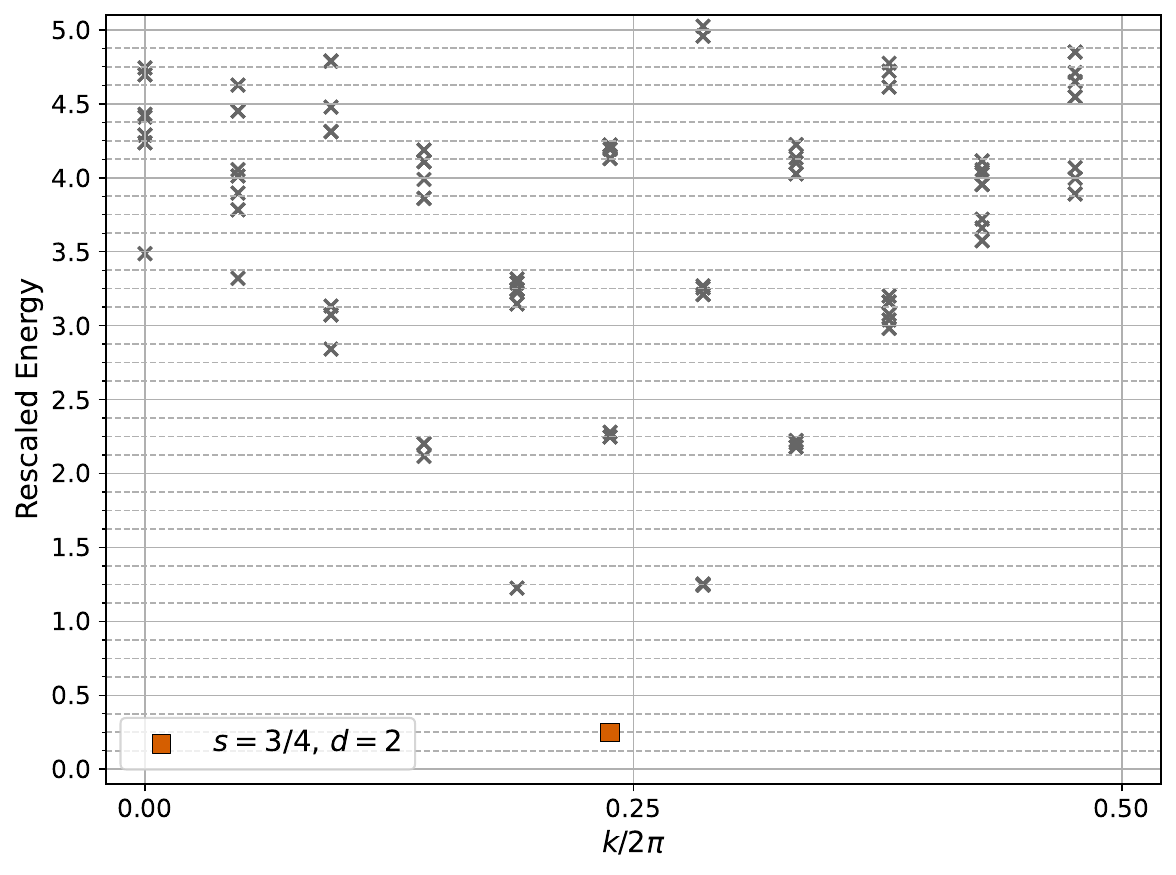}
}\hfill \subfigure[$\mathlarger{\mathlarger{s_{t}}}$: $L=21$]{
\includegraphics[width=0.48\textwidth]{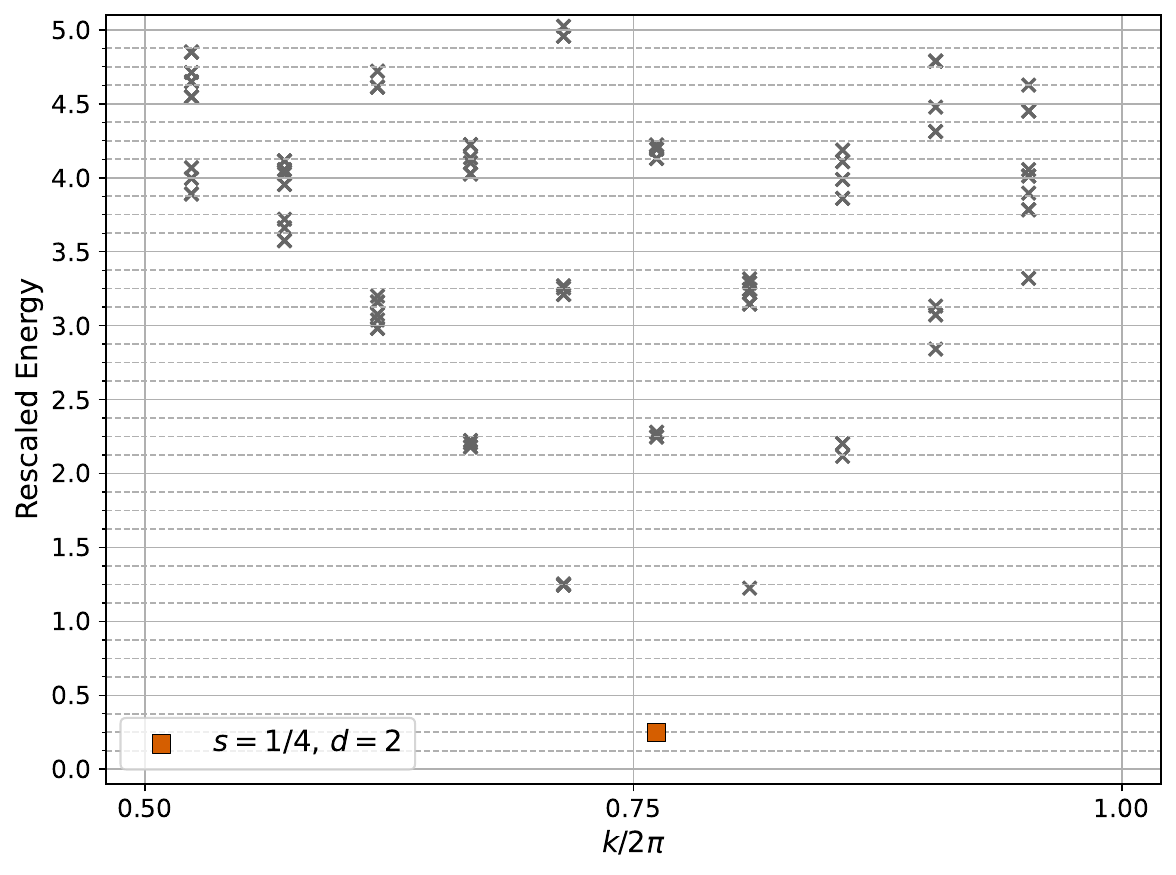} }
\caption{The low-energy spectra for the flux sectors of the diagonal subgroup
$\mathbb{Z}^t_2$. (a) shows the momentum sector at $k = \tfrac \pi 2 \pmod{2
\pi}$; (b) shows the momentum sector at $k = \tfrac{3 \pi}{2} \pmod{2 \pi}$.
Both anyons are the flux anyons for $\mathbb{Z}^t_2$.} \label{fig:t-anyon}
\end{figure*}

We first focus on the cases where either one or two of the $\mathbb{Z}_2$
components are twisted. Owing to the full $S_3$ permutation symmetry of the
spin axes, it suffices to study the subgroups
$\mathbb{Z}^z_2$ and $\mathbb{Z}^{zt}_2$. Note that a $\mathbb{Z}^{z}_2$ twist
corresponds to twisting by $\mathbb{Z}^z_2$ for even $L$ while a
$\mathbb{Z}^{zt}_2$ twist corresponds to twisting by $\mathbb{Z}^z_2$
for odd $L$.

To access the flux sector of $\mathbb{Z}^z_2$, we compute the low-energy
spectrum with $\mathbb{Z}_2^z$-twisted boundary conditions at lattice size $L =
0 \pmod{4}$.  In this case, the emanant symmetry group is $\mathbb{Z}^z_2
\times (\mathbb{Z}^x_2 \times_\om \mathbb{Z}^{zt}_2)$, which has two
2-dimensional irreducible representations with $R_z = \pm 1$.  The
$R_z = 1$ representation corresponds to a pure flux, which gives
rise to the anyon $m_z$.  The $R_z = -1$ representation corresponds
to a dyon, a bound state of $R_z$ chage and flux,
which gives rise to the anyon $f_z$.

Again, in principle we can also compute low-energy spectrum on lattice of size
$L = 2  \pmod{4}$. The resulting anyons are the same, except some of the
corresponding quantum numbers may be flipped (as shown in
Table~\ref{tab:flux-anyon}).

To access the flux sector of $\mathbb{Z}^{zt}_2$, we compute the low-energy
spectrum with $\mathbb{Z}_2^z$-twisted boundary conditions at lattice size $L =
1 \pmod{4}$.  In this case, the emanant symmetry group is $\mathbb{Z}^{zt}_2
\times (\mathbb{Z}^x_2 \times_\om \mathbb{Z}^{z}_2)$, which has two
2-dimensional irreducible representations with $R_z T_z = \pm 1$.
The $R_z T_z = 1$ representation corresponds to a pure flux, which
gives rise to the anyon $m_{zt}$.  The $R_z T_z = -1$
representation corresponds to a dyon, which gives rise to the anyon $f_{zt}$.

The corresponding spectra for the $\mathbb{Z}^z_2$ and $\mathbb{Z}^{zt}_2$
twists are shown in Fig.~\ref{fig:fluxes-dyons}, where the energy levels are
exactlytwofold degenerate, suggesting the correspond anyons all have quantum
dimension $d = 2$.  Extracting $s$ using Eq.~\eqref{eq:topo-spin}, we find that
flux anyons, $m_z,\ m_{zt}$, have topological spin $s = 0$,  while the dyons,
$f_z,\ f_{zt}$, have topological spin $s = \frac12$.

Permuting $x \leftrightarrow y \leftrightarrow z$ gives the remaining four flux
anyons $m_y,\ m_{yt},\ m_z,\ m_{zt}$, and four dyons $f_y,\ f_{yt},\ f_z,\
f_{zt}$. The low-energy spectra are identical due to the exact
$S_3$ symmetry of the Hamiltonian \eqref{eq:h-pbc}.  The
relevant data for the flux and dyons are summarized in
Table~\ref{tab:flux-anyon} and Table~\ref{tab:dyons}.

\subsubsection{Fluxes and dyons for pure translation symmetry twists}

\begin{table}[t] \centering \resizebox{\columnwidth}{!} {
\begin{tabular}{c|c|c|c|c|c|c|c|c} \toprule Anyon & $s$ & $d$ & $L \mod 4$ &
$R_x$ & $R_y$ &
$R_z$ & $\ee^{\ii k}$ & irrep \\ \hline $s_t$ & $\tfrac 1 4$ &
2 & 1 (or 3) & na & na & $\pm \ii$ & $+\ii$ (or $-\ii$) & $\pi (\mathbb{Z}^{x}_2 \times_\om
\mathbb{Z}^{z}_2)$ \\ \hline $s_{t}'$ & $\tfrac 3 4$ & 2 & 1 (or 3) & na & na &
$\pm \ii$ & $-\ii$ (or $+\ii$)& $\pi (\mathbb{Z}^{x}_2 \times_\om \mathbb{Z}^{z}_2)$ \\
\bottomrule \end{tabular} } \caption{Relevant data for flux anyons of
$\mathbb{Z}^t_2$.} \label{tab:t-anyon} \end{table}

\begin{table*}[t] \centering
{ \begin{tabular}{c|c|c|c|c|c|c|c|c|c|c|c|c|c|c|c|c|c|c|c|c|c|c} \toprule
Anyon & $\mathbb{1}$ & $e_x$ & $e_y$ & $e_z$ & $e_t$ & $e_{tx}$ & $e_{ty}$ &
$e_{tz}$ & $m_x$ & $m_y$ & $m_z$ & $m_{xt}$ & $m_{yt}$ & $m_{zt}$ & $f_x$ &
$f_y$ & $f_z$ & $f_{xt}$ & $f_{yt}$ & $f_{zt}$ & $s_t$ & $s_{t}'$ \\ \hline $s$
& 0 & 0 & 0 & 0 & 0 & 0 & 0 & 0 & 0 & 0 & 0 & 0 & 0 & 0 & $\tfrac 1 2$ &
$\tfrac 1 2$ & $\tfrac 1 2$ & $\tfrac 1 2$ & $\tfrac 1 2$ & $\tfrac 1 2$ &
$\tfrac 1 4$ & $\tfrac 3 4$ \\ \hline $d$ & 1 & 1 & 1 & 1 & 1 & 1 & 1 & 1 & 2 &
2 & 2 & 2 & 2 & 2 & 2 & 2 & 2 & 2 & 2 & 2 & 2 & 2 \\ \bottomrule \end{tabular}
} \caption{Anyon data summary.} \label{tab:d8-anyon} \end{table*}

Finally, we twist by the translation symmetry. This is done so by computing
the low-energy spectrum for odd $L$ with PBCs (see Fig.~\ref{fig:t-anyon}),
which we interpret as the $\mathbb{Z}^t_2$ flux sector. Again we choose  $L = 1 \pmod{4}$ without loss of generality. All  energy levels
are twofold degenerate (since they all carry half-integer spins).

The $\mathbb{Z}^t_2$ flux sector can be further divided into two momentum
sectors, located at $k = \tfrac \pi 2$ and $\tfrac{3 \pi}{2} \pmod{2 \pi}$.
Each sector has twofold degenerate groundstates, that differ
in relative momenta from the reference point $k_{\text{ref}}$. As a
result, the two anyons from the $\mathbb{Z}^t_2$ flux sector, both with $d =
2$, but one with $s = \tfrac 1 4$, the other with $s = \tfrac 3 4$. We label
the one with $s = \tfrac 1 4$ as $s_t$ and the one with $s = \tfrac 3 4$ as
$s_{t}'$. This convention is adopted since their topological spin suggests that
they correspond to anyons with semionic statistics.
The relevant data for the flux anyons are
summarized in Table~\ref{tab:t-anyon}.

This completes our classifications of all sectors from irreducible
representations and twists of the symmetry.  We have identified all possibly
distinct anyon types from the low-energy spectra.
The relevant data for anyons and their sectors are shown in
Tables~\ref{tab:charge-anyon}--\ref{tab:t-anyon}. For these tables, the column
$\ee^{\ii k}$ indicates the momentum sector, instead of the
momentum of a specific state.
We next assemble this data to pin down the
symTO.

\subsection{Matching with SymTO $\cD(D_8)$} \label{subsec:match-d8}

We tabulate all 22 low-energy sectors and their corresponding anyon labels
in Table~\ref{tab:d8-anyon}.
To identify the corresponding topological order, we
note that any qualifying topological order should have 8 Abelian anyons that correspond to the
charges of $G_\text{IR} = \mathbb{Z}^x_2 \times \mathbb{Z}^z_2 \times
\mathbb{Z}^{xzt}_2$.
Here, it is more convenient to choose the generators of $G_\text{IR}$
to be those of
$\mathbb{Z}^x_2$, $\mathbb{Z}^z_2$,  and $\mathbb{Z}^{xzt}_2$
subgrorups rather than
$\mathbb{Z}^x_2$, $\mathbb{Z}^z_2$,  and $\mathbb{Z}^{xzt}_2$,
because  as we shall see in Sec.~\ref{sec:d8} there will be a
useful symTO automorphism permuting the corresponding flux anyons
$m_x,m_z,m_{yt}$.
Those 8 Abelian anyons form a symmetric fusion category.
We know that the quantum double of $G_\text{IR} = \mathbb{Z}^x_2 \times
\mathbb{Z}^z_2 \times \mathbb{Z}^{xzt}_2$ will have  8 Abelian anyons that form
a symmetric fusion category $\cRep(\mathbb{Z}^x_2 \times \mathbb{Z}^z_2 \times
\mathbb{Z}^{xzt}_2)$.  But such a quantum double has 64 anyons instead of the expected 22.
The quantum double of $G_\text{IR}$ twisted by a
cocycle \begin{equation} [\nu]\in H^3(\mathbb{Z}^x_2 \times \mathbb{Z}^z_2
\times \mathbb{Z}^{xzt}_2, U(1)) \end{equation} also has  8 Abelian anyons that
form a symmetric fusion category $\cRep(\mathbb{Z}^x_2 \times \mathbb{Z}^z_2
\times \mathbb{Z}^{xzt}_2)$.  We thus observe that  the anyon data above is precisely matching
that of the twisted quantum double $\eD^\nu(\mathbb{Z}^x_2 \times
\mathbb{Z}^z_2 \times \mathbb{Z}^{xzt}_2)$, defined by the type-III cocycle, or
equivalently, the quantum double of $D_8$ group (\ie the group of all
symmetries of the square), denoted by $\cD(D_8)$.

Having established that $G_\text{IR} = \mathbb{Z}^x_2 \times \mathbb{Z}^z_2
\times \mathbb{Z}^{xzt}_2$ is endowed with a type-III anomaly among the three
$\mathbb{Z}_2$ components, we now clarify how this anomaly fixes the emanant
symmetry groups and their irreps in the flux sectors.
The type-III anomaly in $G_\text{IR}$ implies that whenever a single
$\mathbb{Z}_2$ flux is threaded, the remaining $\mathbb{Z}_2 \times
\mathbb{Z}_2$ subgroup carries a 2D projective representation (rep) within the
flux sector.  This matches the result from Section~\ref{typeIII}.  We denote such
a group by the shorthand $\mathbb{Z}_2 \times_\om \mathbb{Z}_2$.
The algebra $\mathbb{Z}_2 \times_\om \mathbb{Z}_2$ admits exactly one irrep that is
compatible with the nontrivial projective phase $\omega$, \ie, a 2D projective irrep
in which the generators act as Pauli matrices.

This statement can be repackaged equivalently as follows.
The cocycle $\omega$ is
trivialized by the minimal central extension
\begin{equation} 1 \rightarrow
\mathbb{Z}_2 \rightarrow D_8 \rightarrow \mathbb{Z}_2 \times \mathbb{Z}_2
\rightarrow 1,
\end{equation}
so that projective irrep of $\mathbb{Z}_2 \times \mathbb{Z}_2$ is nothing but the
unfaithful realization of the 2D irrep for group $D_8$. Since $D_8$ is not
represented faithfully, none of its 1D irreps appear. This corresponds to the
fact that in the low-energy spectrum, every energy level is at least twofold
degenerate. In other words, when a non-trivial $\mathbb{Z}_2$ flux is inserted,
the entire spectrum can be organized into the 2D projective reps of the
remaining $\mathbb{Z}_2 \times \mathbb{Z}_2$ symmetry.

In Tables~\ref{tab:flux-anyon}--\ref{tab:t-anyon}, we use $\pi (\mathbb{Z}^i_2
\times_\om \mathbb{Z}^j_2)$ to denote the projective irrep of the remaining
$\mathbb{Z}_2 \times \mathbb{Z}_2$ after the flux of a third $\mathbb{Z}_2$ is
threaded.

\section{SymTO $\cD(D_8)$ and nearby gapped phases}
\label{sec:d8}

Having identified the emanant symTO of the gapless spin-$1/2$ Heisenberg chain
as the quantum double $\mathcal{D}(D_8)$, we now employ this  symTO to constrain the neighboring gapped phases.
To this end, we first review the properties of the $D_8$ quantum double.

\begin{widetext}
\subsection{$D_8$ quantum double $\eD(D_8)$}

The $D_8$ quantum double $\eD(D_8)$ has 22 types of anyons, whose quantum
dimensions and topological spins are given in Table \ref{tab:d8-anyon}.
It is non-chiral, i.e., its chiral central charge is $c = 0$, and its total quantum dimension is $D^2 =
\sum_i d_i^2 = 64$.  Every anyon is self-conjugate.  The $S$-matrix and the
fusion rule of $D_8$ quantum double are given in Appendix \ref{fusionD8}.

The automorphism group of $D_8$ quantum double from anyon permutation
is given by $S_4$, which contains the following 24 permutations of anyon types
\begingroup
\allowdisplaybreaks
\begin{align}
\label{eq:S4 automorphisms}
\begin{split}
& ( ),
\\
& (\aY ,\aZ )(\aV ,\aW )(\aJ ,\aK )(\aM ,\aN )(\ay ,\az )(\av ,\aw ),
\\
& (\aY ,\aV )(\aZ ,\aW )(\aJ ,\aM )(\aK ,\aN )(\ak ,\au )(\av ,\aw ),
\\
& (\aY ,\aW )(\aZ ,\aV )(\aJ ,\aN )(\aK ,\aM )(\ay ,\az )(\ak ,\au ), 
\\
& (\aX ,\aY )(\aU ,\aV )(\aI ,\aJ )(\aL ,\aM )(\ax ,\ay )(\au ,\av ),
\\
& (\aX ,\aY ,\aZ )(\aU ,\aV ,\aW )(\aI ,\aJ ,\aK )(\aL ,\aM ,\aN )(\ax ,\ay ,\az )(\au ,\av ,\aw ),
\\
& (\aX ,\aY ,\aU ,\aV )(\aZ ,\aW )(\aI ,\aJ ,\aL ,\aM )(\aK ,\aN )(\ax ,\ay )(\ak ,\av ,\aw ,\au ),
\\
& (\aX ,\aY ,\aW )(\aZ ,\aU ,\aV )(\aI ,\aJ ,\aN )(\aK ,\aL ,\aM )(\ax ,\ay ,\az )(\ak ,\av ,\au ),
\\
& (\aX ,\aZ ,\aY )(\aU ,\aW ,\aV )(\aI ,\aK ,\aJ )(\aL ,\aN ,\aM )(\ax ,\az ,\ay )(\au ,\aw ,\av ),
\\
& (\aX ,\aZ )(\aU ,\aW )(\aI ,\aK )(\aL ,\aN )(\ax ,\az )(\au ,\aw ),
\\
& (\aX ,\aZ ,\aV )(\aY ,\aU ,\aW )(\aI ,\aK ,\aM )(\aJ ,\aL ,\aN )(\ax ,\az ,\ay )(\ak ,\aw ,\au ),
\\
& (\aX ,\aZ ,\aU ,\aW )(\aY ,\aV )(\aI ,\aK ,\aL ,\aN )(\aJ ,\aM )(\ax ,\az )(\ak ,\aw ,\av ,\au )
\\
& (\aX ,\aU )(\aZ ,\aW )(\aI ,\aL )(\aK ,\aN )(\ak ,\av )(\au ,\aw )
\\
& (\aX ,\aU )(\aY ,\aZ ,\aV ,\aW )(\aI ,\aL )(\aJ ,\aK ,\aM ,\aN )(\ay ,\az )(\ak ,\aw ,\au ,\av ),
\\
& (\aX ,\aU )(\aY ,\aV )(\aI ,\aL )(\aJ ,\aM )(\ak ,\aw )(\au ,\av ),
\\
& (\aX ,\aU )(\aY ,\aW ,\aV ,\aZ )(\aI ,\aL )(\aJ ,\aN ,\aM ,\aK )(\ay ,\az )(\ak ,\av ,\au ,\aw ),
\\
& (\aX ,\aV ,\aU ,\aY )(\aZ ,\aW )(\aI ,\aM ,\aL ,\aJ )(\aK ,\aN )(\ax ,\ay )(\ak ,\au ,\aw ,\av ),
\\
& (\aX ,\aV ,\aW )(\aY ,\aZ ,\aU )(\aI ,\aM ,\aN )(\aJ ,\aK ,\aL )(\ax ,\ay ,\az )(\ak ,\aw ,\av ),
\\
& (\aX ,\aV )(\aY ,\aU )(\aI ,\aM )(\aJ ,\aL )(\ax ,\ay )(\ak ,\aw ),
\\
& (\aX ,\aV ,\aZ )(\aY ,\aW ,\aU )(\aI ,\aM ,\aK )(\aJ ,\aN ,\aL )(\ax ,\ay ,\az )(\ak ,\au ,\aw ),
\\
& (\aX ,\aW ,\aY )(\aZ ,\aV ,\aU )(\aI ,\aN ,\aJ )(\aK ,\aM ,\aL )(\ax ,\az ,\ay )(\ak ,\au ,\av )
\\
& (\aX ,\aW )(\aZ ,\aU )(\aI ,\aN )(\aK ,\aL )(\ax ,\az )(\ak ,\av ), 
\\
& (\aX ,\aW ,\aV )(\aY ,\aU ,\aZ )(\aI ,\aN ,\aM )(\aJ ,\aL ,\aK )(\ax ,\az ,\ay )(\ak ,\av ,\aw ),
 \\
& (\aX ,\aW ,\aU ,\aZ )(\aY ,\aV )(\aI ,\aN ,\aL ,\aK )(\aJ ,\aM )(\ax ,\az )(\ak ,\au ,\av ,\aw ),
\end{split}
\end{align}
\endgroup
\end{widetext}
where each line corresponds to an element of $S_4$ and each parenthesis
corresponds to a cyclic permutation of anyon types.
The automorphism of the symTO in general does not correspond to a symmetry
of the Hamiltonian \eqref{eq:Ham Heisenberg}, or the perturbations 
\eqref{eq:NNN deformation} and \eqref{eq:pert-neel-1}.
For example, in the XYZ limit with various anisotropy terms nonvanishing,
$S_4$ symmetry is fully broken.  However, we will later
argue that when we restore the full $SO(3)$ spin rotation symmetry, the
$S_4$  automorphism becomes an emergent symmetry.  In fact, this emergent
$S_4$ symmetry is a subgroup of the larger emergent $SO(4)$ symmetry.

The $D_8$ quantum double has 11 Lagrangian condensable algebras given by the
following composite anyons:
\begingroup
\allowdisplaybreaks
\begin{align}
\label{eq:D8 condensable algebras}
& \cA_{1,1} =  \onebb \oplus \ax \oplus \ay \oplus \az \oplus \ak \oplus \au \oplus \av \oplus \aw
\nonumber \\
& \cA_{2,1} =  \onebb \oplus \aX \oplus \aY \oplus \aZ \oplus \ak
\nonumber \\
& \cA_{2,2} =  \onebb \oplus \aX \oplus \aV \oplus \aW \oplus \au
\nonumber \\
& \cA_{2,3} =  \onebb \oplus \aY \oplus \aU \oplus \aW \oplus \av
\nonumber \\
& \cA_{2,4} =  \onebb \oplus \aZ \oplus \aU \oplus \aV \oplus \aw
\nonumber \\
& \cA_{3,1} =  \onebb \oplus 2\aX \oplus \ax \oplus \ak \oplus \au
\nonumber \\
& \cA_{3,2} =  \onebb \oplus 2\aY \oplus \ay \oplus \ak \oplus \av
\nonumber \\
& \cA_{3,3} =  \onebb \oplus 2\aZ \oplus \az \oplus \ak \oplus \aw
\nonumber \\
& \cA_{3,4} =  \onebb \oplus 2\aU \oplus \ax \oplus \av \oplus \aw
\nonumber \\
& \cA_{3,5} =  \onebb \oplus 2\aV \oplus \ay \oplus \au \oplus \aw
\nonumber \\
& \cA_{3,6} =  \onebb \oplus 2\aW \oplus \az \oplus \au \oplus \av
\end{align}
The Lagrangian condensable algebras are divided into three classes according to their orbits
under the $S_4$ permutations \eqref{eq:S4 automorphisms}, i.e., the
condensable algebras in the same class are connected by the $S_4$
automorphisms. The first digit in their subscript labels the orbit while
 the second labels elements within it.

\begin{figure*}[htbp] 
\centering 
\subfigure[]{
\includegraphics[width=0.3\textwidth]{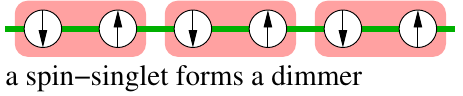}
\label{SpinDimmer}
}
\hfill 
\subfigure[]{
\includegraphics[width=0.3\textwidth]{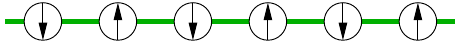}
\label{SpinAF}
}
\subfigure[]{
\includegraphics[width=0.3\textwidth]{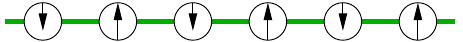}
\label{SpinAFF}
}
\subfigure[]{
\includegraphics[width=0.3\textwidth]{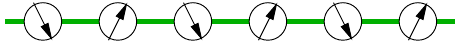}
\label{SpinAFCant}
}
\subfigure[]{
\includegraphics[width=0.3\textwidth]{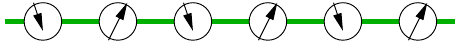}
\label{SpinAFCantF}
}
\caption{
(a) The dimer phase of a spin chain (from condensable algebra
$\cA_{2,1}$, where each pair of spins forms a spin-singlet \textit{dimer} shown in pink.
(b) The N\'eel phase of a spin chain (from condensable algebras
$ \cA_{2,2}, \cA_{2,3}, \cA_{2,4} $). 
(c) A collinear ferromagnetic phase
of a spin chain, that also breaks translation symmetry
(from condensable algebras
$ \cA_{3,1}, \cA_{3,2}, \cA_{3,3} $.
(d) A phase of a spin chain,  with ferromagnetic order for spins in
$x$-direction and anti-ferromagnetic order for spins in $y$-direction (from
condensable algebras $ \cA_{3,4}, \cA_{3,5}, \cA_{3,6} $). 
(e) A ferromagnetic phase of a spin chain, with non-collinear spins which
also breaks the translation symmetry (from condensable algebra $ \cA_{1,1}$).
} 
\end{figure*}

The Heisenberg chain\eqref{eq:Ham Heisenberg} has a $SO(3)$ spin rotation plus a lattice translation
symmetry, with each lattice site carrying a spin-1/2 representation
of $SO(3)$.  If we restrict ourselves to the $\mathbb{Z}_2^x \times \mathbb{Z}_2^z$
subgroup of the full $SO(3)$ symmetry and the lattice translation symmetry, the emanant symTO
of the Heisenberg chain is found to be the quantum double of $D_8$, with the
topological symmetry boundary induced by the
Lagrangian condensable algebra:
\begin{align}
\label{eq:Sym boundary Heisenberg}
\cA_{1,1} =  \onebb \oplus \ax \oplus \ay \oplus \az \oplus \ak \oplus \au \oplus \av \oplus \aw .
\end{align}
The 22 anyons in the $D_8$ quantum double correspond to 22 low-energy sectors
of the Heisenberg chain, which are obtained via the irreps of emanant symmetry
groups for various symmetry twists. Following and reiterating the discussion in 
Section \ref{sec:symto}, choosing the Lagrangian algebra \eqref{eq:Sym boundary Heisenberg}
as the topological symmetry boundary leads to the following identifications of the 
anyons in $\eD_8$ with the emanant symmetry charges and defects.
\begin{enumerate}

\item
The anyon $\onebb$ describes trivial excitation that carry no charge or flux. 
In the ED spectra, it corresponds to $R_x=R_y=R_z=1$ and 
quasi-momentum $k=0$ sector on a ring of size $L=0$ mod 4 (see Fig.\
\ref{fig:charge-anyons}(a)). We identify 
with $\onebb$ the ground state without any charges 
and symmetry twists.

\item
The anyons $\ax$, $\ay$, $\az$ carry the $\Z_2$ charges of $\pi$
spin-rotations, implemented by the operators $R_a$, $a=x,y,z$.
In the ED spectra they correspond to the sectors with $R_x=\pm1$, $R_y\pm1$,
$R_z=\pm 1$ and quasi-momentum $k=\pm 2\pi/L$  
on a ring of size $L=0$ mod 4 (see Fig.\ \ref{fig:charge-anyons}(b,c,d)).

Since the three $\Z_2$ spin-rotations are not independent, i.e., 
$R_x\,R_y = (-1)^L R_z $ and $R_x\,R_y = (-1)^LR_y\,R_x$, 
each $e_a$ anyon carries charge under two $\Z_2$ rotations. 
For example,  $\ax$ anyon carries
non-trivial $\Z_2$ charges under $R_y$ and $R_z$. 
In the Heisenberg model \eqref{eq:Ham Heisenberg}, 
the excitations corresponding to the
$\ax$, $\ay$, $\az$ anyons can be created by 
translationally-invariant linear combinations of the spin
operators $S^x_j$, $S^y_j$, $S^z_j$, respectively.   
As a result, for instance, condensation of $\ax$, 
stabilizes a ferromagnetic phase with order parameter
$\<\sum_j  S^x_j\> \neq 0$.

\item
The anyon $\ak$ carries $-1$ charge under the  $\Z^t_2$ that emanate from 
lattice translations. In the ED spectra it corresponds to the sector 
with $R_x=R_y=R_z=1$ and quasi-momentum $k=\pi$ on a ring of size $L=0$ mod 4
(see Fig.\ \ref{fig:charge-anyons}(e)).

In the Heisenberg model \eqref{eq:Ham Heisenberg}, 
the excitations corresponding to $e_t$ It is created by linear combinations of local
spin-singlet operators weighted by phase factors $(-1)^j$ (where $j$ labels
lattice sites).  If the anyon $\ak$ condenses, the translation by one sites $j\to j+1$, 
and thus the emanant $\Z^t_2$ symmetry, is spontaneously broken while 
the translation by two sites $j\to j+2$ is preserved.

\item
The anyons $\au$, $\av$, $\aw$ carry the $\Z_2$ charges of $\pi$
spin-rotations as well as $-1$ charge under $\Z^t_2$ symmetry. They 
can be obtained by fusing the anyons $\ax$, $\ay$, $\az$ with $\ak$,
respectively. In the ED spectra, they 
correspond to the sectors with $R_x=\pm1$, $R_y\pm1$,
$R_z=\pm 1$ and quasi-momentum $k=\pi$  on a ring of size $L=0$ mod 4
(see Fig.\  \ref{fig:charge-anyons}(f,g,h)).

In the Heisenberg model \eqref{eq:Ham Heisenberg},
the excitations corresponding to 
the $\au$, $\av$, $\aw$ anyons are created by staggered linear combinations of 
spin operators $S^x_j$, $S^y_j$, $S^z_j$, repectively.
For example, if the $\au$ anyon condenses, it stabilizes an antiferromagetic phase with
$\<\sum_j (-1)^j S^x_j\> \neq 0$.

As expected in the symTO framework, the topological symmetry boundary \eqref{eq:Sym boundary Heisenberg} 
is obtained by condensing all the symmetry charges, the ``$e$-anyons", 
of the emanant $G_\text{IR}=\Z^x_2\times\Z^z_2
\times Z^{xzt}_2$ symmetry.

\item
The anyons $\aX$, $\aY$, $\aZ$ are the fluxes of pin-rotations by $\pi$ around  
$S_x$, $S_y$, and $S_y$, respectively. One way to see this is to note that these anyons
have non-trivial mutual statistics with the anyons
$\ax$, $\ay$, $\az$ while having trivial statistics with the anyon $\ak$
according to the S-matrix (see Appendix \ref{fusionD8}), which are the charges of the $\Z^x_2\times\Z^z_2$ symmetry.
In the ED spectra, they correspond to 
the ground state sectors with twisted quasi-momentum $k=0$ on a ring of size $L=0$ mod 4,
when the $\Z^x_2$-, $Z^y_2$-, and $Z^z_2$-twisted boundary conditions are imposed,
respectively (see Fig.~\ref{fig:fluxes-dyons}(a)). 

Because of the type-III mixed anomaly of the emanant $G_\text{IR}$ symmetry, 
a flux of $\Z^a_2$ ($a=x,y,z$) symmetry realizes a two-dimensional projective representation of the $G_\text{IR}/\Z^a_2$. 
Insertion of two symmetry fluxes results in four one-dimensional irreps of $G_\text{IR}/\Z^a_2$.
This is reflected in the fusion rules for example, 
$m_x\otimes m_x= \onebb\oplus e_z\oplus e_t\oplus e_{zt}$ where
bound state of two $S_x$ $\pi$-rotation symmetry flux is decomposed
into one-dimensional states carrying the pure charges of the remaining 
$G_\textbf{IR}/Z^x_2 = \Z^z_2\times \Z^{zt}_2$ symmetry.

\item 
The anyons $\aT$, $\aD$ are the fluxes of emanant $\Z^t_2$ symmetry.
These anyons have non-trivial mutual statistics with the anyon $\ak$ while
having trivial mutual statistics with the anyons $\ax$, $\ay$, $\az$, see Appendix
\ref{fusionD8}. In the ED spectra,
they correspond to the ground state sectors with
quasi-momentum momentum $k=\mp \pi/2 \pm {2\pi}/{8L}$, respectively, 
on a ring of size $L=\pm1$ mod 4
(see Fig.\ \ref{fig:t-anyon}). 

The excitations corresponding to the two anyons $\aT$ and $\aD$ 
are centered around momenta that differ by $\pi$,
which is a pure charge of emanant $\Z^t_2$ symmetry. This is consistent with the fusion rule
$\aT\otimes \ak = \aD$. The fact that there are two anyons with fractional topological spins 
that realize the pure flux of $\Z^t_2$ symmetry is a reflection of the fact that this emanant $\Z_2$
symmetry is anomalous and the corresponding fluxes carry semionic statistics.
A translationally-invariant gapped state must involve $\aT$ and $\aD$ condensations.
However, because of they are semions these anyons cannot condense and, thus,
any gapped states must break translation symmetry. This is an alternative 
way to say that the emanant $\Z^t_2$ symmetry is anomalous.

Because of the type-III anomaly of the emanant $G_\text{IR}$
symmetry, fluxes of $\Z^t_2$ carry a 2D projective representation of 
$G_\text{IR}/\Z^t_2 = \Z^x_2\times\Z^z_2$ symmetry. Hence, the two anyons $\aT$ and $\aD$ have quantum dimensions
2. From the point of view of the lattice model such a flux is implemented by adding a new site, $L\mapsto L+1$,
i.e., an additional spin-1/2 degree of freedom which indeed carries the 2D projective representation of 
spin-rotation symmetry. 


\item 
The anyons $\aL$, $\aM$, $\aN$ are the fluxes of the $\Z^a_2$ symmetry
that is also charged under the same symmetry for $a=x,y,z$, respectively. 
They are obtained by appropriately fusion of pure fluxes and charges. 
For instance, the anyon $\aN$ can be obtained by $\ax\otimes \aZ = \aN$
or $\ay\otimes \aZ = \aN$, see Appendix \ref{fusionD8}.
In the ED spectra, they correspond to the excited states that carry $\Z^a_2$ charge
on a ring of size $L=0$ mod 4,
when the $\Z^a_2$-twisted boundary conditions are imposed (see Fig.\ \ref{fig:fluxes-dyons}(b)). 

\item
The remaining anyons $\aU$, $\aV$, $\aW$ correspond to the fluxes of 
non-anomalous diagonal subgroups $\Z^{at}_2$ ($a=x,y,z$), while 
the anyons $\aI$, $\aJ$, $\aK$ are the bound states of these pure fluxes with appropriate pure
$\Z^x_2 \times Z^z_2$ charges. In the ED spectra, the anyons $m_{at}$ ($f_{at}$)
correspond to the sectors with  twisted momentum $k=0$ ($k=\pi$)
on a ring of size $L=\pm 1$ mod 4 with $\Z^a_2$-twisted boundary conditions, see Fig.\ 
\ref{fig:fluxes-dyons}(c)  (Fig.\ \ref{fig:fluxes-dyons}(d)).
\end{enumerate}

The gapped phases of Heisenberg chain with $\Z^x_2\times Z^z_2$
$\pi$-rotation and $\Z_L$ translation symmetries are classified by the 11 Lagrangian condensable
algebras \eqref{eq:D8 condensable algebras} of the symTO $\eD(D_8)$.  
These 11 Lagrangian algebras can be organized into three classes according to the $S_4$ automorphisms
\eqref{eq:S4 automorphisms} of the symTO $\eD(D_8)$\footnote{Here, we use the 
notation introduced in Ref.\ \cite{AW250321764} to organize the gapped phases according to
the automorphisms of the corresponding symTO.}:
\begin{align}
&\Big\{ \big(  \cA_{2,1}, \cA_{2,2}, \cA_{2,3}, \cA_{2,4}\big)^\text{SSB}_{2}\Big\},
\nonumber\\
&\Big\{ \big(  \cA_{3,1}, \cA_{3,2}, \cA_{3,3}, \cA_{3,4}, \cA_{3,5}, \cA_{3,6}\big)^\text{SSB}_{4}\Big\},
\nonumber\\
&\Big\{ \big(  \cA_{1,1}\big)^\text{SSB}_{8}\Big\}.
\end{align}
Each Lagrangian condensable algebra describe a gapped phase with 
spontaneous symmetry breaking (SSB), while the ones in the same class share the same 
ground state degeneracy that is given in the subscript.

The first class consisting of Lagrangian condensable algebras $(\cA_{2, \times})$
describes SSB states with twofold degeneracy. Each of these condensable algebras consist of 
three pure fluxes and a single pure charge anyon. Namely, the Lagrangian condensable algebras
$\cA_{2,1}$, $\cA_{2,2}$, $\cA_{2,3}$, and $\cA_{2,4}$ are obtained by condensing 
$\ak$, $\au$, $\av$ and $\aw$ anyons, respectively. Each of these anyons carry $\Z^t_2$ charge. 
Consequently, all these phases spontaneously break translation symmetry by one lattice site. 

The Lagrangian condensable algebra $\cA_{2,1}$ consists of condensations of $\aX$, $\aY$, $\aZ$. 
Hence, it preserves the $\Z^x_2\times Z^z_2$ symmetry while breaking  the translation symmetry. It describes 
the spin-singlet dimer phase (see Fig.\ \ref{SpinDimmer}). The remaining three Langrangian algebras 
involve condensation of anyons that preserve $\pi$ spin-rotation symmetry only along one axis. In addition, they preserve the symmetry
generated by $\pi$ spin-rotations along the other two axes followed by translation. 
For example, $\cA_{2,2}$ consists of $m_x$, $m_{yt}$, $m_{zt}$ which preserves $\Z^x_2\times \Z^{zt}_2$
subgroup. Hence, the three Langrangian condensable algebras $\cA_{2,2}$, $\cA_{2,3}$, and $\cA_{2,4}$ all describe
N\'eel phases (see Fig.\ \ref{SpinAF}) in which the staggered magnetizations $\sum_{j}(-1)^j S^x_j$, $\sum_{j}(-1)^j S^y_j$,
and $\sum_{j}(-1)^j S^z_j$ acquire non-vanishing expectation values, respectively.

The second class of Lagrangian condensable algebras $(\cA_{3, \times})$
describe SSB phases with fourfold degeneracy.  These Lagrangian condensable algebras corresponding
preserving 6 non-anomalous $\Z_2$ subgroups of the emanant $G_\text{IR}$ symmetry. 
The preserved symmetry for each Lagrangian condensable algebra is the one for which the corresponding 
flux anyon is condensed. For example, the Lagrangian condensable algebra $\cA_{3,1}$ realizes the 
gapped phase where only the $\Z^x_2$ symmetry is preserved. This ordered phase can be obtained from the
N\'eel phase along $x$-direction, realized by Lagrangian algebra $\cA_{2,2,}$ by condensing the charge anyon $\ax$. 
Because the anyon $\ax$ condenses there is an additional ferromagnetic ordering along $x$-direction. 
Hence, both uniform and staggered magnetizations, $\sum_{j} S^x_j$ and $\sum_{j}(-1)^j S^x_j$, 
acquire non-vanishing expectation values, i.e., this is a ferromagnetic phase 
which also breaks the translations by one site (see Fig.\ \ref{SpinAFF}).

The phases realized by condensable algebras $\cA_{3,4}, \cA_{3,5}, \cA_{3,6} $
break the spin-rotation symmetry fully while retaining a combination of $\pi$ spin-rotations and 
translation by one site as a symmetry. For example, the condensable algebra $\cA_{3,4}$ corresponds 
to a gapped phase that preserve only $\Z^{xt}_2$ subgroup that is generated by 
$\pi$ spin-rotation along $x$-axis followed by translations. This phase corresponds 
supports a ferromagnetic order in $S^x_j$ and antiferromagnetic order for 
$S^y_j$ and $S^z_j$ (see Fig.\ \ref{SpinAFCant}).

The remaining condensable algebra $\cA_{1,1}$ coincides the topological symmetry boundary
\eqref{eq:Sym boundary Heisenberg} and consists of condensing all pure symmetry charges. 
The phase induced by Lagrangian condensable algebra $\cA_{1,1}$ has eightfold degenerate ground states 
that break the emanant$G_\text{IR}$ symmetry completely. 
Since all charges are condensed, both uniform and staggered magnetizations
along all three spin directions,  $\sum_{j} S^a_j$ and $\sum_{j}(-1)^j S^a_j$ with
$a=x,y,z$, acquire non-vanishing expectation values.  Thus, the gapped phases realized by
condensable algebra $\cA_{1,1}$ is a ferromagnetic phase with non-collinear spins which also
breaks the translation symmetry (see Fig.\ \ref{SpinAFCantF}).

\subsection{Gapless CFT boundary of symTO $\eD(D_8)$}

We have so far only considered the gapped boundary conditions of the symTO
$\eD(D_8)$. One may wonder if a gapless boundary condition for this symTO is 
consistent with the Heisenberg chain which is described at low energies by
$SU(2)_1$ WZW theory. 

Let us consider the gapless boundary of symTO $\eD(D_8)$ that is obtained by 
condensing only the identity anyon~$\onebb$. This gapless boundary 
is described by multi component partition function with characters $\chi_\alpha^{\eD(D_8)}(\tau,\bar\tau)$, where
the subscript 
$\alpha$ label the 22 types of anyons of the quantum double $\eD(D_8)$.
Characters $\chi_\alpha^{\eD(D_8)}(\tau,\bar\tau)$ transform
according to the $S$ and $T$ matrices of $\eD(D_8)$
\begin{align}
\begin{split}
\chi_\alpha^{\eD(D_8)}(-1/\tau,-1/\bar\tau) &= D^{-1} S^{\eD(D_8)}_{\alpha\beta}\chi_\beta^{\eD(D_8)}(\tau,\bar\tau)
\\
\chi_\alpha^{\eD(D_8)}(\tau+1,\bar\tau+1) &= T^{\eD(D_8)}_{\alpha\beta}\chi_\beta^{\eD(D_8)}(\tau,\bar\tau)
\end{split}
\end{align}
We seek to find a CFT that gives rise to a partition function that is built out of characters
$\chi_\alpha^{\eD(D_8)}(\tau,\bar\tau)$ and all of its components are non-vanishing.

The low-energy degrees of freedom of the Heisenberg chain \eqref{eq:Ham Heisenberg}
is described by the ${c=1}$ CFT with $SU(2)$ level-1 Kac-Moody algebra $\mathfrak{su}(2)_1\times
\overline{\mathfrak{su}(2)}_1$.  This CFT is described by a 4-component partition function
with characters
$\chi_\al^{\mathfrak{su}(2)_1\times
\overline{\mathfrak{su}(2)}_1}(\tau,\bar\tau)$. The components are labeled by the four anyons   
$\al\in \{ (\onebb,\onebb), (\onebb,\bar s), (s,\onebb), (s,\bar s) \}$
of the double-semion topological order
and must transform according to the corresponding $S$ and $T$ matrices. 
We find that the components 
$\chi_\alpha^{\eD(D_8)}(\tau,\bar\tau)$
and
$\chi_\al^{\mathfrak{su}(2)_1\times\overline{\mathfrak{su}(2)}_1}(\tau,\bar\tau)$ 
of the two partition functions must have the following relation
\begin{align}
	\small
\label{su2D8}
\begin{split}
\chi_{(\onebb,\onebb)}^{\mathfrak{su}(2)_1\times
\overline{\mathfrak{su}(2)}_1} 
&=
\chi^{\eD(D_8)}_{\mathbb{1}} 
+  
\chi^{\eD(D_8)}_{\ax} 
+  
\chi^{\eD(D_8)}_{\ay} 
+  
\chi^{\eD(D_8)}_{\az}
\\
\chi_{(\onebb,\bar{s})}^{\mathfrak{su}(2)_1\times
\overline{\mathfrak{su}(2)}_1} &= 2\chi^{\eD(D_8)}_{\aD}
\\
\chi_{(s,\onebb)}^{\mathfrak{su}(2)_1\times
\overline{\mathfrak{su}(2)}_1} &= 2\chi^{\eD(D_8)}_{\aT}
\\
\chi_{(s,\bar{s})}^{\mathfrak{su}(2)_1\times
\overline{\mathfrak{su}(2)}_1} &= \chi^{\eD(D_8)}_{\ak} +  \chi^{\eD(D_8)}_{\au} +  \chi^{\eD(D_8)}_{\av} +  \chi^{\eD(D_8)}_{\aw}
\end{split}
\end{align}
In other words, the modular covariance of the characters $\chi_\alpha^{\eD(D_8)}(\tau,\bar\tau)$
ensure the modular covariance of the characters 
$\chi_\al^{\mathfrak{su}(2)_1\times\overline{\mathfrak{su}(2)}_1}(\tau,\bar\tau)$. 
The relation \eqref{su2D8} is the only where all
components $\chi_\al^{\mathfrak{su}(2)_1\times\overline{\mathfrak{su}(2)}_1}(\tau,\bar\tau)$ 
are non-vanishing.  The existence of the solution \eqref{su2D8} indicates that the
quantum double $\eD(D_8)$ is consistent as a symTO with the $\mathfrak{su}(2)_1\times
\overline{\mathfrak{su}(2)}_1$ CFT as its gapless boundary without non-trivial condensation.

\section{$SO(4)$ representation of $2d$ SPT and its connection to $1d$ Spin
Chain}

Our numerical calculation for spin-$\frac12$ Heisenberg chain reveals an
emergent $S_4$ symmetry since the $S_4$ automorphism of $D_8$ quantum double which 
leaves the symmetry boundary \eqref{eq:Sym boundary Heisenberg}.  This emergent $S_4$ symmetry suggests an emergent
$SO(4)$ symmetry of the spin-$\frac12$ Heisenberg chain.

To see the emergent $SO(4)$ symmetry, here we mostly follow the nonlinear
$\si$-model (NL$\si$M) description of a class of SPT states in two-dimensional ($2d$) space,
see Refs.~\cite{ashvinsenthil,xusenthil,xuclass}. For these SPT states, the 
NL$\si$M description has a maximal ``parent" $SO(4)$ symmetry such that 
various SPT states can be viewed as its descendants. 
We will connect the one-dimensional ($1d$) boundary of the parent SPT 
state to the spin-1/2 Heisenberg chain. 

The $2d$ bulk description of the $SO(4)$ SPT state is 
\begin{align}
 \cS_{2d} = \int d\tau d^2x \ \frac{ (\partial \vec{\phi})^2}{g^2} + \frac{\ii 2\pi}{\Omega_3} \epsilon_{abcd} \phi_a  \partial_\tau \phi_b \partial_x \phi_c \partial_y \phi_d. \label{bulk} 
\end{align}
 Here $\vec{\phi} = (\phi_1, \phi_2, \phi_3, \phi_4)$ is a unit ${SO}(4)$ vector. The $1d$ boundary of this state is (see also \cite{AffleckHaldane87})
\begin{align}
 \cS_{1d} &= \int d\tau dx \ \frac{1}{g} (\partial \vec{\phi})^2 
\nonumber\\
&\ \ \ \
+ \int_{u = 0}^1 du \ \frac{\ii 2\pi}{\Omega_3} \epsilon_{abcd} \phi_a  \partial_\tau \phi_b \partial_x \phi_c \partial_u \phi_d. 
\label{boundary} 
 \end{align}
This boundary theory has an $SO(4)$ anomaly $(k_R,k_L)=(1,-1)$.
It is known that the $(1+1)$D NL$\si$M with a WZW term flows to a CFT fixed
point $g = g^\ast$. At this fixed point, the $SO(4)$ symmetry factorizes into
SU(2)$_L$ and SU(2)$_R$ symmetries \cite{AffleckHaldane87}. To see the
$SU(2)_L \times SU(2)_R$ symmetry, we can define a matrix-valued field
$\Phi$ as 
\begin{subequations}
\label{sym} 
\begin{align}
 \Phi = \phi_4 \sigma^0 + \ii \phi_1 \sigma^1 + \ii \phi_2 \sigma^2 + \ii \phi_3 \sigma^3, 
\end{align}
where $\sigma^i$ with $i=1,2,3$ are Pauli matrices and $\sigma^0$ is the identity matrix. 
The $SU(2)_L \times SU(2)_R $ symmetry acts on the field $\Phi$ as
\begin{align}
SU(2)_L \times SU(2)_R : \ \ \Phi \to U_L \Phi U_R^\dagger,
\end{align}
\end{subequations}
where $U_L$ and $U_R$ are both $SU(2)$ matrices. 

Let us consider the SPT state realized by the CZX model \cite{CLW1141} as an
example. This state corresponds to the NL$\si$M model \eqref{boundary}, with the 
$\Z_2$ symmetry
\begin{align}
  \Z_2: \ \ \vec{\phi} \to - \vec{\phi},
\end{align}
which is the $\Z_2$ center of $SU(2)_R$.\ Although the $1d$ boundary theory is nonchiral, 
only the right-moving modes carry the $\Z_2$ charge. 
Hence, this $\Z_2$ CZX symmetry is anomalous and 
\emph{plays the same role as the translation of the spin-1/2 chain}. 

The field theory in Eq.~\eqref{boundary} also describes the $1d$ spin-1/2 chain, 
with the following identifications between the order parameter and the components of $\vec{\phi}$
\begin{align}
 \vec{\phi} \sim (n_x, n_y, n_z, V), 
\end{align}
 where $\vec{n} = (n_x, n_y, n_z)$ describes the three Neel order parameters, 
 while $V$ is the VBS order parameter. 
 Under translation by one lattice constant, both Neel and VBS order parameter change sign, hence $T: \vec{\phi} \to - \vec{\phi}$. 

There are multiple $\Z_2$ subgroups of ${SO}(4)$
whose generators act as
\begin{align}
 \Z_2^z :& \  n_x \to - n_x,  \ n_y \to - n_y,  \ n_z \to n_z, \ \ V \to V,
\nonumber \\
 \Z_2^x :& \  n_x \to n_x, \  n_y \to - n_y, \  n_z \to - n_z, \ V \to V,
\nonumber \\
 \Z_2^{yt} :& \  n_x \to n_x, \  n_y \to -n_y, \  n_z \to n_z,  \ V \to - V. 
\label{Z2s} 
\end{align}
Here,  $\Z_2^z$ and $\Z_2^x$ are subgroup of $SO(3)$ spin-rotation symmetry,
i.e., the $\pi$-rotations around  the $z$ and $x$ axis. The subgroup
$\Z_2^{yt}$ is the combination of translation $T$ and $\Z_2^y$,
the $\pi$-rotation around $y$ axis. The advantage of grouping symmetries this way is
that, the three $\Z_2$ subgroups are related by the $SO(4)$ symmetry, as well as an
automorphism of quantum double $\eD(D_8)$: 
\begin{align}
&(\aX ,\aZ ,\aV )(\aY ,\aU ,\aW )(\aI ,\aK ,\aM )(\aJ ,\aL ,\aN )
\nonumber\\
&(\ax ,\az ,\ay )(\ak ,\aw ,\au ).
\end{align}
In other words, the $\Z_2^z$, $\Z_2^x$, and $\Z_2^{yt}$ symmetry transformations
correspond to the symmetry-defects $\aX ,\aZ ,\aV$.  We remark that each of these
$\Z_2$ subgroups are non-anomalous, but there is type-III mixed anomaly between
them. If we view translation $T$ as a $\Z_2$ symmetry, it will have
self-anomaly, as it plays the same role as the $\Z_2$ symmetry in the CZX model. 

Similarly, the four $\Z_2$ subgroups that are related by the $SO(4)$ symmetry
\begin{align}
 \Z_2^z :& \  n_x \to - n_x,  \ n_y \to - n_y,  \ n_z \to n_z, \ \ V \to V, 
\nonumber \\
 \Z_2^x :& \  n_x \to n_x, \  n_y \to - n_y, \  n_z \to - n_z, \ V \to V,
\nonumber \\
 \Z_2^{zt} :& \  n_x \to n_x, \  n_y \to n_y, \  n_z \to -n_z,  \ V \to - V, 
\nonumber \\
 \Z_2^{xt} :& \  n_x \to -n_x, \  n_y \to n_y, \  n_z \to n_z,  \ V \to - V, 
\label{Z2s4} 
\end{align}
are also related by another automorphism of $\eD(D_8)$:
\begin{align}
& (\aX ,\aW ,\aU ,\aZ )(\aY ,\aV )(\aI ,\aN ,\aL ,\aK )
\nonumber\\
& (\aJ ,\aM )(\ax ,\az )(\ak ,\au ,\av ,\aw )
\end{align}
Thus the subgroups $ \Z_2^z $, $ \Z_2^x$, $ \Z_2^{zt}$, and $ \Z_2^{xt}$
correspond to the symmetry defects $\aX ,\aW ,\aU ,\aZ$.  We observe that the $S_4$
automorphism of the quantum double $\eD(D_8)$ is a subgroup of the $SO(4)$ symmetry.

In the field theory language, we  can consider a defect of translation symmetry $T$. 
This defect carries a spin-1/2 excitation 
(compatible with the numerical result of the difference between odd
and even lengths), which is precisely the CP$^1$ field $z_\alpha$: 
\begin{align}
 \vec{n} = z^\dagger \vec{\sigma} z. 
\end{align}
 Under translation $T$ and $\Z_2^{yt}$, $z$ transforms as 
\begin{align}
 T: \ \ z \to \ii \sigma^y z^\ast, \ \ \ \Z_2^{yt} : \ \ z \to  z^\ast. 
\end{align}
When acting on the spinon field $z_\alpha$, one has $T^2 = -1$, which is
compatible with the numerical result that the spin-1/2 excitation is at
quasi-momentum $\pi/2$.

The $\Z_2$ subgroups in Eq.~\eqref{Z2s} do not permute 
the components  of $\vec{\phi}$. One can also consider
other subgroups of $SO(4)$ that do permute them. 
For example, one such subgroup is the $\Z_4$ symmetry
with the action
\begin{align}
	\Z_4 & : & \phi_1 \to \phi_4, \ \ \phi_2 \to \phi_3, \ \ \phi_3 \to - \phi_2, \ \ \phi_4 \to - \phi_1.
\end{align}
 This $\Z_4$ subgroup swaps the $\Z_2^z$ and $\Z_2^{yt}$ subgroups, and 
 interestingly its generator squares to  $T$.

\section{Neighboring states  of spin-$1/2$ Heisenberg chain}
\label{neighbor}

In this section, we explore the neighboring phases to the
$SO(3)$-symmetric Heisenberg chain \eqref{eq:Ham Heisenberg}
whose low-energy description is given by the $\mathfrak{su}(2)_1\times
\overline{\mathfrak{su}(2)}_1$ CFT.  By neighboring phases, we mean only those that are induced by
interactions of the low-energy excitations, i.e., we assume that the high-energy excitations,
at the scale of microscopic exchange coupling, remain unchanged as we tune the interactions
between the low-energy excitations. What type of phases can be stabilized by tuning the interactions 
between the low-energy excitations on the antiferromagnetic spin-$1/2$ Heisenberg chain?

As shown in Sec.\ \ref{sec:symto}, when the full $SO(3)$ spin-rotation 
symmetry of the Hamiltonian \eqref{eq:Ham Heisenberg} is lowered down to 
discrete $\Z^x_2\times\Z^z_2$ subgroup, the emanant symmetries of the 
low-energy theory is described by the symTO $\eD(D_8)$. We expect that the emanant symmetries
to be sill described by the same symTO even if we tune the interactions between the low-energy excitations
as long as the high-energy excitations are not affected. Thus, the possible neighboring phases of the Heisenberg chain
are controlled by the emanant symTO $\eD(D_8)$.

The gaped phases with the emanant symmetry 
are classified by the Lagrangian condensable algebras of the symTO $\eD(D_8)$.  
We find that there are 11 gapped 
phases if we reduce $SO(3)$ spin-rotation symmetry down to $\Z_2^x\times \Z_2^z$ subgroup, recall Sec.\ \ref{sec:d8}. 
Using these 11  gapped phases with $\Z_2^x\times \Z_2^z$ symmetry,
it is possible to infer the neighboring phases for the Heisenberg model with the full $SO(3)$
spin-rotation symmetry. We shall keep in mind that in this case 
there is an emergent $SO(4)$ symmetry of
the low-energy theory described by the $\mathfrak{su}(2)_1\times
\overline{\mathfrak{su}(2)}_1$ CFT.
This emergent $SO(4)$ symmetry contains 
the exact emanant symTO $\eD(D_8)$ as well as the 
$SO(3)$ spin-rotation symmetry.  
In particular,  the $S_4$ automorphism of the  symTO $\eD(D_8)$ matches the $S_4$
subgroup of the emergent $SO(4)$ symmetry.

The three N\'eel phases corresponding to the Lagrangian condensable algebras 
$\cA_{2,2}$, $\cA_{2,3}$, $\cA_{2,4}$ are mapped to each other 
by the $SO(3)$ spin-rotation symmetry. 
When the full $SO(3)$ symmetry is present, 
the three would-be  phases corresponding to the $SO(3)$-symmetrization of 
these condensable algebras must 
spontaneously break a continuous $SO(3)$ symmetry
(without ferromagnetic order), as well as the emergent $SO(4)$ symmetry.  
However, this is not possible in
1+1D as the quantum fluctuations always restore the symmetry.
Hence, restoring $SO(3)$ spin-rotation symmetry gives rise to
a gapless phase described by the $\mathfrak{su}(2)_1\times
\overline{\mathfrak{su}(2)}_1$ CFT.  In other
words, the three gapped N\'eel phases for
reduced symmetry become a single gapless phase for the restored $SO(3)$
symmetry. 
Such a gapless phase has the exact emergent SO(4) symmetry in zero
energy limit. This is precisely the gapless phase 
realized by the Heisenberg model \eqref{eq:Ham Heisenberg}.

The gapped dimer phase realized by the 
Lagrangian condensable algebra $\cA_{2,1}$ 
remains to be a gapped phase when
the full $SO(3)$ spin-rotation symmetry is restored(see Fig.\  \ref{SpinDimmer}).
This condensable algebra is related to the condensable algebras 
$\cA_{2,2}$, $\cA_{2,3}$, and $\cA_{2,4}$ by the $S_4$ automorphism of 
the symTO  $\eD(D_8)$. Thus, when the full $SO(3)$ 
symmetry is restored, we expect them to be connected by
an emergent $SO(4)$ symmetry.  Therefore, the gapped  dimer phase 
spontaneously breaks the emergent $SO(4)$ symmetry.  
The dimer phase would 
have a linearly dispersing gapless mode if
spontaneous $SO(4)$ symmetry breaking if the symmetry was exact.  
However, the gapless mode can consistently acquire a gap 
since at the gap scale, the $SO(4)$ symmetry becomes approximate.

Can we get a gapped symmetric state, with the full $SO(3)$ and translation 
symmetry? The answer is no.
Since the $SO(3)$ symmetric model has at least the reduced spin
rotation symmetry and translation symmetry, we have showed that the gapped
symmetric state with the translation symmetry is impossible. This is the 
restatement of the LSM Theorem~\cite{lieb-1961, oshikawa-2000}
within the symTO framework (see also Ref.\ \cite{PAL250702036} for a more general treatment of 
LSM-type anomalies within the symTO construction). However, if the
gapped state has only spin rotation symmetry but breaks the translation
symmetry,then such a gapped is possible.

The three co-linear ferromagnetic phases that break the translation symmetry
realized by the Lagrangian condensable algebras 
$\cA_{3,1}$, $\cA_{3,2}$, $\cA_{3,3}$  (see Fig.\ \ref{SpinAFF}),
which are connected by the $SO(3)$ spin-rotation symmetry.  
When the full $SO(3)$ symmetry is restored,
these phases spontaneously break a continuous $SO(3)$ symmetry with a
ferromagnetic order.  In this case, the quantum fluctuations do not restore the
$SO(3)$ symmetry, which leads  to a $SO(3)$ SSB phase supporting gapless
spin waves with quadratic dispersion $\om \sim k^2$. In other words, the
three gapped phases $\cA_{3,1}$, $\cA_{3,2}$, $\cA_{3,3}$ 
for reduced symmetry become a single gapless ferromagnetic 
phase that breaks the translation and $SO(3)$ symmetries.

The three canted anti-ferromagnetic phases realized by 
Lagrangian condensable algebras $\cA_{3,4}$, $\cA_{3,5}$,
$\cA_{3,6}$ (where the canting induce a ferromagnetic order, see Fig.
\ref{SpinAFCant}) again are connected by the $SO(3)$ spin-rotation
symmetry.  To understand the dynamics of such a phase in the presence of the
full $SO(3)$ symmetry, consider a magnetic field to pin the ferromagnetic spin
direction.  In this case, we still have an anti-ferromagnetic order for spin 
degrees of freedom in the transverse direction and 
the $SO(3)$ symmetry is  explicitly broken down to $U(1)$ symmetry.  
The anti-ferromagnetic order for the transverse spin would break the
$U(1)$ symmetry. However, the quantum fluctuations restore 
the $U(1)$ symmetry back and lead to a gapless translation symmetric phase 
described by $\mathfrak{u}(1) \times
\overline{\mathfrak{u}(1)}$ CFT with linear dispersion $\om \sim |k|$.  
We stress that such a  gapless phase must be commensurate due to the 
anti-ferromagnetic order in the transverse direction~\footnote{
Here we use the following definitions for commensurate and incommensurate phases of \emph{translationally invariant models}~\cite{PokrovskyTalapov1979,OshikawaYamanakaAffleck1997,Totsuka1998}.
Naively, if a gapless mode in a phase carry crystal momentum that is a
rational fraction of Brillouin zone size, then we say 
that the phase is commensurate.
Otherwise, we say that the phase is incommensurate.  
More precisely, we consider the
crystal momenta of low-energy 
many-body excitations created by local operators.
If the crystal momenta of such excitations  are commensurate with the Brillouin zone size, then the
state is commensurate.  Otherwise,  the state is incommensurate.  An example
of an incommensurate state is a lattice gas of bosons with fixed number and 
incommensurate density.  
Another example is a 1+1D Fermi liquid which is incommensurate 
when $k_F\, \mathfrak{a}/2\pi$
is irrational, where $\mathrm{a}$ is the unit cell size.  
Gapped phases of translationally invariant Hamiltonians  
are always commensurate.}.

Now, if we remove the pinning magnetic field, 
the ferromagnetic spin order spontaneously breaks the 
$SO(3)$ symmetry down to a $U(1)$ symmetry.  In this
case, we expect to have two gapless modes: (i) 
an anti-ferromagnetic transverse
spin mode with  linear dispersion $\om \sim |k|$, (ii) 
and a ferromagnetic spin mode
with quadratic dispersion $\om \sim k^2$.

We turn to the remaining non-co-linear ferromagnetic phase with translation 
symmetry breaking (see Fig.\  \ref{SpinAFCantF}),
realized by the Lagrangian condensable algebra $\cA_{1,1}$. 
In the presence of the full $SO(3)$ symmetry, this phase  spontaneously 
breaks the continuous $SO(3)$ symmetry, and must be gapless.  
To understand such a phase, we again
consider a magnetic field to pin the ferromagnetic spin direction
which explicitly breaks the $SO(3)$ symmetry down to $U(1)$.
In this case,
we may have an ferromagnetic order for spin degrees of freedom in the 
transverse direction.  
The ferromagnetic order in the transverse direction spontaneously would break the 
$U(1)$ symmetry.  However,  the quantum fluctuations
restore the $U(1)$ symmetry and leads to a gapless translation symmetric 
phase described by $\mathfrak{u}(1) \times
\overline{\mathfrak{u}(1)}$ CFT with linear dispersion $\om \sim |k|$.  
We stress that such a gapless phase is an incommensurate phase due to
the ferromagnetic order in the transverse direction.  
We note that fluctuations of the ferromagnetic order in the transverse direction are described by a ferromagnetic XY
model with non-zero expectation value for the spin operator $\<S^z\>$.  
A ferromagnetic XY model is equivalent to a gas of conserved bosons, 
where the  boson density corresponds to average magnetization
$\<S^z\>$, which gives rise to an incommensurate phase.

Now, if we remove the pining magnetic field, the ferromagnetic spin order
spontaneously breaks the $SO(3)$ symmetry down to a $U(1)$ symmetry.  
In this case, we expect to have two gapless modes: (i) 
a ferromagnetic transverse spin mode with linear dispersion $\om \sim |k|$, 
(ii) and a ferromagnetic spin mode with quadratic dispersion $\om \sim k^2$.  
Such a gapless phase is an incommensurate
phase, due to the ferromagnetic order in the transverse direction.

When we only consider the $\Z^x_2\times \Z^z_2\subset SO(3)$,
Hamiltonian \eqref{eq:Ham Heisenberg} and its deformations 
also realize various gapless states. These states can be characterized
by the \emph{non-Lagrangian} condensable algebras of the $\eD(D_8)$-symTO.  Those
gapless states describe the critical points for the continuous phase
transitions between the 11 gapped phases with internal 
$\Z^x_2\times\Z^z_2$ and translation symmetries.  
Similarly, after when the full $SO(3)$ symmetry is recovered, 
the gapless state of the spin-$1/2$
Heisenberg chain also has various gapless neighboring states, describing the
critical points for the continuous phase transitions between the
 gapped and gapless phases discussed so far.
 
To summarize, the gapless phase of Heisenberg model \eqref{eq:Ham Heisenberg} has the
following neighboring phases or critical states:
\begin{enumerate}

\item
A gapped commensurate dimer phase from the condensable algebra
$\cA_{2,1}$, see Fig.\ \ref{SpinDimmer}.

\item
A gapless commensurate ferromagnetic phase with $\om \sim k^2$ that breaks
the translation symmetry from the condensable algebras 
$ \cA_{3,1}$, $\cA_{3,2}$, and $\cA_{3,3} $, see Fig.\ \ref{SpinAFF} for its classical picture.

\item
A gapless translationally-symmetric incommensurate 
ferromagnetic phase that supports a mode with 
linear dispersion $\om \sim
|k|$ and a mode  with quadratic dispersion $\om \sim k^2$ 
from the condensable algebras $\cA_{3,4}$, $\cA_{3,5}$, and  $\cA_{3,6}$,
see Fig.\ \ref{SpinAFCant} for its classical picture.

\item
A gapless incommensurate 
ferromagnetic phase that supports a mode with 
linear dispersion $\om \sim|k|$ 
and a mode with quadratic dispersion $\om \sim k^2$, 
as well as translation symmetry breaking
from the condensable algebra $ \cA_{1,1}$,
see Fig.\ \ref{SpinAFCantF} for its classical picture.

\item
Gapless states, describing the critical points for the stable or unstable
continuous phase transitions between the above mentioned gapped/gapless phases. 
In general, the critical points between commensurate phases support linear
dispersion $\om \sim |k|$ and are commensurate states.  
In contrast, the critical points
between commensurate phases and incommensurate phases support
non-linear dispersion $\om \sim |k|^\ga$, $\ga> 1$ 
and are commensurate states.

\end{enumerate}
In this paper, the term ``phase'' denotes states whose properties remain robust
under all symmetric perturbations. A stable continuous phase transition
corresponds to a critical point with only one symmetric relevant operator.
Conversely, an unstable continuous phase transition refers to a multi-critical
point with two or more symmetric relevant operators.

We remark that the dimer phase (Fig.\ \ref{SpinDimmer}) and the
anti-ferromagnetic phase (Fig.\ \ref{SpinAF}) are connected by the $S_4$
automorphisms.  Also, the ferromagnetic phase that breaks the translation
symmetry (Fig.\ \ref{SpinAFF}) and the ferromagnetic phase with
anti-ferromagnetic order for transverse spin (Fig.\ \ref{SpinAFCant}) are
connected by the $S_4$ automorphisms.  Those phases are also connected by the
approximate emergent $SO(4)$ symmetry, since the $S_4$ automorphisms is a part
of the  $SO(4)$ symmetry.

Since the $S_3$ subgroup of the $S_4$ automorphisms permute $x,y,z$, 
it is actually a part of $SO(3)$ spin rotation symmetry.  The three
spectra in Fig.\ \ref{fig:charge-anyons}(b,c,d) are connected by the $S_3$
automorphisms, and thus are identical due the $SO(3)$ spin rotation symmetry.
Similarly, the three spectra in Fig.\ \ref{fig:charge-anyons}(f,g,h) are
identical since they are connected by the  $S_3$ automorphisms.

The spectrum in Fig.\ \ref{fig:charge-anyons}(e) and the three spectra in
Fig.~\ref{fig:charge-anyons}(f,g,h) are connected by the $S_4$ automorphisms.
From the ED calculation, we find that they are identical at low energies only.  This
indicates that the $S_4$ automorphisms become
low-energy emergent symmetries of the spin-$\frac12$
Heisenberg chain.

We stress that for systems described by a symTO, the
automorphisms of the symTO, in general, are not the symmetry of the systems.
However, we can fine tune the systems to make the automorphisms to be the
symmetry of fine-tuned systems.  Also, if some of those systems are gapless,
the automorphisms can be the emergent symmetry of those gapless systems.
In our case, the gapless spin-$\frac12$ Heisenberg chain has an emanant symmetry
described by $\eD(D_8)$ symTO.  Our numerical calculation suggests that the
$S_4$ automorphism of $\eD(D_8)$ symTO is an emergent symmetry of the gapless
spin-$\frac12$ Heisenberg chain.

\section{Conclusion \& Outlook} \label{sec:conclusion}

We have shown that the antiferromagnetic Heisenberg chain has a low-energy
emanant symmetry described by $SO(3)\times \Z_2^t$ with a mixed anomaly.
Can this emanant symmetry be captured holographically by a 2+1D symTO? Guided
by the finite-group case, one might guess that the relevant symTO is simply a
2+1D $SO(3)\times \Z_2^t$ gauge theory with a topological term twisted by a
cocycle in $H^3(SO(3)\times \Z_2^t;\R/\Z)$:
\begin{align}
\ee^{\ii S}
\;=\;
\ee^{\ii\pi \!\int_{M_{2+1}} w_2 \smile a  + a\smile a \smile a  },
\end{align}
where $a\in Z^1(M^{2+1},\Z_2)$ is a $1$-cocycle describing the flat connection
of the $\Z_2^t$ gauge field, and $w_2\in H^2(M^{2+1},\Z_2)$ is the second
Stiefel-Whitney class of the $SO(3)$ bundle (equivalently the pullback of the
universal class in $H^2(BSO(3),\Z_2)$).

However, because $SO(3)$ is a \emph{continuous} group, the $SO(3)\times \Z_2^t$
gauge theory is gapless in the flat-connection limit; it is not a topological
field theory and therefore does not realize a topological order. From a purely
categorical viewpoint, one might nevertheless attempt to define a twisted
quantum double $\eD^{\om}_{SO(3)\times \Z_2^t}$ (a braided tensor category) for
the compact group $SO(3)\times \Z_2^t$, with $[\om]\in H^3(SO(3)\times
\Z_2^t;\R/\Z)$. If an appropriate notion of \emph{condensable algebra} can be
formulated for $\eD^{\om}_{SO(3)\times \Z_2^t}$, then its condensable algebras
could be used to compute and classify the phases compatible with the emanant
symTO $\eD^{\om}_{SO(3)\times \Z_2^t}$.

A pragmatic way to avoid continuous groups is to replace them by suitable
\emph{finite} subgroups of the continuous symmetry. This yields symTOs
described by non-degenerate braided fusion categories in the trivial Witt
class. After determining the phases via condensable algebras for these finite
symTOs, one can then enlarge the symmetry and study how the phases evolve under
symmetry enhancement. This is the route taken in this paper.

\section*{Acknowledgments}

We thank Leon Balents and Christopher Mudry for insightful discussions.
This research was supported in part by grant no.\ NSF PHY-2309135 to the Kavli Institute for Theoretical Physics (KITP).
This work was partially supported by NSF grant DMR2022428 and by the Simons Collaboration on Ultra-Quantum Matter, which is a grant from the Simons Foundation (651446, XGW).
ZJC acknowledges support from Kurt Forrest Foundation Fellowship. ZJC thanks IJM for support and encouragement during her undergraduate studies and beyond.
{\"OMA} is also supported by Swiss National Science Foundation (SNSF)
under grant no.\ P500PT-214429. CX is supported by the Simons Foundation through the Simons Investigator program.
%


\appendix

\begin{widetext}
\section{$S$-matrix and fusion rules of $D_8$ quantum double $\eD(D_8)$}
\label{fusionD8}

We present in this appendix the $S$ matrix and the fusion rules of the quantum double 
$\eD(D_8$). The $S$-matrix is given by
{
\begin{align}
S=
\left(
\begin{array}{cccc|cccc|ccc|ccc|ccc|ccc|cc}
1
& 1
& 1
& 1
& 1
& 1
& 1
& 1
& 2
& 2
& 2
& 2
& 2
& 2
& 2
& 2
& 2
& 2
& 2
& 2
& 2
& 2 \\ 
1
& 1
& 1
& 1
& 1
& 1
& 1
& 1
& 2
& -2
& -2
& 2
& -2
& -2
& 2
& -2
& -2
& 2
& -2
& -2
& 2
& 2 \\ 
1
& 1
& 1
& 1
& 1
& 1
& 1
& 1
& -2
& 2
& -2
& -2
& 2
& -2
& -2
& 2
& -2
& -2
& 2
& -2
& 2
& 2 \\ 
1
& 1
& 1
& 1
& 1
& 1
& 1
& 1
& -2
& -2
& 2
& -2
& -2
& 2
& -2
& -2
& 2
& -2
& -2
& 2
& 2
& 2 \\ 
\hline
1
& 1
& 1
& 1
& 1
& 1
& 1
& 1
& 2
& 2
& 2
& -2
& -2
& -2
& -2
& -2
& -2
& 2
& 2
& 2
& -2
& -2 \\ 
1
& 1
& 1
& 1
& 1
& 1
& 1
& 1
& 2
& -2
& -2
& -2
& 2
& 2
& -2
& 2
& 2
& 2
& -2
& -2
& -2
& -2 \\ 
1
& 1
& 1
& 1
& 1
& 1
& 1
& 1
& -2
& 2
& -2
& 2
& -2
& 2
& 2
& -2
& 2
& -2
& 2
& -2
& -2
& -2 \\ 
1
& 1
& 1
& 1
& 1
& 1
& 1
& 1
& -2
& -2
& 2
& 2
& 2
& -2
& 2
& 2
& -2
& -2
& -2
& 2
& -2
& -2 \\ 
\hline
2
& 2
& -2
& -2
& 2
& 2
& -2
& -2
& 4
&  
&  
&  
&  
&  
&  
&  
&  
& -4
&  
&  
&  
&   \\ 
2
& -2
& 2
& -2
& 2
& -2
& 2
& -2
&  
& 4
&  
&  
&  
&  
&  
&  
&  
&  
& -4
&  
&  
&   \\ 
2
& -2
& -2
& 2
& 2
& -2
& -2
& 2
&  
&  
& 4
&  
&  
&  
&  
&  
&  
&  
&  
& -4
&  
&   \\ 
\hline
2
& 2
& -2
& -2
& -2
& -2
& 2
& 2
&  
&  
&  
& 4
&  
&  
& -4
&  
&  
&  
&  
&  
&  
&   \\ 
2
& -2
& 2
& -2
& -2
& 2
& -2
& 2
&  
&  
&  
&  
& 4
&  
&  
& -4
&  
&  
&  
&  
&  
&   \\ 
2
& -2
& -2
& 2
& -2
& 2
& 2
& -2
&  
&  
&  
&  
&  
& 4
&  
&  
& -4
&  
&  
&  
&  
&   \\ 
\hline
2
& 2
& -2
& -2
& -2
& -2
& 2
& 2
&  
&  
&  
& -4
&  
&  
& 4
&  
&  
&  
&  
&  
&  
&   \\ 
2
& -2
& 2
& -2
& -2
& 2
& -2
& 2
&  
&  
&  
&  
& -4
&  
&  
& 4
&  
&  
&  
&  
&  
&   \\ 
2
& -2
& -2
& 2
& -2
& 2
& 2
& -2
&  
&  
&  
&  
&  
& -4
&  
&  
& 4
&  
&  
&  
&  
&   \\ 
\hline
2
& 2
& -2
& -2
& 2
& 2
& -2
& -2
& -4
&  
&  
&  
&  
&  
&  
&  
&  
& 4
&  
&  
&  
&   \\ 
2
& -2
& 2
& -2
& 2
& -2
& 2
& -2
&  
& -4
&  
&  
&  
&  
&  
&  
&  
&  
& 4
&  
&  
&   \\ 
2
& -2
& -2
& 2
& 2
& -2
& -2
& 2
&  
&  
& -4
&  
&  
&  
&  
&  
&  
&  
&  
& 4
&  
&   \\ 
\hline
2
& 2
& 2
& 2
& -2
& -2
& -2
& -2
&  
&  
&  
&  
&  
&  
&  
&  
&  
&  
&  
&  
& -4
& 4 \\ 
2
& 2
& 2
& 2
& -2
& -2
& -2
& -2
&  
&  
&  
&  
&  
&  
&  
&  
&  
&  
&  
&  
& 4
& -4 \\ 
\end{array}
\right).
\end{align}
}

The fusion rules are given by the following table.

\setlength\tabcolsep{1pt}
\setlength\extrarowheight{1pt}
\centerline{
\begin{adjustbox}{angle=90}

\scriptsize
\begin{tabular}{ |c||c|c|c|c|c|c|c|c|p{12mm}|p{12mm}|p{12mm}|p{12mm}|p{12mm}|p{12mm}|p{12mm}|p{12mm}|p{12mm}|p{12mm}|p{12mm}|p{12mm}|p{12mm}|p{12mm}|}
 \hline 
$\otimes$  & $\onebb$  & $\ax$  & $\ay$  & $\az$  & $\ak$  & $\au$  & $\av$  & $\aw$  & $\aX$  & $\aY$  & $\aZ$  & $\aU$  & $\aV$  & $\aW$  & $\aI$  & $\aJ$  & $\aK$  & $\aL$  & $\aM$  & $\aN$  & $\aT$  & $\aD$ \\ 
\hline 
 \hline 
$\onebb$  & $ \onebb$  & $ \ax$  & $ \ay$  & $ \az$  & $ \ak$  & $ \au$  & $ \av$  & $ \aw$  & $ \aX$  & $ \aY$  & $ \aZ$  & $ \aU$  & $ \aV$  & $ \aW$  & $ \aI$  & $ \aJ$  & $ \aK$  & $ \aL$  & $ \aM$  & $ \aN$  & $ \aT$  & $ \aD$  \\ 
 \hline 
$\ax$  & $ \ax$  & $ \onebb$  & $ \az$  & $ \ay$  & $ \au$  & $ \ak$  & $ \aw$  & $ \av$  & $ \aX$  & $ \aM$  & $ \aN$  & $ \aU$  & $ \aJ$  & $ \aK$  & $ \aI$  & $ \aV$  & $ \aW$  & $ \aL$  & $ \aY$  & $ \aZ$  & $ \aT$  & $ \aD$  \\ 
 \hline 
$\ay$  & $ \ay$  & $ \az$  & $ \onebb$  & $ \ax$  & $ \av$  & $ \aw$  & $ \ak$  & $ \au$  & $ \aL$  & $ \aY$  & $ \aN$  & $ \aI$  & $ \aV$  & $ \aK$  & $ \aU$  & $ \aJ$  & $ \aW$  & $ \aX$  & $ \aM$  & $ \aZ$  & $ \aT$  & $ \aD$  \\ 
 \hline 
$\az$  & $ \az$  & $ \ay$  & $ \ax$  & $ \onebb$  & $ \aw$  & $ \av$  & $ \au$  & $ \ak$  & $ \aL$  & $ \aM$  & $ \aZ$  & $ \aI$  & $ \aJ$  & $ \aW$  & $ \aU$  & $ \aV$  & $ \aK$  & $ \aX$  & $ \aY$  & $ \aN$  & $ \aT$  & $ \aD$  \\ 
 \hline 
$\ak$  & $ \ak$  & $ \au$  & $ \av$  & $ \aw$  & $ \onebb$  & $ \ax$  & $ \ay$  & $ \az$  & $ \aX$  & $ \aY$  & $ \aZ$  & $ \aI$  & $ \aJ$  & $ \aK$  & $ \aU$  & $ \aV$  & $ \aW$  & $ \aL$  & $ \aM$  & $ \aN$  & $ \aD$  & $ \aT$  \\ 
 \hline 
$\au$  & $ \au$  & $ \ak$  & $ \aw$  & $ \av$  & $ \ax$  & $ \onebb$  & $ \az$  & $ \ay$  & $ \aX$  & $ \aM$  & $ \aN$  & $ \aI$  & $ \aV$  & $ \aW$  & $ \aU$  & $ \aJ$  & $ \aK$  & $ \aL$  & $ \aY$  & $ \aZ$  & $ \aD$  & $ \aT$  \\ 
 \hline 
$\av$  & $ \av$  & $ \aw$  & $ \ak$  & $ \au$  & $ \ay$  & $ \az$  & $ \onebb$  & $ \ax$  & $ \aL$  & $ \aY$  & $ \aN$  & $ \aU$  & $ \aJ$  & $ \aW$  & $ \aI$  & $ \aV$  & $ \aK$  & $ \aX$  & $ \aM$  & $ \aZ$  & $ \aD$  & $ \aT$  \\ 
 \hline 
$\aw$  & $ \aw$  & $ \av$  & $ \au$  & $ \ak$  & $ \az$  & $ \ay$  & $ \ax$  & $ \onebb$  & $ \aL$  & $ \aM$  & $ \aZ$  & $ \aU$  & $ \aV$  & $ \aK$  & $ \aI$  & $ \aJ$  & $ \aW$  & $ \aX$  & $ \aY$  & $ \aN$  & $ \aD$  & $ \aT$  \\ 
 \hline 
$\aX$  & $ \aX$  & $ \aX$  & $ \aL$  & $ \aL$  & $ \aX$  & $ \aX$  & $ \aL$  & $ \aL$  & $ \onebb \oplus \ax \oplus \ak \oplus \au$  & $ \aZ \oplus \aN$  & $ \aY \oplus \aM$  & $ \aT \oplus \aD$  & $ \aW \oplus \aK$  & $ \aV \oplus \aJ$  & $ \aT \oplus \aD$  & $ \aW \oplus \aK$  & $ \aV \oplus \aJ$  & $ \ay \oplus \az \oplus \av \oplus \aw$  & $ \aZ \oplus \aN$  & $ \aY \oplus \aM$  & $ \aU \oplus \aI$  & $ \aU \oplus \aI$  \\ 
 \hline 
$\aY$  & $ \aY$  & $ \aM$  & $ \aY$  & $ \aM$  & $ \aY$  & $ \aM$  & $ \aY$  & $ \aM$  & $ \aZ \oplus \aN$  & $ \onebb \oplus \ay \oplus \ak \oplus \av$  & $ \aX \oplus \aL$  & $ \aW \oplus \aK$  & $ \aT \oplus \aD$  & $ \aU \oplus \aI$  & $ \aW \oplus \aK$  & $ \aT \oplus \aD$  & $ \aU \oplus \aI$  & $ \aZ \oplus \aN$  & $ \ax \oplus \az \oplus \au \oplus \aw$  & $ \aX \oplus \aL$  & $ \aV \oplus \aJ$  & $ \aV \oplus \aJ$  \\ 
 \hline 
$\aZ$  & $ \aZ$  & $ \aN$  & $ \aN$  & $ \aZ$  & $ \aZ$  & $ \aN$  & $ \aN$  & $ \aZ$  & $ \aY \oplus \aM$  & $ \aX \oplus \aL$  & $ \onebb \oplus \az \oplus \ak \oplus \aw$  & $ \aV \oplus \aJ$  & $ \aU \oplus \aI$  & $ \aT \oplus \aD$  & $ \aV \oplus \aJ$  & $ \aU \oplus \aI$  & $ \aT \oplus \aD$  & $ \aY \oplus \aM$  & $ \aX \oplus \aL$  & $ \ax \oplus \ay \oplus \au \oplus \av$  & $ \aW \oplus \aK$  & $ \aW \oplus \aK$  \\ 
 \hline 
$\aU$  & $ \aU$  & $ \aU$  & $ \aI$  & $ \aI$  & $ \aI$  & $ \aI$  & $ \aU$  & $ \aU$  & $ \aT \oplus \aD$  & $ \aW \oplus \aK$  & $ \aV \oplus \aJ$  & $ \onebb \oplus \ax \oplus \av \oplus \aw$  & $ \aZ \oplus \aN$  & $ \aY \oplus \aM$  & $ \ay \oplus \az \oplus \ak \oplus \au$  & $ \aZ \oplus \aN$  & $ \aY \oplus \aM$  & $ \aT \oplus \aD$  & $ \aW \oplus \aK$  & $ \aV \oplus \aJ$  & $ \aX \oplus \aL$  & $ \aX \oplus \aL$  \\ 
 \hline 
$\aV$  & $ \aV$  & $ \aJ$  & $ \aV$  & $ \aJ$  & $ \aJ$  & $ \aV$  & $ \aJ$  & $ \aV$  & $ \aW \oplus \aK$  & $ \aT \oplus \aD$  & $ \aU \oplus \aI$  & $ \aZ \oplus \aN$  & $ \onebb \oplus \ay \oplus \au \oplus \aw$  & $ \aX \oplus \aL$  & $ \aZ \oplus \aN$  & $ \ax \oplus \az \oplus \ak \oplus \av$  & $ \aX \oplus \aL$  & $ \aW \oplus \aK$  & $ \aT \oplus \aD$  & $ \aU \oplus \aI$  & $ \aY \oplus \aM$  & $ \aY \oplus \aM$  \\ 
 \hline 
$\aW$  & $ \aW$  & $ \aK$  & $ \aK$  & $ \aW$  & $ \aK$  & $ \aW$  & $ \aW$  & $ \aK$  & $ \aV \oplus \aJ$  & $ \aU \oplus \aI$  & $ \aT \oplus \aD$  & $ \aY \oplus \aM$  & $ \aX \oplus \aL$  & $ \onebb \oplus \az \oplus \au \oplus \av$  & $ \aY \oplus \aM$  & $ \aX \oplus \aL$  & $ \ax \oplus \ay \oplus \ak \oplus \aw$  & $ \aV \oplus \aJ$  & $ \aU \oplus \aI$  & $ \aT \oplus \aD$  & $ \aZ \oplus \aN$  & $ \aZ \oplus \aN$  \\ 
 \hline 
$\aI$  & $ \aI$  & $ \aI$  & $ \aU$  & $ \aU$  & $ \aU$  & $ \aU$  & $ \aI$  & $ \aI$  & $ \aT \oplus \aD$  & $ \aW \oplus \aK$  & $ \aV \oplus \aJ$  & $ \ay \oplus \az \oplus \ak \oplus \au$  & $ \aZ \oplus \aN$  & $ \aY \oplus \aM$  & $ \onebb \oplus \ax \oplus \av \oplus \aw$  & $ \aZ \oplus \aN$  & $ \aY \oplus \aM$  & $ \aT \oplus \aD$  & $ \aW \oplus \aK$  & $ \aV \oplus \aJ$  & $ \aX \oplus \aL$  & $ \aX \oplus \aL$  \\ 
 \hline 
$\aJ$  & $ \aJ$  & $ \aV$  & $ \aJ$  & $ \aV$  & $ \aV$  & $ \aJ$  & $ \aV$  & $ \aJ$  & $ \aW \oplus \aK$  & $ \aT \oplus \aD$  & $ \aU \oplus \aI$  & $ \aZ \oplus \aN$  & $ \ax \oplus \az \oplus \ak \oplus \av$  & $ \aX \oplus \aL$  & $ \aZ \oplus \aN$  & $ \onebb \oplus \ay \oplus \au \oplus \aw$  & $ \aX \oplus \aL$  & $ \aW \oplus \aK$  & $ \aT \oplus \aD$  & $ \aU \oplus \aI$  & $ \aY \oplus \aM$  & $ \aY \oplus \aM$  \\ 
 \hline 
$\aK$  & $ \aK$  & $ \aW$  & $ \aW$  & $ \aK$  & $ \aW$  & $ \aK$  & $ \aK$  & $ \aW$  & $ \aV \oplus \aJ$  & $ \aU \oplus \aI$  & $ \aT \oplus \aD$  & $ \aY \oplus \aM$  & $ \aX \oplus \aL$  & $ \ax \oplus \ay \oplus \ak \oplus \aw$  & $ \aY \oplus \aM$  & $ \aX \oplus \aL$  & $ \onebb \oplus \az \oplus \au \oplus \av$  & $ \aV \oplus \aJ$  & $ \aU \oplus \aI$  & $ \aT \oplus \aD$  & $ \aZ \oplus \aN$  & $ \aZ \oplus \aN$  \\ 
 \hline 
$\aL$  & $ \aL$  & $ \aL$  & $ \aX$  & $ \aX$  & $ \aL$  & $ \aL$  & $ \aX$  & $ \aX$  & $ \ay \oplus \az \oplus \av \oplus \aw$  & $ \aZ \oplus \aN$  & $ \aY \oplus \aM$  & $ \aT \oplus \aD$  & $ \aW \oplus \aK$  & $ \aV \oplus \aJ$  & $ \aT \oplus \aD$  & $ \aW \oplus \aK$  & $ \aV \oplus \aJ$  & $ \onebb \oplus \ax \oplus \ak \oplus \au$  & $ \aZ \oplus \aN$  & $ \aY \oplus \aM$  & $ \aU \oplus \aI$  & $ \aU \oplus \aI$  \\ 
 \hline 
$\aM$  & $ \aM$  & $ \aY$  & $ \aM$  & $ \aY$  & $ \aM$  & $ \aY$  & $ \aM$  & $ \aY$  & $ \aZ \oplus \aN$  & $ \ax \oplus \az \oplus \au \oplus \aw$  & $ \aX \oplus \aL$  & $ \aW \oplus \aK$  & $ \aT \oplus \aD$  & $ \aU \oplus \aI$  & $ \aW \oplus \aK$  & $ \aT \oplus \aD$  & $ \aU \oplus \aI$  & $ \aZ \oplus \aN$  & $ \onebb \oplus \ay \oplus \ak \oplus \av$  & $ \aX \oplus \aL$  & $ \aV \oplus \aJ$  & $ \aV \oplus \aJ$  \\ 
 \hline 
$\aN$  & $ \aN$  & $ \aZ$  & $ \aZ$  & $ \aN$  & $ \aN$  & $ \aZ$  & $ \aZ$  & $ \aN$  & $ \aY \oplus \aM$  & $ \aX \oplus \aL$  & $ \ax \oplus \ay \oplus \au \oplus \av$  & $ \aV \oplus \aJ$  & $ \aU \oplus \aI$  & $ \aT \oplus \aD$  & $ \aV \oplus \aJ$  & $ \aU \oplus \aI$  & $ \aT \oplus \aD$  & $ \aY \oplus \aM$  & $ \aX \oplus \aL$  & $ \onebb \oplus \az \oplus \ak \oplus \aw$  & $ \aW \oplus \aK$  & $ \aW \oplus \aK$  \\ 
 \hline 
$\aT$  & $ \aT$  & $ \aT$  & $ \aT$  & $ \aT$  & $ \aD$  & $ \aD$  & $ \aD$  & $ \aD$  & $ \aU \oplus \aI$  & $ \aV \oplus \aJ$  & $ \aW \oplus \aK$  & $ \aX \oplus \aL$  & $ \aY \oplus \aM$  & $ \aZ \oplus \aN$  & $ \aX \oplus \aL$  & $ \aY \oplus \aM$  & $ \aZ \oplus \aN$  & $ \aU \oplus \aI$  & $ \aV \oplus \aJ$  & $ \aW \oplus \aK$  & $ \onebb \oplus \ax \oplus \ay \oplus \az$  & $ \ak \oplus \au \oplus \av \oplus \aw$  \\ 
 \hline 
$\aD$  & $ \aD$  & $ \aD$  & $ \aD$  & $ \aD$  & $ \aT$  & $ \aT$  & $ \aT$  & $ \aT$  & $ \aU \oplus \aI$  & $ \aV \oplus \aJ$  & $ \aW \oplus \aK$  & $ \aX \oplus \aL$  & $ \aY \oplus \aM$  & $ \aZ \oplus \aN$  & $ \aX \oplus \aL$  & $ \aY \oplus \aM$  & $ \aZ \oplus \aN$  & $ \aU \oplus \aI$  & $ \aV \oplus \aJ$  & $ \aW \oplus \aK$  & $ \ak \oplus \au \oplus \av \oplus \aw$  & $ \onebb \oplus \ax \oplus \ay \oplus \az$  \\ 
 \hline 
\end{tabular}
\end{adjustbox}
}
\end{widetext}

\bibliography{../../../bib/all,../../../bib/publst,refs}

\end{document}